\pdfoutput=1
\documentclass[cernpreprint, atlasdraft=false, texlive=2023, UKenglish, texmf, orcidlogo]{atlasdoc}
 
\usepackage{atlaspackage}
\usepackage{atlasbiblatex}
 
\usepackage{atlasphysics}

\usepackage{amsmath}
\usepackage{rotating}
\usepackage{graphicx}
\usepackage{multirow}

\newcolumntype{~}{>{\global\let\currentrowstyle\relax}}
\newcolumntype{^}{>{\currentrowstyle}}
\newcommand{\rowstyle}[1]{\gdef\currentrowstyle{#1}
#1\ignorespaces
}
 
\mathchardef\mhyphen="2D
 
\graphicspath{{logos/}{figures/}}
 
\usepackage{ANA-TOPQ-2019-30-PAPER-defs}

\AtlasTitle{Measurement of the total and differential cross-sections of $t\bar{t}W$ production in $pp$ collisions at $\sqrt{s}=13$~TeV with the ATLAS detector}

\AtlasAbstract{
Measurements of inclusive and differential production cross-sections of a top-quark--top-antiquark pair in association with a $W$ boson ($t\bar{t}W$) are presented. They are performed by targeting final states with two same-sign or three isolated leptons (electrons or muons) and are based on $\sqrt{s} = 13\,\mathrm{TeV}$ proton--proton collision data with an integrated luminosity of $140\,\mathrm{fb}^{-1}$, recorded from $2015$ to $2018$ with the ATLAS detector at the Large Hadron Collider. The inclusive $t\bar{t}W$ production cross-section is measured to be $880 \pm 80$~fb, compared to a reference theoretical prediction of $745 \pm 50\,\textrm{(scale)} \pm 13\,\textrm{(2-loop approx.)} \pm 19\,\textrm{(PDF, \alphas)}$~fb. Differential \xs measurements characterise this process in detail for the first time. Several particle-level observables are compared with a variety of theoretical predictions, which generally agree well with the normalised differential \xs results. Additionally, the relative charge asymmetry of ${\ttW}^{+}$ and ${\ttW}^{-}$ is measured inclusively to be $\rqasym = 0.33 \pm 0.05$, in very good agreement with the theoretical prediction of $0.322 \pm 0.003\,\mathrm{(scale)} \pm 0.007\,\mathrm{(PDF)}$, as well as differentially.
}
 
\author{The ATLAS Collaboration}
 
\AtlasRefCode{TOPQ-2019-30}
 
\PreprintIdNumber{CERN-EP-2023-260}

\AtlasJournal{JHEP}
\AtlasJournalRef{\JHEP 05 (2024) 131}
\AtlasDOI{10.1007/JHEP05(2024)131}

\hypersetup{pdftitle={ATLAS document},pdfauthor={The ATLAS Collaboration}}
 
\begin{document}
 
\maketitle

\section{Introduction}
\label{sec:intro}

Two of the primary goals of the Large Hadron Collider (LHC) physics
programme are to look for signs of new physics beyond the Standard
Model (SM) and to determine the nature of electroweak (EWK) symmetry
breaking. Despite thorough scrutiny, no significant deviation from the
SM has been observed at the LHC. This strongly motivates
further testing of the SM by measuring rare processes more precisely,
examining final states where only a few SM processes contribute, and
probing the Higgs boson in more detail. The study of
associated production of a top--antitop quark pair and a
$W$ boson, \ttW, is strongly connected to each of these
goals. With one of the heaviest SM final states accessible at the LHC, and a
correspondingly small production \xs, this process was first observed
in \RunOne~\cite{TOPQ-2013-05,CMS-TOP-14-021} but a precise inclusive
measurement and the first differential measurements became possible
only with the full \RunTwo dataset.
 
Beyond the inherent interest in measuring such a rare process more
precisely, better understanding of \ttW production is also important
because it is a key background in many searches and other measurements at
the LHC. It is one of the few SM processes that is an irreducible source of
same-sign dilepton pairs.  The rarity of these final states is exploited in many
searches for new physics in extensions of the SM~\cite{EXOT-2016-16,SUSY-2018-09,CMS-SUS-19-008}
where \ttW is a major background.
Moreover, \ttW production is the dominant background in many
measurements of other rare processes, such as \ttH and \tttt production,
that are important when probing the top-quark Yukawa coupling and EWK symmetry breaking.
The precision, accuracy and understanding of \ttW modelling is one of the main limitations on
the sensitivity of these measurements.
 
Further motivation for precise measurements of \ttW production comes
from tensions between the data and SM predictions observed in previous LHC
measurements. Early measurements of \ttW production by the ATLAS and
CMS Collaborations at $\sqrt{s}=8$~TeV~\cite{TOPQ-2013-05,CMS-TOP-14-021} were in agreement with the SM,
but are less precise. More
recent direct measurements of \ttW production at
$\sqrt{s}=13$~TeV~\cite{TOPQ-2015-22,TOPQ-2016-11,CMS-TOP-17-005,CMS-TOP-21-011},
and indirect measurements of \ttW in analyses
targeting \ttH~\cite{CMS-HIG-19-008}
and \tttt~\cite{TOPQ-2021-08,CMS-TOP-22-013} production, have
consistently observed \ttW yields larger than the SM predictions.
 
\begin{figure}[t]
\centering
\subfloat[]{\includegraphics[width=0.24\textwidth]{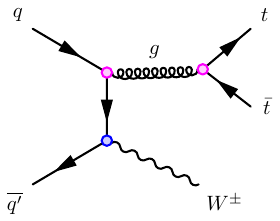}\label{fig:ttw-lo}}
\subfloat[]{\includegraphics[width=0.24\textwidth]{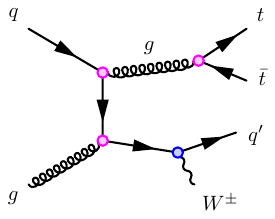}\label{fig:ttw-nlo1}}
\subfloat[]{\includegraphics[width=0.24\textwidth]{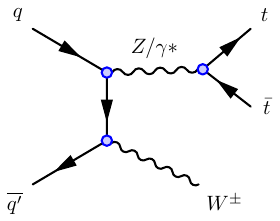}\label{fig:ttw-lo3}}
\subfloat[]{\includegraphics[width=0.24\textwidth]{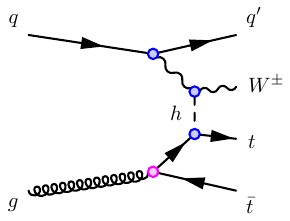}\label{fig:ttw-nlo3}}
\caption{
Illustrative Feynman diagrams for the dominant production modes of \ttW: (a) the LO contribution ($\alpha\alphas^2$), (b) a real emission diagram from the NLO QCD contribution ($\alpha\alphas^3$), (c) the tree-level EWK contribution ($\alpha^3$), and (d) a representative diagram of the combined NLO QCD and EWK contributions ($\alpha^3\alphas$). The pink circles correspond to QCD couplings and the blue circles correspond to EWK couplings. }
\label{fig:feynmandiagrams}
\end{figure}

Although \ttW production superficially appears to be a simple process,
and one governed by well-known SM couplings, the quark--antiquark
initial state at leading order (LO) in the strong coupling constant leads to
a rich array of phenomena. In particular, the process features
charge-asymmetric production (of $\ttW^+$ and $\ttW^-$) from the
parton distribution functions and unusually complex higher-order
quantum chromodynamics (QCD) and EWK corrections.  One consequence of
the complex EWK corrections is the additional production mechanisms
that they enable, such as $tW$-scattering~\cite{Dror:2015nkp}. This
arises through EWK \ttW{}+jet production, which has embedded within it
an effective $tW \to tW$ vertex that is interesting to understand in
the SM but can also be probed for sensitivity to physics beyond the SM,
for example in the context of effective field theory operators~\cite{Maltoni:2019aot}.
Figure~\ref{fig:feynmandiagrams} shows the dominant production modes for the
\ttW process, including additional QCD and EWK corrections.
Therefore, measurements of \ttW production are key in validating and improving
the challenging theoretical predictions of this process, which has
attracted much recent interest from the theoretical community, as
described in Section~\ref{sec:datamc}.

A theoretical prediction at next-to-next-to-leading order (NNLO) in
QCD, including next-to-leading order (NLO) EWK corrections, from
Ref.~\cite{Buonocore_2023} of $\sigma(\ttW) = 745\pm50\;\textrm{(scale)}\pm 13\;\textrm{(2-loop approx.)}\pm 19\;\textrm{(PDF, \alphas)}$~fb, is used as a reference \xs. It supersedes previous calculations, Ref.~\cite{YR4,Frederix:2017wme,Kulesza:2018tqz,Broggio:2019ewu,Kulesza:2020nfh,Bevilacqua:2020pzy,Denner:2020hgg,Bevilacqua:2020srb,FebresCordero:2021kcc,Denner:2021hqi,Bevilacqua:2021tzp,Frederix:2021agh,Ferencz:2023fso}.

The different contributions to the total \ttW \xs are shown in
Table~\ref{tab:intro_ttW}. Considering only QCD corrections, the
predicted \xs is $\sigma(\ttW)^{\textrm{QCD}} =
711\;^{+35}_{-46}\,\textrm{(scale)} \pm 14\,\textrm{(2-loop
approx.)}$~fb, while NLO EWK corrections give a total positive
contribution of 4.9\%~\cite{Buonocore_2023}. Virtual EWK
corrections are negative and 2.4\% of the
$\sigma(\ttW)^{\textrm{QCD}}$ value, and the remaining EWK contribution
(including $tW$-scattering, as shown in
Figure~\ref{fig:feynmandiagrams}(d)) gives a 6.9\% positive correction
to the $\sigma(\ttW)^{\textrm{QCD}}$
calculation~\cite{Frederix:2021agh}. The NLO multileg-merged FxFx
prediction from Ref.~\cite{Frederix:2021agh} captures a significant
fraction of the NNLO QCD contributions to the \xs.

\begin{table}[htbp]
\begin{center}
\caption{Summary of theoretical predictions with NNLO precision in the strong coupling~\cite{Buonocore_2023} and using FxFx NLO multijet merging~\cite{Frederix:2021agh}, both including NLO EWK corrections. The first uncertainty is due to variations of the chosen renormalisation and factorisation scales. Where there is a second contribution to the uncertainty, this corresponds to the approximation used in the 2-loop calculation. Uncertainties due to the choice of PDF and \alphas are omitted. \label{tab:intro_ttW}}
\renewcommand*\arraystretch{1.3}
\begin{tabular}{c|c|c|c}
\hline \hline
\multicolumn{2}{c|}{NNLO~\cite{Buonocore_2023}} & \multicolumn{2}{c}{FxFx~\cite{Frederix:2021agh}}  \\
\hline
Order & $\sigma$~[fb]  &   Order & $\sigma$~[fb]  \\
\hline
LO QCD:   $\alphas^2\alpha$    & $420\;^{+106}_{-79}$ & & \\
NLO QCD:  $+\,\alphas^3\alpha$ & $622\;^{+79}_{-72}$ & \ttW{}+0,1,2j@NLO & $691\;^{+66}_{-74}$\\
NNLO QCD: $+\,\alphas^4\alpha$ & $711\;^{+35}_{-46}\pm 14$ & &  \\
\hline
\multirow{ 2}{*}{NLO EWK: $+\,\alphas\alpha^3+ \alphas^2\alpha^2 + \alpha^4$} & \multirow{ 2}{*}{$745 \pm 50 \pm 13$} & $+\,\alphas\alpha^3$   & $739\;^{+75}_{-81}$ \\
& & $+\,\alphas^2\alpha^2 + \alpha^4$                                                     & $722\;^{+70}_{-78}$ \\
\hline  \hline
\end{tabular}
\end{center}
\end{table}

This paper presents measurements of inclusive and differential \xs{}s of \ttW production
at $\sqrt{s}=13$~TeV, including measurements of the \ttW relative charge asymmetry.
This is the first measurement of differential \xs{}s of \ttW production at the LHC.
The measurements are performed by analysing final states with two same-sign electric-charge leptons (\llSS)
and three leptons (\lll), where $\ell$ denotes electrons or muons. The rest of the paper is
structured as follows. Section~\ref{sec:detector} describes the ATLAS
detector, Section~\ref{sec:datamc} provides an overview of the data
and simulated samples used in the measurements, Section~\ref{sec:objects}
details the object reconstruction and selection and
Section~\ref{sec:events} defines the analysis strategy.
Section~\ref{sec:background} provides a description of the strategy used to
estimate non-prompt-lepton backgrounds. An overview of the systematic
uncertainties is given in Section~\ref{sec:systematics}. The results
are presented in Section~\ref{sec:results} and conclusions are given
in Section~\ref{sec:conclusion}.


\section{ATLAS detector}
\label{sec:detector}

\newcommand{\AtlasCoordFootnote}{
ATLAS uses a right-handed coordinate system with its origin at the nominal interaction point (IP)
in the centre of the detector and the \(z\)-axis along the beam pipe.
The \(x\)-axis points from the IP to the centre of the LHC ring,
and the \(y\)-axis points upwards.
Cylindrical coordinates \((r,\phi)\) are used in the transverse plane,
\(\phi\) being the azimuthal angle around the \(z\)-axis.
The pseudorapidity is defined in terms of the polar angle \(\theta\) as \(\eta = -\ln \tan(\theta/2)\).
Angular distance is measured in units of \(\Delta R \equiv \sqrt{(\Delta\eta)^{2} + (\Delta\phi)^{2}}\).}
 
The ATLAS detector~\cite{PERF-2007-01} at the LHC covers nearly the entire solid angle around the collision point.\footnote{\AtlasCoordFootnote}
It consists of an inner tracking detector surrounded by a thin superconducting solenoid, electromagnetic and hadron calorimeters,
and a muon spectrometer incorporating three large superconducting air-core toroidal magnets.
 
The inner-detector system (ID) is immersed in a \qty{2}{\tesla} axial magnetic field
and provides charged-particle tracking in the range \(|\eta| < 2.5\).
The high-granularity silicon pixel detector covers the vertex region and typically provides four measurements per track,
the first hit normally being in the insertable B-layer installed before Run~2~\cite{ATLAS-TDR-19,PIX-2018-001}.
It is followed by the silicon microstrip tracker, which usually provides eight measurements per track.
These silicon detectors are complemented by the transition radiation tracker (TRT),
which enables radially extended track reconstruction up to \(|\eta| = 2.0\).
The TRT also provides electron identification information
based on the fraction of hits (typically 30 in total) above a higher energy-deposit threshold corresponding to transition radiation.
 
The calorimeter system covers the pseudorapidity range \(|\eta| < 4.9\).
Within the region \(|\eta|< 3.2\), electromagnetic calorimetry is provided by barrel and
endcap high-granularity lead/liquid-argon (LAr) calorimeters,
with an additional thin LAr presampler covering \(|\eta| < 1.8\)
to correct for energy loss in material upstream of the calorimeters.
Hadron calorimetry is provided by the steel/scintillator-tile calorimeter,
segmented into three barrel structures within \(|\eta| < 1.7\), and two copper/LAr hadron endcap calorimeters.
The solid angle coverage is completed with forward copper/LAr and tungsten/LAr calorimeter modules
optimised for electromagnetic and hadronic energy measurements respectively.
 
The muon spectrometer (MS) comprises separate trigger and
high-precision tracking chambers measuring the deflection of muons in a magnetic field generated by the superconducting air-core toroidal magnets.
The field integral of the toroids ranges between \num{2.0} and \qty{6.0}{\tesla\metre}
across most of the detector.
Three layers of precision chambers, each consisting of layers of monitored drift tubes, cover the region \(|\eta| < 2.7\),
complemented by cathode-strip chambers in the forward region, where the background is highest.
The muon trigger system covers the range \(|\eta| < 2.4\) with resistive-plate chambers in the barrel, and thin-gap chambers in the endcap regions.
 
Interesting events are selected by the first-level trigger system implemented in custom hardware,
followed by selections made by algorithms implemented in software in the high-level trigger~\cite{TRIG-2016-01}.
The first-level trigger accepts events from the \qty{40}{\MHz} bunch crossings at a rate below \qty{100}{\kHz},
which the high-level trigger reduces in order to record events to disk at about \qty{1}{\kHz}.
 
An extensive software suite~\cite{ATL-SOFT-PUB-2021-001} is used in data simulation, in the reconstruction
and analysis of real and simulated data, in detector operations, and in the trigger and data acquisition
systems of the experiment.


\section{Data and simulated event samples}
\label{sec:datamc}

 
This analysis uses proton--proton ($pp$) collision data recorded with the ATLAS detector
from 2015 to 2018 at a centre-of-mass energy of $\sqrt{s}=13$~TeV. The
LHC conditions required to achieve high instantaneous luminosity
lead to additional collisions in the same and
neighbouring proton bunch crossings (in-time and out-of-time \pileup). During the
data-taking period considered, the number of collisions per bunch
crossing ranged from about 8 to 70, with an average of 34.
After the application of data-quality
requirements~\cite{DAPR-2018-01}, the data sample corresponds to an
integrated luminosity of 140~\ifb.
 
Theoretical predictions of the \ttW process with higher-order
corrections in the QCD and EWK couplings (\alphas and $\alpha$) are very
challenging. At LO in \alphas there are complications because
\ttW is a $q\bar{q}$-initiated process. The radiation of
the $W$ boson from one of the initial-state quarks polarises the
incoming quark in the associated \ttbar production, so it is important to correctly account for spin
correlations~\cite{Maltoni:2014zpa}.
Initial calculations of \ttW production at
NLO in QCD at fixed order~\cite{Campbell:2012dh} and later
matched to a parton shower~\cite{Garzelli:2012bn,Maltoni:2015ena} were
also augmented with NLO EWK corrections (of order
$\alpha^2\alphas^2$)~\cite{Frixione:2015zaa} to provide the
higher-order \xs{}s used across the LHC programme for a number of
years~\cite{YR4}. Full NLO calculations including
fixed-order corrections matched to the parton shower (PS) in the \POWHEGBOX
framework and accounting for LO spin-correlations of the decay products
were provided in Ref.~\cite{FebresCordero:2021kcc}.
There has also been significant theoretical progress in
calculating more complex and precise predictions. Higher-order QCD
corrections including \ttW production with additional partons (of order
$\alpha\alphas^3$) open gluon-initiated production modes that
contribute significantly to the total \xs. Recent studies show that these
contributions also have large NLO corrections, which require NLO-merged
calculations for such effects to be included properly~\cite{Frederix:2021agh}.
Furthermore, beyond the traditionally \enquote{leading} NLO EWK corrections
(of order $\alpha^2\alphas^2$) there are even larger contributions from
traditionally \enquote{sub-leading} NLO corrections (of order
$\alpha^3\alphas$)~\cite{Dror:2015nkp,Frederix:2017wme,FebresCordero:2021kcc}
because of the presence of $tW$ scattering contributions embedded in
the \ttW{}$j$ process.
Calculations at NLO in QCD accounting for next-to-next-to-leading-logarithm
(NNLL) effects are also available~\cite{Kulesza:2018tqz}
and so are recent predictions at NLO+NNLL in QCD, also with NLO EWK
corrections~\cite{Broggio:2019ewu,Kulesza:2020nfh}.
The NNLL corrections can be as large as 7\% of the NLO QCD \xs,
depending on the calculation and the functional form chosen for the scales.
Full off-shell calculations at NLO in
QCD~\cite{Bevilacqua:2020pzy,Denner:2020hgg,Bevilacqua:2020srb} have
also become available, and more recently the NLO EWK corrections were
incorporated~\cite{Denner:2021hqi} into these calculations, along
with the development of procedures to apply the off-shell corrections
to NLO+PS simulations~\cite{Bevilacqua:2021tzp}.
 
Most recently, a NNLO calculation in Ref.~\cite{Buonocore_2023}
gives the state-of-the-art prediction for the
inclusive \ttW \xs. This calculation is exact except for the finite
part of the two-loop virtual corrections, which is computed using two
different approaches with a corresponding uncertainty. The NNLO QCD
result is combined with complete NLO EWK corrections.

Samples of simulated events were produced using Monte Carlo (MC)
techniques to model the different SM processes. In particular, the
estimation of the \ttW signal and several backgrounds is
based upon these simulations. More details of the event generation are
given in the following paragraphs and are summarised in Table~\ref{tab:mcconfig},
with the samples in black, or in parentheses and in grey, indicating those used as the
nominal configurations, or to estimate the systematic uncertainties, respectively.
Several data-driven corrections are applied to
the estimated backgrounds and detailed in Section~\ref{sec:background}.
 
After generating events for each sample's process of interest, the
detector response was modelled by either a full simulation of the
ATLAS detector based on \GEANT~\cite{SOFT-2010-01,Agostinelli:2002hh}
or a fast simulation~(\textsc{AtlFast~2}) relying on parameterised
showers in the
calorimeter~\cite{ATL-PHYS-PUB-2010-013,ATL-SOFT-PUB-2018-002}.
The latter was only used for the modelling of some rare background processes, such as $tHjb$ and $WtH$,
and the modelling of the \ttZ and \ttH processes with alternative generators (see Section~\ref{sec:systematics}).
Additional simulated $pp$ collisions, generated with \Pythia[8.186]~\cite{Sjostrand:2007gs} using
parameter values from the A3 tune~\cite{ATL-PHYS-PUB-2011-014},
were overlaid to model the effects of both in-time and
out-of-time \pileup. The \pileup distribution was reweighted to
reflect the number of additional interactions observed in
data. All simulated events are processed using the same
reconstruction algorithms and analysis chain as the data. Corrections are applied so that the object reconstruction and
identification efficiencies, energy scales and energy resolutions
in simulation match those determined from data, with uncertainties described in Section~\ref{sec:inst_syst}.

Several generators were studied for the simulation of \ttW
production, and multiple samples were generated to model this
process. The events in the nominal \ttW sample were simulated using
the \Sherpa[2.2.10]~\cite{Bothmann:2019yzt,Gleisberg:2008ta} generator
with the \NNPDF[3.0nnlo] PDF set~\cite{Ball:2014uwa}.  The matrix
element (ME) was calculated for up to one additional parton at NLO and
up to two partons at LO using \textsc{Comix}~\cite{Gleisberg:2008fv}
and \OPENLOOPS~\cite{Cascioli:2011va,Denner:2016kdg,Buccioni:2019sur}, and merged with the \Sherpa
parton shower~\cite{Schumann:2007mg} using the \MEPSatNLO
prescription~\cite{Hoeche:2012yf} with a merging scale of 30~GeV. The
renormalisation and factorisation scales are chosen to be $\muR = \muF =
\HT/2$, where \HT is defined as the scalar sum of
the transverse masses $\sqrt{\pTsq+m^2}$ of all final-state
particles with transverse momentum \pT and invariant mass $m$.
Top quarks were decayed at LO using the Sherpa particle decay program to preserve spin correlations.
In addition to this nominal prediction at NLO in the
strong coupling, higher-order corrections related to EWK
contributions were also included in two ways.  First,
event-by-event correction factors were applied; these provide virtual NLO
EWK corrections of order $\alpha^{2}\alphas^{2}$ derived using the
formalism described in Ref.~\cite{Kallweit:2015dum} along with LO
corrections of order $\alpha^3$, reducing the cross-section by 3.9\% relative to NLO QCD.
Both were implemented using the prescription described in
Refs.~\cite{Bothmann:2019yzt,Gutschow:2018tuk}. Second, the real emission contributions of the sub-leading EWK
corrections at order $\alpha^{3}\alphas$~\cite{Frederix:2017wme} were accounted for via the
addition of an independent \Sherpa[2.2.10] sample produced at LO in
QCD.
The combination of contributions from NLO QCD and NLO EWK effects taken
from this \Sherpa configuration closely follows the
strategy described in Ref.~\cite{Frederix:2021agh} and results in a total \xs of $\sigma(\ttW) = 614.7$~fb.

In this measurement of \ttW production \xs{}s, the signal process is defined to include
the combined QCD and EWK corrections, as described for the nominal \Sherpa sample.
The measured differential \xs{}s are compared with theoretical predictions
obtained from various generators, including the nominal \Sherpa one.
For one of the alternative predictions,
the \MGNLO[2.9.3] program~\cite{Alwall:2014hca} generated \ttW events with up to one
additional parton in the final state at NLO accuracy in the strong coupling, using the \NNPDF[3.1nnlo]~\cite{NNPDF:2017mvq} PDF set.
The showering and subsequent hadronisation was performed
using \PYTHIA[8.245] with the A14 tune~\cite{ATL-PHYS-PUB-2014-021},
and the \NNPDF[2.3lo]~\cite{Ball:2014uwa} PDF set with $\alphas =
0.130$. The decays of bottom and charm hadrons were performed
by \EVTGEN[1.6.0]~\cite{Lange:2001uf}. The different jet
multiplicities were merged using the FxFx NLO matrix-element and
parton-shower merging prescription~\cite{Frederix:2012ps} with a
merging scale of 30~GeV. This prediction uses a different merging scale
and differs significantly from that in
Ref.~\cite{Frederix:2021agh}, which is also based on the FxFx
prescription. The top quarks were decayed at LO using the \MADSPIN~\cite{Frixione:2007zp,Artoisenet:2012st} particle decay program to preserve spin correlations.
Several other predictions were made inclusively at
NLO in \alphas to investigate the effect of different
additional-jet modelling. These come from (i) the \MGNLO[2.3.3] generator,
which used the same scale choice and PDF set as the nominal \SHERPA sample and
was interfaced to \PYTHIA[8.210] in combination with the A14 tune,
(ii) the \POWHEGBOX~\cite{Frixione:2007nw,Nason:2004rx,Frixione:2007vw,Alioli:2010xd,Hartanto:2015uka,FebresCordero:2021kcc} generator interfaced to \PYTHIA[8],
and (iii) the \POWHEGBOX generator interfaced to \HERWIG[7]~\cite{Bahr:2008pv,Bellm:2015jjp}.
The top quarks were decayed at LO withing the \POWHEGBOX program to preserve spin correlations.
A final prediction, used to assess the impact of off-shell
contributions to \ttW production, was calculated at fixed order and
includes full NLO off-shell effects using the procedure outlined in
Ref.~\cite{Bevilacqua:2020srb}.
 
The production of \ttH events was modelled using the
\POWHEGBOX[v2]
generator at NLO with the \NNPDF[3.0nlo]
PDF set.
The events were interfaced to \PYTHIA[8.230]
using the A14 tune and the \NNPDF[2.3lo]
PDF set. The decays of bottom and charm hadrons
were performed by \EVTGEN[1.6.0].
The \hdamp parameter\footnote{The \hdamp parameter is a resummation
damping factor and one of the parameters that controls the matching
of \POWHEG matrix elements to the parton shower and thus effectively
regulates the high-\pT radiation against which the hard-process system
recoils.} was set to $0.75 \times (m_t + m_{\bar{t}} + m_{H}) =
352.5\,\gev$.
The \xs was calculated at NLO QCD and NLO EWK accuracy using
\MGNLO as reported in Ref.~\cite{YR4}.
The predicted value at \(\rts = \SI{13}{\TeV}\) is
\(507^{+35}_{-50}\,\si{\fb}\),
where the uncertainties were estimated from the combined PDF+\alphas uncertainties and variations of the renormalisation and factorisation scales.
 
Background events from $t\bar{t}Z/\gamma^\ast$ production were simulated using
the \MGNLO[2.8.1] generator at NLO in \alphas with the \NNPDF[3.0nlo] PDF set. The functional form of the renormalisation and factorisation scales (\muR, \muF) was set to the default scale $0.5 \times \sum_i \sqrt{m^2_i+p^{2}_{\textrm{T},i}}$, where the sum runs over all the particles generated from the matrix element calculation.
Top quarks were decayed at LO using \MADSPIN~\cite{Frixione:2007zp,Artoisenet:2012st} to preserve all spin correlations.
The showering and subsequent hadronisation was performed
using \PYTHIA[8.244] with the A14 tune, and the \NNPDF[2.3lo] PDF set with $\alphas = 0.130$. The
decays of bottom and charm hadrons were performed
by \EVTGEN[1.7.0].
The $t\bar{t}Z/\gamma^\ast(\to \ell^+\ell^-)$ prediction was normalised to the calculation at NLO QCD and NLO EWK accuracy reported in
Ref.~\cite{YR4} for an on-shell $Z$ boson, scaled to the leptonic contributions including off-shell $\gamma^\ast\to\ell^+\ell^-$ contributions with a correction estimated at one-loop level in \alphas. The resulting $\ttbar\ell^+\ell^-$ \xs, with $m(\ell^+\ell^-) > 1~\gev$, is $162 \pm 21$~fb.
A dedicated \ttbar sample including rare $t \to Wb\gamma^\ast(\to \ell^+\ell^-)$ radiative decays, i.e.\ $t\bar{t} \to W^+bW^-\bar{b}\ell^+\ell^-$, was
generated using a LO ME and requiring $m(\ell^+\ell^-) > 1$~GeV. In this
sample the photon can be radiated from the top quark, the $W$ boson,
or the $b$-quark. The $t\bar{t}Z/\gamma^\ast$ and $t\bar{t} \to W^+bW^-\bar{b}\ell^+\ell^-$ samples were combined and together form the
\enquote{$t\bar{t}Z/\gamma^\ast$} sample. The contribution
from internal photon conversions $(\gamma^\ast \to \ell^+\ell^-)$ with
$m(\ell^+\ell^-) < 1$~GeV was modelled with QED multiphoton radiation via the
parton shower in an inclusive \ttbar sample.
Care was taken to avoid both double-counting of contributions and uncovered regions of phase
space when combining the different simulated event samples.  The LO \xs for the $t\bar{t} \to W^+bW^-\bar{b}\ell^+\ell^-$ sample was scaled to
match the higher-order \xs used for \ttbar production in the
phase space they overlap, and was assigned a 50\% normalisation
uncertainty.

The production of \ttbar and single-top-quark events was modelled using the
\POWHEGBOX[v2]~\cite{Frixione:2007nw,Nason:2004rx,Frixione:2007vw,Alioli:2010xd,Re:2010bp,Frederix:2012dh,Alioli:2009je}~
generator at NLO with the \NNPDF[3.0nlo]
PDF set. The events were interfaced
to \PYTHIA[8.230]
to model the parton shower,
hadronisation, and underlying event, using
the A14 tune and the \NNPDF[2.3lo]
set of PDFs.
The decays of bottom and charm hadrons
were performed by \EVTGEN[1.6.0].
The \ttbar process was modelled with the \hdamp parameter set
to 1.5\,$m_t$~\cite{ATL-PHYS-PUB-2016-020}.
The production of a top quark in association with a \(W\) boson (\(tW\)) was
modelled using the five-flavour scheme. The diagram removal scheme~\cite{Frixione:2008yi} was used to remove
interference and overlap with \ttbar production.
Single-top \(s\)- and \(t\)-channel production was modelled using the
\POWHEGBOX[v2]~\cite{Frederix:2012dh,Alioli:2009je} generator at NLO in QCD using the five- and four-flavour schemes, respectively, and the
corresponding \NNPDF[3.0nlo] set of PDFs.
The \ttbar sample was normalised to the \xs prediction at NNLO
in QCD including the resummation of NNLL soft-gluon terms calculated using
\TOPpp[2.0]~\cite{Beneke:2011mq,Cacciari:2011hy,Baernreuther:2012ws,Czakon:2012zr,Czakon:2012pz,Czakon:2013goa-fixed,Czakon:2011xx}.
This \xs is \(\sigma(\ttbar)_{\text{NNLO+NNLL}} = 832 \pm 51\,\si{\pb}\).
The single-top-quark \(tW\)-channel inclusive \xs was corrected to
the theory prediction calculated at approximate NNLO in QCD with NNLL soft-gluon
corrections~\cite{Kidonakis:2010ux,Kidonakis:2013zqa}: \(\sigma(t,tW)_{\text{aNNLO+NNLL}} = 71.7 \pm 3.8\)~pb. The single-top-quark \(s\)- and \(t\)-channel inclusive \xs{}s are corrected to theory predictions calculated at NLO in QCD~\cite{Aliev:2010zk,Kant:2014oha}:
\(\sigma(t,s\text{-chan})_{\text{NLO}} = 6.35^{+0.23}_{-0.20}\)~pb and
\(\sigma(t,t\text{-chan})_\text{NLO}= 216.97^{+9.46}_{-8.18}\)~pb.

Diboson events with fully leptonic decays (yielding between zero and
four charged leptons) were generated by \SHERPA[2.2.2] with up to
one additional parton at NLO and three additional partons at LO.
\SHERPA[2.2.1] was used for diboson semileptonic decays at the same
accuracy. The EWK $VVjj$ ($V=W,Z$) process was simulated for fully
leptonic decays with one additional parton at LO using \SHERPA[2.2.2].
Events of vector boson production in association with additional jets,
$V$+jets, were generated using \SHERPA[2.2.1] with two additional
partons at NLO and four additional partons at LO. The EWK $VVjj$
($V=W,Z$) process was simulated for fully leptonic decays with one
additional parton at LO using \SHERPA[2.2.2].

Rare background contributions ($tZ$, $t\bar{t}t\bar{t}$, $ttWW$,
$WtZ$,
triboson, $t\ttbar$, $tHjb$ and $WtH$) were normalised using their
NLO theoretical \xs{}s.
The production of four top quarks, $t\bar{t}t\bar{t}$, was normalised to the theoretical prediction of $\sigma(t\bar{t}t\bar{t})=13.37$~fb in Ref.~\cite{vanbeekveld:2022fourtops},
with an assumed 100\% normalisation uncertainty based on the larger \xs measured recently
for this process by ATLAS~\cite{TOPQ-2021-08}.
A 30\% uncertainty~\cite{HIGG-2018-51} was assumed for triboson production,
while a 5\% uncertainty was used for $tZ$ production, and all other rare backgrounds were assigned a 50\%
normalisation uncertainty.

\begin{table}[htb]
\begin{center}
\caption{
The configurations used to generate events from signal and background processes.
The samples used to estimate the systematic uncertainties are indicated in parentheses and in grey.
}
\label{tab:mcconfig}
\vspace{0.25cm}
\small
\setlength\tabcolsep{1.5pt}
\begin{tabular}{~l^l^l^l^l^l}
\hline\hline
Process & Generator & ME order & Parton shower & PDF & Tune  \\
\hline
$\ttbar W$             & \Sherpa[2.2.10] & \MEPSatNLO & \Sherpa &  \NNPDF[3.0nnlo] & \Sherpa default \\
\rowstyle{\color{gray}}
& (MG5\_aMC)  & (FxFx NLO) & (\PYTHIA[8]) & (\NNPDF[3.0nnlo]) & (A14)   \\
\rowstyle{\color{gray}}
& (\POWHEG)  & (NLO) & (\PYTHIA[8]) & (\NNPDF[3.0nlo]) & (A14)   \\
\rowstyle{\color{gray}}
& (\POWHEG)  & (NLO) & (\HERWIG[7]) & (\NNPDF[3.0nlo]) & (H7-UE-MMHT)   \\
$\ttbar W$ (EW)        & \Sherpa[2.2.10] & LO & \Sherpa &  \NNPDF[3.0nnlo] & \Sherpa default \\ 
\rowstyle{\color{gray}}
& (MG5\_aMC)  & (LO) & (\PYTHIA[8]) & (\NNPDF[3.0nlo]) & (A14)   \\
\ttH                   & \POWHEGBOX  & NLO & \PYTHIA[8]\ & \NNPDF[3.0nlo] & A14 \\
\rowstyle{\color{gray}}
& (\POWHEGBOX) & (NLO) & (\Herwig[7.0.4]) &  (\NNPDF[3.0nlo]) &  (H7-UE-MMHT) \\
\rowstyle{\color{gray}}
& (MG5\_aMC)  & (NLO) & (\PYTHIA[8]) & (\NNPDF[3.0nlo]) & (A14)   \\
\ttll                  & MG5\_aMC & NLO & \PYTHIA[8] & \NNPDF[3.0nlo] & A14  \\
\rowstyle{\color{gray}}
& (MG5\_aMC) & (NLO) & (\PYTHIA[8]) & (\NNPDF[3.0nlo]) & (A14 Var3c)  \\
\rowstyle{\color{gray}}
& (MG5\_aMC) & (NLO) & (\HERWIG[7]) & (\NNPDF[3.0nlo]) & (H7-UE-MMHT)  \\
$\ttbar \to W^+bW^-\bar{b}\ell^{+}\ell^{-}$ & MG5\_aMC & LO & \PYTHIA[8] & \NNPDF[3.0lo] & A14   \\
$t\bar t t\bar t$      & MG5\_aMC & NLO & \PYTHIA[8] & \NNPDF[3.1nlo] & A14 \\
$t\bar t WW, WtZ, t\bar t t, tHjb, WtH$ & MG5\_aMC & LO & \PYTHIA[8] & \NNPDF[2.3lo] & A14 \\
$tZ$                  & MG5\_aMC & LO & \PYTHIA[8] & \NNPDF[3.1nlo] & A14 \\
$\ttbar$               & \POWHEGBOX & NLO & \PYTHIA[8] & \NNPDF[3.0nlo] & A14  \\
\rowstyle{\color{gray}}
& (\POWHEGBOX) & (NLO) & (\Herwig[7.1.3]) & (\NNPDF[3.0nlo]) &  (H7-UE-MMHT) \\
Single top             & \POWHEGBOX & NLO & \PYTHIA[8] & \NNPDF[3.0nlo] & A14   \\
($t$-, $Wt$-, $s$-channel) &  &  &                                     &  &  \\
$VV$, $qqVV$, $VVV$    & \Sherpa[2.2.2]  & \MEPSatNLO & \Sherpa & \NNPDF[3.0nnlo] & \Sherpa default  \\
$Z \to \ell^{+}\ell^{-}$         & \Sherpa[2.2.1] & \MEPSatNLO & \Sherpa & \NNPDF[3.0nlo] & \Sherpa default \\ 
$Z \to \ell^{+}\ell^{-}\gamma(\to e^{+}e^{-})$ & \POWHEGBOX & NLO & \PYTHIA[8] & \CTEQ[6L1nlo] & A14 \\
$Z \to \ell^{+}\ell^{-}\gamma^*(\to e^{+}e^{-})$ & \POWHEGBOX & NLO & \PYTHIA[8] & \CTEQ[6L1nlo] & A14 \\
\hline\hline
\end{tabular}
\end{center}
\end{table}


\section{Event reconstruction and object identification}
\label{sec:objects}

Interaction vertices from the $pp$ collisions are reconstructed from at least two tracks with $\pt$ larger than $500~\mev$ that are consistent with originating from the beam collision region in the $x$--$y$ plane. If more than one primary vertex candidate is found in the event, the candidate for which the associated tracks form the largest sum of squared $\pt$ is selected as the hard-scatter primary vertex~\cite{ATL-PHYS-PUB-2015-026}.
 
Electron candidates are reconstructed from energy clusters in the electromagnetic calorimeter matched to a track in the ID~\cite{EGAM-2018-01}. They are required to satisfy $\pt>10$~GeV and $|\eta_{\mathrm{cluster}}|<2.47$, excluding the transition region between the endcap and barrel calorimeters ($1.37<|\eta_{\mathrm{cluster}}|<1.52$).
\enquote{Loose} and \enquote{tight} electron identification working points are used, based on a likelihood discriminant employing calorimeter, tracking and combined variables that provide separation between electrons and jets.
 
Muon candidates are reconstructed by performing a combined fit of the track information from the ID and MS~\cite{MUON-2018-03}. The resulting muon candidates are re-fitted using the complete track information from both detector systems. They are required to satisfy $\pt > 10~\gev$ and $|\eta| < 2.5$. \enquote{Loose} and \enquote{medium} muon identification working points are used.
 
Electron (muon) candidates are matched to the primary vertex by
requiring that the significance of their transverse impact parameter,
$d_0$,\footnote{The transverse impact parameter, $d_0$, is
defined in the $x$--$y$~plane as the distance of closest approach of the
track to the beamline.} satisfies $|d_0/\sigma(d_0)|<5\,(3)$, where
$\sigma(d_0)$ is the measured uncertainty in $d_0$, and by requiring
that their longitudinal impact parameter, $z_0$,\footnote{The longitudinal impact parameter, $z_0$, is defined as the distance in $z$ between the primary vertex and the point on the track used to evaluate $d_0$.}
satisfies
$|z_0 \sin\theta|<0.5$~mm.
 
To further suppress leptons from heavy-flavour hadron decays, misidentified jets, or photon conversions (collectively referred to as \enquote{non-prompt leptons}), lepton candidates are also required to be isolated in the tracker and in the calorimeter. The track-based lepton isolation criterion is based on the quantity $I_R = \sum \pt^{\textrm{trk}}$, where the scalar sum includes all tracks (excluding the lepton candidate itself) within a cone of size $\Delta R<R_{\textrm{cut}}$ around the direction of the lepton.  The value of $R_{\textrm{cut}}$ is the smaller of $r_{\textrm{min}}$ and $10~\gev/\pt^\ell$, where $r_{\textrm{min}}$ is set to 0.2 (0.3) for electron (muon) candidates and $\pt^\ell$ is the lepton's $\pt$. All lepton candidates must satisfy $I_R/\pt^\ell < 0.15$.
They are also required to satisfy a calorimeter-based isolation criterion: the sum of the transverse energy within a cone of size $\Delta R=0.2$ around the lepton, after subtracting contributions from \pileup and the energy deposit of the lepton itself, is required to be less than 20\% (30\%) of the electron's (muon's) $\pt^\ell$.
 
These selection criteria largely suppress the contribution from non-prompt leptons. However, several channels considered in this measurement have additional suppression requirements targeting the main types of non-prompt leptons. Non-prompt leptons from hadron decays that contain bottom- or charm-quarks (referred to as \enquote{heavy-flavour (HF) non-prompt leptons}) are further rejected using a boosted decision tree (BDT) discriminant, referred to as the non-prompt-lepton BDT~\cite{HIGG-2017-02}, based on isolation and lifetime information about a track-jet that matches the selected electron or muon (referred to as a \enquote{light lepton}). Three working points (WPs) based on the non-prompt-lepton BDT are used: \textit{Tight}, \textit{VeryTight}, and \textit{Tight--not-VeryTight}.
The \textit{Tight} WP allows prompt muons (barrel/endcap electrons) satisfying the calorimeter- and track-based isolation criteria to be selected with an efficiency that is about 60\% ($60\%/70\%$) for $\pt \sim 20~\gev$ and reaches a plateau of 95\% ($95\%/90\%$) for $\pt \sim 40\ (40/65)~\gev$.
The prompt-lepton efficiency of the \textit{VeryTight} WP for muons (barrel/endcap electrons) that satisfy the calorimeter- and track-based isolation criteria is about 55\% ($55\%/60\%$) for $\pt \sim 20~\gev$ and reaches a plateau of 90\% ($85\%/83\%$) for $\pt \sim 40\ (40/65)~\gev$. The corresponding rejection factor\footnote{The rejection factor is defined as the reciprocal of the efficiency.} for muons (electrons) from the decay of $b$-hadrons ranges from 33 to 50 (20 to 50) for the \textit{Tight} WP, and from 50 to 100 (33 to 66) for the \textit{VeryTight} WP, depending on \pt and $\eta$, after resolving ambiguities between overlapping reconstructed objects.
The \textit{Tight--not-VeryTight} WP allows the selection of non-prompt leptons and is part of the event selection for control regions enriched in HF non-prompt-lepton background, as described in Section~\ref{sec:background}.
 
In order to further suppress electrons with incorrect charge assignment, a BDT discriminant based on calorimeter and tracking quantities~\cite{PERF-2017-01} is used. 
An efficiency of approximately 96\% in the barrel region and 81\% in the endcaps is obtained, with rejection factors of 19 in the barrel region and 40 in the endcaps.
The electron candidates are separated into three classes: \enquote{material conversion}, \enquote{internal conversion}, and \enquote{non-conversion} electron candidates.
Most electrons arising from material conversions, i.e.\ from photon conversions in the detector material, are rejected by the standard electron identification selection, but additional requirements are imposed to remove residual material-conversion candidates.
These candidates have a reconstructed displaced vertex with radius $r > 20$~mm that includes the track associated with the electron.\footnote{The beampipe and insertable B-layer inner radii are 23.5~mm and 33~mm, respectively.} The invariant mass of the associated track and the closest (in $\Delta\eta$) opposite-charge track reconstructed in the silicon detector, calculated at the conversion vertex, is required to be $<100~\mev$. Internal conversion candidates, which correspond to the internal photon conversions (see Section~\ref{sec:datamc}), must fail the requirements for material conversions, and the di-track invariant mass, calculated here at the primary vertex, is also required to be $<100~\mev$.
 
The various lepton working points used in this analysis are summarised in Table~\ref{tbl:tightleps}. After the initial categorisation based on \enquote{loose} leptons (corresponding to $L$), the best lepton working point to further optimise the event selection is chosen; it depends on the main background processes and available amount of data in each category. The various choices for the signal and control regions are described in Section~\ref{sec:events}.
 
\begin{table}
\begin{center}
\caption{\label{tbl:tightleps} Description of the loose inclusive ($L$), medium inclusive ($M$), medium exclusive ($M_{\mathrm{ex}}$), and tight ($T$)  lepton definitions. The electron $e^{*}$ is required to fulfil, in addition to the corresponding lepton definition requirements, those corresponding to an internal or material conversion candidate.}
\resizebox{\textwidth}{!}{
\begin{tabular}{l||c|c|c|c||c|c|c|c}
\hline\hline
& \multicolumn{4}{c||}{Electron} & \multicolumn{4}{c}{Muon} \\
\hline
Lepton definition                  & $L$ & $M$ & $M_{\mathrm{ex}}$ & $T$ &  $L$ & $M$ & $M_{\mathrm{ex}}$ & $T$ \\
\hline
Identification                     & Loose & \multicolumn{3}{c||}{Tight} & Loose & \multicolumn{3}{c}{Medium} \\
\hline
Transverse impact parameter        &  \multicolumn{4}{c||}{$<5$} & \multicolumn{4}{c}{$<3$ } \\
significance $|d_0|/\sigma_{d_0}$  & \multicolumn{4}{c||}{} &  \multicolumn{4}{c}{}  \\
\hline
Longitudinal impact parameter $z_0$     &  \multicolumn{8}{c}{$|z_0 \sin \theta|< 0.5$~mm}  \\
\hline
Isolation                          &  \multicolumn{4}{c||}{Yes} & \multicolumn{4}{c}{Yes} \\
\hline
Non-prompt lepton WP               &  No & \textit{Tight} & \textit{Tight--not-}  & \textit{VeryTight} &  No & \textit{Tight} & \textit{Tight--not-}  & \textit{VeryTight}  \\
&  &   & \textit{VeryTight} & &  &  & \textit{VeryTight} &  \\
\hline
Electron charge-misassignment veto &  No & \multicolumn{3}{c||}{Yes} & \multicolumn{4}{c}{--} \\
Electron conversion candidate veto &  No & \multicolumn{3}{c||}{Yes (except $e^{*}$)} & \multicolumn{4}{c}{--} \\
\hline\hline
\end{tabular}
}
\end{center}
\end{table}

The constituents for jet reconstruction are identified by combining measurements from both the ID and the calorimeter using a particle flow (PFlow) algorithm~\cite{PERF-2015-09}. Jet candidates are reconstructed from these PFlow objects using the \antikt algorithm~\cite{Cacciari:2008gp,Cacciari:2011ma} with a radius parameter of $R=0.4$. They are calibrated using simulation with corrections obtained by using in situ techniques in data~\cite{JETM-2018-05}. Only jet candidates with $\pT> 25$~\GeV\ and within $|\eta|<2.5$ are selected. In order to reduce the effect of \pileup, each jet with $\pT<60$~\GeV\ and $|\eta|<2.4$ must satisfy the \enquote{Tight}  working point of the Jet Vertex Tagger (JVT)~\cite{PERF-2014-03} criteria used to identify jets that originate from the primary vertex. A set of quality criteria is also applied to reject events containing at least one jet arising from non-collision sources or detector noise~\cite{ATLAS-CONF-2015-029}.
 
Jets containing $b$-hadrons are identified ($b$-tagged) via an algorithm~\cite{FTAG-2019-07} that uses a deep-learning neural network based on the distinctive features of $b$-hadron decays, primarily the impact parameters of tracks and the displaced vertices reconstructed in the ID. Additional input to this network is provided by discriminating variables constructed by a recurrent neutral network~\cite{ATL-PHYS-PUB-2017-003}, which exploits the spatial and kinematic correlations between tracks originating from the same $b$-hadron. A multivariate $b$-tagging discriminant value is calculated for each jet.
In this measurement, a jet is considered $b$-tagged if it passes the working point corresponding to 85\%, 77\%, 70\%, or 60\% average expected efficiency to tag a $b$-quark jet,
with a light-jet\footnote{`Light jet' refers to a jet originating from the hadronisation of a light quark ($u$, $d$, $s$) or a gluon.} rejection factor of about 40 to 2500, and a charm-jet ($c$-jet) rejection factor of about 3 to 40, as determined for jets with $\pt >20~\gev$ and $|\eta|<2.5$ in simulated $\ttbar$ events.
The notation \bveryloose, \bloose, \bmedium, and \btight is used to denote a $b$-tagged jet (\bjet) that passed the corresponding working point.
Correction factors derived from dedicated calibration samples enriched in \bjet{}s, $c$-tagged jets, or light-tagged jets, are applied to the simulated event samples~\cite{FTAG-2018-01,FTAG-2020-08,FTAG-2019-02}.

Ambiguities between independently reconstructed electrons, muons and jets can arise. A sequential \enquote{overlap removal} procedure is performed to resolve these ambiguities and thus avoid double counting of candidates.
This procedure is applied to leptons satisfying the $L$ criteria.
If two electrons are separated by $\Delta R<0.1$, only the one with the higher \pt\ is kept. If an electron and a muon overlap within $\Delta R<0.1$, the muon is removed if it is reconstructed only from an ID track and calorimeter energy deposits consistent with a minimum-ionising particle (i.e.\ if it is \enquote{calo-tagged}), otherwise the electron is removed. If an electron and a selected jet are found within $\Delta R < 0.2$, the jet is removed if it is not a \bmedium jet\footnote{For the overlap removal, a jet is considered $b$-tagged if it passes the 70\% working point. However, the choice of $b$-tagging working point does not have a sizeable impact on the signal acceptance.} or if it has $\pt > 200~\gev$. Muons ghost-associated~\cite{Cacciari:2008gn} with an $R=0.4$ jet must satisfy a jet--muon separation of $\Delta R < 0.4$.
If the overlapping jet is not a \bmedium jet and contains less than three tracks with $\pt > 500~\mev$, the jet is removed, otherwise the muon is rejected. Thus, if the overlapping jet is a \bmedium jet, the muon is rejected. For each lepton in the event a lepton-\pt-dependent variable-size cone of maximum size $\Delta R=0.4$ is defined. If a selected jet, surviving all previous overlap criteria, is found in this cone, the lepton is rejected.
 
The missing transverse momentum $\vec{p}_{\mathrm{T}}^{\mathrm{miss}}$ (with magnitude $\met$) is defined as the negative vector sum of the $\pt$ of all selected and calibrated objects in the event, including a term to account for the momenta of soft particles that are not associated with any of the selected objects~\cite{PERF-2016-07}.
This soft term is calculated from inner-detector tracks matched to the primary vertex, which makes it more resilient to contamination from \pileup interactions.


\section{Analysis strategy}
\label{sec:events}

The events used in the analysis were selected with high efficiency using dilepton triggers~\cite{TRIG-2018-01,TRIG-2018-05}, based on electron and muon signatures.
The dielectron triggers required two electrons that satisfied loose identification criteria with $\pt$ thresholds that were raised as the luminosity increased during \RunTwo:
12~$\gev$ in 2015, 17~$\gev$ in 2016, and 24~$\gev$ in 2017--2018.
Dimuon triggers utilised asymmetric $\pt$ thresholds for leading (sub-leading) muons: 18~(8)~$\gev$ in 2015 and  22~(8)~$\gev$ in 2016--2018.
An electron+muon trigger required events to have an electron candidate satisfying loose identification with a 17~$\gev$ threshold and a muon candidate passing a 14~$\gev$ threshold for all data-taking periods.
 
Events selected by the trigger are required to satisfy basic preselection requirements. They must have at least one primary vertex candidate.
Events are required to contain either two light leptons with the same electric charge or three light leptons, based on the loose lepton definition $L$ introduced in Section~\ref{sec:objects}. This event categorisation into \llSS and \lll reduces the migration of events from \lll \ttW decays into the \ll channel, which can occur, for example, due to improper identification of leptons (i.e.\ reconstruction inefficiencies).
 
The selected light leptons are required to match, with $\Delta R < 0.15$, the corresponding leptons reconstructed by the trigger and to have a $\pt$ exceeding the trigger $\pt$ threshold by 1~$\gev$ or 2~$\gev$ (depending on the lepton trigger, lepton multiplicity criteria, and data-taking conditions).
The trigger requirement has an efficiency of about 82\% for signal events with at least two light leptons satisfying the preselection requirements.
 
Once the events are separated into the \llSS and \lll channels, control regions (CRs) and signal regions (SRs) are defined by using different lepton and event requirements to enhance the fraction of certain backgrounds or the \ttW signal, respectively, in those regions.
 
The signal regions are split into \llSS and \lll categories. The \lll events are required to have the sum of charges of selected leptons (total lepton charge) to be exactly $\pm 1$. Events are required to contain at least one \btight jet, or at least two \bloose jets. If \lll events contain pairs of opposite-sign charge and same-flavour (OS-SF) leptons, all such pairs must satisfy requirements on the dilepton system mass of $\mllOSSF > 12$ $\GeV$ and $|\mllOSSF - m_Z| > 10$ \GeV. Additionally, \lll events must satisfy $|\mlll - m_Z| > 10$ \GeV.
Leptons are ordered in \pt in the \llSS regions. In the \lll regions the lepton with opposite-sign charge is called the first lepton, followed by the two same-sign leptons.
The \pt and identification requirements for each lepton in each category are optimised, accounting for the different likelihood of being non-prompt, and are summarised in Table~\ref{tab:selectionSR}.
 
Events passing the \llSS and \lll signal region selections are separated into subcategories depending on whether the inclusive or differential \ttW cross-sections are being measured, as shown in Table~\ref{tab:selectionSR}.
For the \ttW differential measurements, two \llSS and six \lll signal categories, based on the total lepton charge of the event and on the number of OS-SF lepton pairs in the \lll channel, are defined in Section~\ref{sec:diffxsec}. The inclusive \ttW measurement (see Section~\ref{sec:inclxsec}) is improved by assigning the preselected events to subcategories based on the total lepton charge, the lepton flavour, and the jet and $b$-jet multiplicities (\njets and \nbjets, respectively), where \nbjets is defined to be 2 or more if the events contain at least two \bloose jets, or 1 if the events contain exactly one \btight jet. As a result, the \ttW inclusive measurement is performed in 48 \llSS and 8 \lll signal categories.

\begin{table}[htb]
\begin{center}
\caption{Event selection summary in the signal regions.
Leptons are ordered in \pt in the \llSS regions. In the \lll regions the lepton with opposite-sign charge is called the first lepton, followed by the two same-sign leptons.
In the lepton selection, \textit{T} and \textit{L} stand for tight and loose lepton definitions (see Table~\ref{tbl:tightleps}).
In the \bjet multiplicity categorisation, \nbjets is defined to be 2 or more if the events contain at least two \bloose jets, or 1 if the events contain exactly one \btight jet.
For each of the \llSS and \lll signal regions in the inclusive and differential measurements, each categorisation split affects all regions above.
}
 
\def\arraystretch{1.4}
\begin{tabular}{l|c|c}
\hline\hline
Signal region preselection         & \llSS                                 & \lll                       \\
\hline
Lepton definition                  & $TT$                                  & $LTT$                      \\
Lepton \pt[\GeV{}]                  & (20, 20)                              & (10, 20, 20)               \\
\mllOSSF[\GeV{}]          			 & --									 & $> 12$                      \\
$|\mllOSSF - m_Z|$ [\GeV{}]          & --                                    & $> 10$                       \\
$|\mlll - m_Z|$ [\GeV{}]             & --                                    & $> 10$                       \\
\njets                             & \multicolumn{2}{c}{$\ge$ 2 } \\
\nbjets                            & \multicolumn{2}{c}{$\geq$ 1 \btight \textrm{or} $\geq$ 2 \bloose}           \\
\hline\hline
& \multicolumn{2}{c}{\textbf{Inclusive cross-section measurement}}        \\
\hline
Lepton charge split                & ($\ell^+\ell^+, \ell^-\ell^-$)        & ($\ell^+\ell^-\ell^-, \ell^-\ell^+\ell^+$) \\
Lepton flavour split               & ($\mu\mu$, $e\mu$, $\mu e$, $ee$)     & -- \\
Jet multiplicity split             & (3, 4, $\geq$5)                       & (2, $\geq$3) \\
\bjet multiplicity split           & \multicolumn{2}{c}{(1, $\geq$2)} \\
\hline\hline
Total inclusive SRs                & 48                                    & 8 \\
\hline\hline
& \multicolumn{2}{c}{\textbf{Differential cross-section measurement}}        \\
\hline
Lepton charge split                & ($\ell^+\ell^+, \ell^-\ell^-$)        & ($\ell^+\ell^-\ell^-, \ell^-\ell^+\ell^+$) \\
Number of OS-SF pairs split        & --                                    & (0, 1, 2) \\
\hline\hline
Total differential SRs             & 2                                     & 6 \\
\hline\hline
\end{tabular}
\label{tab:selectionSR}
\end{center}
\end{table}
 
Various control regions are defined in order to fit the normalisation of the leading backgrounds. Two regions enriched in either diboson or \ttZ events are defined by requiring one OS-SF lepton pair compatible with a $Z$ boson, $|\mllOSSF - m_Z| < 10$ \GeV, but different jet multiplicities. These regions are labelled as \lllVV and \lllttZ for the diboson and \ttZ CRs, respectively. Two control regions enriched in electron-from-photon conversions from $Z \rightarrow \mu^+ \mu^- \gamma^{(*)} \rightarrow \mu^+\mu^- e^+e^-$, where one of the electrons is not reconstructed, are defined, according to the identification of the electron as a material conversion or internal conversion candidate, and are labelled as \lllmatc and \lllintc, respectively. Finally, six control regions enriched in HF non-prompt leptons are defined, making use of the exclusive lepton identification $M_{\mathrm{ex}}$ in order to be orthogonal to the signal regions. Events fulfilling the criteria $TM_{\mathrm{ex}}$, $M_{\mathrm{ex}}T$, or $M_{\mathrm{ex}}M_{\mathrm{ex}}$ for the leading and sub-leading leptons in $\pt$ are selected and further separated according to the sub-leading lepton flavour. Additionally, the transverse mass of the leading lepton, $\ell_0$, and the missing transverse momentum, defined as $m_{\textrm{T}}(\ell_0, \met)=\sqrt{2 \met p_{\textrm{T}, \ell_0} (1- \cos(\phi_{\mathrm{miss}}-\phi_{\ell_0}))}$, must be less than 250~$\GeV$ in the $TM_{\mathrm{ex}}$ and $M_{\mathrm{ex}}T$ regions, in order to reduce the \ttW contribution in these CRs. The full definition of the kinematic selection applied to events in each control region is given in Table~\ref{tab:selectionCR}.
 
\begin{table}[htb]
\begin{center}
\caption{Event selection summary in the control regions. The lepton types $L$, $M$, $M_{\mathrm{ex}}$, and $T$ correspond to the loose inclusive, medium inclusive, medium exclusive, and tight lepton definitions, as described in Table~\ref{tbl:tightleps}. The notation $e^*$ is used to denote material conversion or internal conversion candidates, as described in Section~\ref{sec:objects}.
}
 
\resizebox{\textwidth}{!}{
\def\arraystretch{1.4}
\begin{tabular}{l|c|c|c|c}
\hline\hline
Control regions for:               & Diboson     & \ttZ     & Conversions & HF non-prompt \\
\hline
Lepton requirement                 & \multicolumn{2}{c|}{\lll}         & $\mu\mu e^*$ & \llSS \\
Lepton definition                  & \multicolumn{3}{c|}{$(L,M,M)$} & {\small$(T,M_{\mathrm{ex}})$ || $(M_{\mathrm{ex}},T)$ || $(M_{\mathrm{ex}},M_{\mathrm{ex}})$} \\
Lepton \pt[\GeV{}]                  & \multicolumn{3}{c|}{(10, 20, 20)}                                 & (20, 20) \\
\mllOSSF[\GeV{}]                    & \multicolumn{2}{c|}{$>$ 12} & $>$ 12 & --\\
$|\mllOSSF - m_Z|$ [\GeV{}]          & \multicolumn{2}{c|}{$<$ 10} & $>$ 10 & --\\
$|\mlll - m_Z|$ [\GeV{}]             & \multicolumn{2}{c|}{$>$ 10}  & $<$ 10 & --\\
$m_{\mathrm{T}}(\ell_0, \met)$ [\GeV{}]       & \multicolumn{3}{c|}{--}  & $<$ 250 for \small$TM_{\mathrm{ex}}$ and $M_{\mathrm{ex}}T$ pairs \\
\njets                             & 2 or 3 & $\geq 4$ &  $\geq 0$           & $\geq 2$ \\
\nbjets                            & 1 \btight & $\geq$ 1 \btight \textrm{or} $\geq$ 2 \bloose & 0 \bloose & 1 \bloose\\
\hline\hline
Region split                       & -- & -- & internal / material & sub-leading $e/\mu$ $\times$ {\small($TM_{\mathrm{ex}}$, $M_{\mathrm{ex}}T$, $M_{\mathrm{ex}}M_{\mathrm{ex}}$)} \\
\hline
Region naming                      & \lllVV  & \lllttZ & \lllintc~/ & \llttetm, \llttemt, \llttemm \\
&         &         & \lllmatc & \llttmutm, \llttmumt, \llttmumm \\
\hline\hline
\end{tabular}
}
\label{tab:selectionCR}
\end{center}
\end{table}

\subsection{Fiducial-volume definition}
The particle-level fiducial volume closely follows the
reconstruction-level selection to avoid the uncertainties in
extrapolating to a more inclusive phase space. Table~\ref{tab:fid_selections}
summarises the object requirements, overlap removal and region
selections applied to define the fiducial regions. The fiducial selections are
the same for the inclusive and differential \xs measurements, except in the
\llSS channel, where a higher jet multiplicity is required for the inclusive measurement.
This is because at detector level the $\njets=2$ bin does not add
significant sensitivity to the inclusive \xs measurement and is
therefore omitted, and the fiducial selection matches the detector-level selection.
 
Particle-level objects in simulated events are defined using
quasi-stable particles with a mean lifetime greater than 30~ps
originating from \pp collisions. They are selected after
hadronisation but before the interaction of these particles with the
detector components or consideration of \pileup effects.
Electrons and muons are required not to have originated from a hadron in the MC generator's event record, either
directly or through a $\tau$-lepton decay, to ensure that they originated from one of the $W$ boson decays.
The leptons are modified (\enquote{dressed}) by adding the four-momenta of
all radiated photons within a cone of size $\Delta R = 0.1$, excluding
photons from hadron decays, to take into account final-state photon radiation.
Particle-level jets are reconstructed with the \antikt~algorithm with a radius parameter
of $R = 0.4$ applied to all quasi-stable particles except the neutrinos originating from
$W$ bosons and the selected electrons, muons and photons
used in the definition of the charged leptons.
If $b$-hadrons with $\pt > 5$~GeV
are found in the MC event record, they are
ghost-associated to jets. Any jets containing one
or more of these $b$-hadrons are considered to originate from a
$b$-quark. The particle-level missing transverse momentum is
calculated from the vector sum of the transverse momenta of all
neutrinos in the event record, excluding those originating from hadron
decays.

\begin{table}[htb]
\begin{center}
\caption{Definition of objects and selections for fiducial regions.}
\begin{tabular}{l|l}
\hline\hline
\multicolumn{2}{c}{Objects} \\
\hline
Electrons &  $\pt \geq 10$~\gev\ and $|\eta| < 2.47$ (excluding the LAr transition region $1.37 < |\eta| < 1.52$) \\
Muons &  $\pt \geq 10$~\gev\ and $|\eta| < 2.5$ \\
Jets & Anti-$k_t$ $R=0.4$ jets with $\pt \geq 25$~\gev\ and $|\eta| < 2.5$ \\
\bjet{}s & Tagged if jet contains a ghost-associated $b$-hadron with $\pT>5$~\gev  \\
\met & Vector sum of $\pT$ for all neutrinos ($\pT(\nu)$) in the event not from hadron decays \\
\hline\hline
\multicolumn{2}{c}{Overlap removal} \\
\hline
Electron--jet & If $\Delta{}R(e,\mathrm{jet}) < 0.2$ (excluding \bjet{}s with $\pt < 200$~\gev) remove jet \\
Jet--lepton & If $\Delta{}R(\ell,\mathrm{jet}) < \mathrm{min}(0.4, 0.04 + 10\,\gev/p_{\mathrm{T},\ell})$ remove lepton \\
\hline\hline
\multicolumn{2}{c}{Selections} \\
\hline
\llSS &  Exactly two leptons with the same charge \\
& Both leptons have $\pt \geq 20$~\gev \\
& $\njets \geq 3$ ($\njets \geq 2$) with at least one \bjet for inclusive (differential) fit \\
& $m_{\ell\ell} > 12$~\gev\ for same-flavour pairs \\
\hline
\lll & Exactly three leptons with a total charge of $\pm 1$ \\
& Both leptons from the same-sign lepton pair are required to have $\pt \geq 20$~\gev \\
& $\njets \geq 1$ with at least one \bjet \\
& $m_{\ell\ell} > 12$~\gev\ \& $|m_{\ell\ell} - m_Z| > 10$~\gev\ (for OS-SF $\ell\ell$)\\
& $|m_{\ell\ell\ell} - m_Z| > 10$~\gev \\
\hline\hline
\end{tabular}
 
\label{tab:fid_selections}
\end{center}
\end{table}


\section{Background estimation}
\label{sec:background}

 
The background processes passing the signal region selections are categorised into irreducible and reducible backgrounds.
Irreducible backgrounds  (Section~\ref{sec:bkg_irreducible}) have only prompt selected leptons, i.e.\ leptons produced in $W/Z$ boson decays, leptonic $\tau$-lepton decays, or internal conversions. Reducible backgrounds (Section~\ref{sec:bkg_reducible}) have prompt leptons with misassigned charge or at least one non-prompt lepton.
 
All backgrounds with the exception of electrons with misassigned charge (denoted by QMisID) are estimated using the simulated event samples described in Section~\ref{sec:datamc}. In some cases, the simulation is improved by applying additional corrections derived from data control samples before the statistical analysis. In particular, the event kinematics of the simulated $\ttbar$ and $\VV$ backgrounds require dedicated corrections to better describe the data. In addition, the yields of some simulated backgrounds, namely $\ttZ$, $\VV$ and non-prompt-lepton backgrounds, are adjusted via normalisation factors that are determined by performing a likelihood fit to data across all event categories in the signal and control regions defined in Tables~\ref{tab:selectionSR} and~\ref{tab:selectionCR}, as discussed in Section~\ref{sec:results}.
 
\subsection{Irreducible backgrounds}
\label{sec:bkg_irreducible}
 
Background contributions with prompt leptons originate from a wide range of physics processes, with the relative importance of individual processes varying by channel. The main irreducible backgrounds originate from $\ttdy$, $\VV$ (especially $WZ$), and $\ttH$ production, having final states and kinematic properties inherently similar to the $\ttW$ signal. Smaller contributions originate from the following rare processes: $\tZ$, $\tW$, $\tWZ$, $\ttWW$, $\VVV$, $\tttt$, and $\ttt$ production. Backgrounds with prompt leptons are estimated from simulation using the samples described in Section~\ref{sec:datamc}.
 
\subsubsection{\VV and \ttdy backgrounds}
 
The \VV simulated event sample does not model the jet multiplicity spectrum in data well. Therefore, a data-driven correction is derived from an inclusive trilepton diboson-enriched region with zero \bveryloose jets.
The events are required to have three leptons passing the ($L$, $M$, $M$) selection.
The $N_\mathrm{jets}$-dependent correction factors are derived using a fit to the ratio of $N_\mathrm{data} - N_{\mathrm{MC,non\mhyphen}VV}$ to $N_{\mathrm{MC},VV}$ as a function of $N_\mathrm{jets}$. The corrections reach about $35\%$ for $N_\mathrm{jets} = 6$.
 
The \lllVV and \lllttZ CRs are used in the likelihood fit to data to improve the predicted background contributions from the $\VV$ and $\ttdy$ processes, respectively.
The numbers of jets and $b$-jets provide good discrimination between these two processes.
Events in the \lllVV CR are required to have 2 or 3 jets, whereas events in the \lllttZ CR must have $\geq$ 4 jets.
Additionally, the number of $b$-jets is fitted in the \lllttZ CR.
The measured normalisation factors are $\lambda_\textrm{\VV} = 0.87 \pm 0.33$ and $\lambda_\textrm{\ttZ} = 1.16 \pm 0.15$.
The normalisation factor for \ttZ controls the dominant contribution from \ttZ in association with light-quark- and gluon-initiated jets. The \ttZ contribution from events produced in association with heavy-quark-initiated jets is normalised to the theoretical \xs and assigned a larger uncertainty as described in Section~\ref{sec:systematics_theory}.
 
Figures~\ref{fig:VVCR_Nbjpost} and~\ref{fig:ttZCR_Nbjpost} show the total yield and $b$-jet multiplicity distribution in the \lllVV and \lllttZ CRs, respectively, before and after the likelihood fit described in Section~\ref{sec:inclxsec}.
 
\begin{figure}[t]
\centering
\subfloat[]{\includegraphics[width=0.40\textwidth]{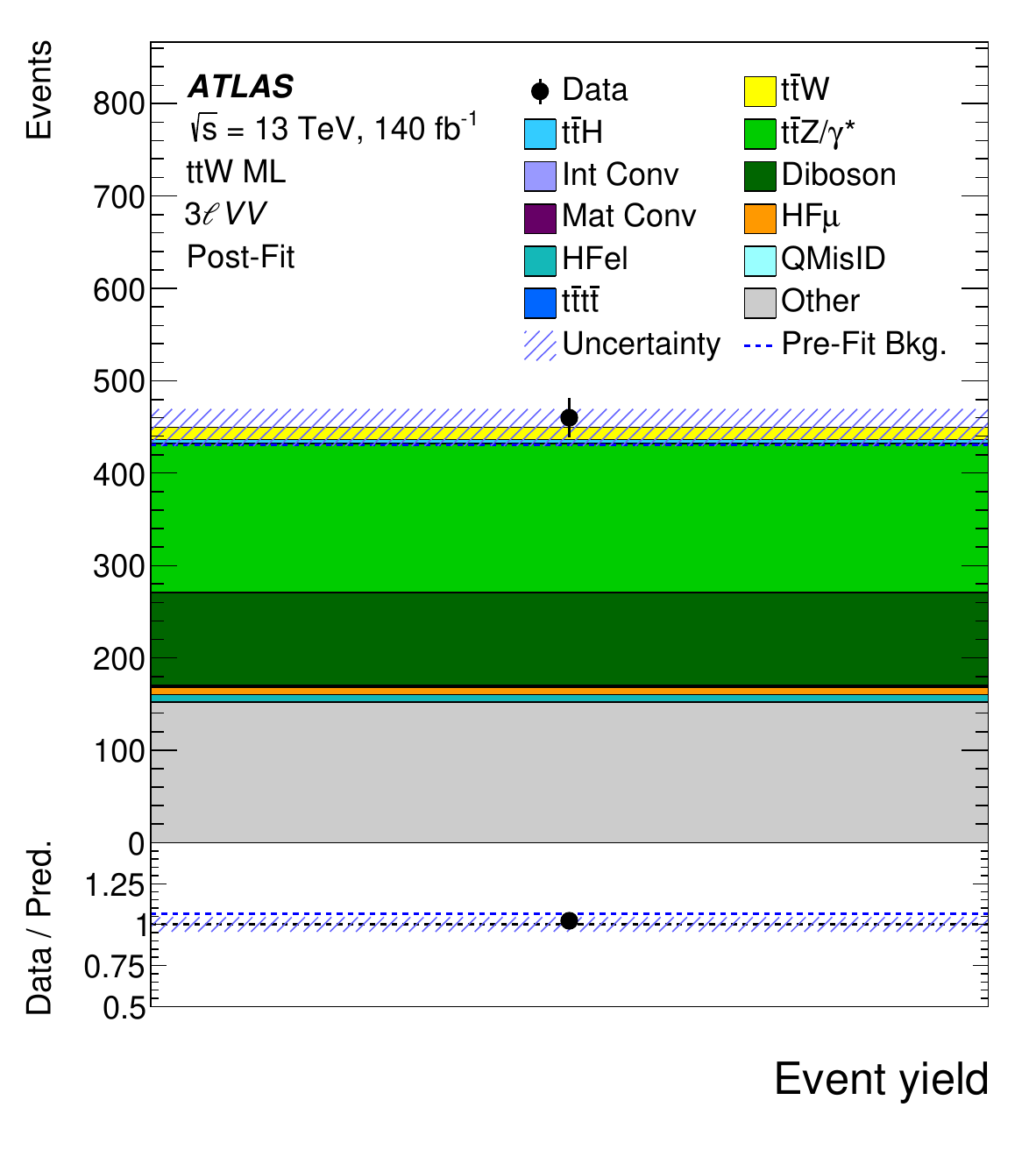}\label{fig:VVCR_Nbjpost}}
\subfloat[]{\includegraphics[width=0.40\textwidth]{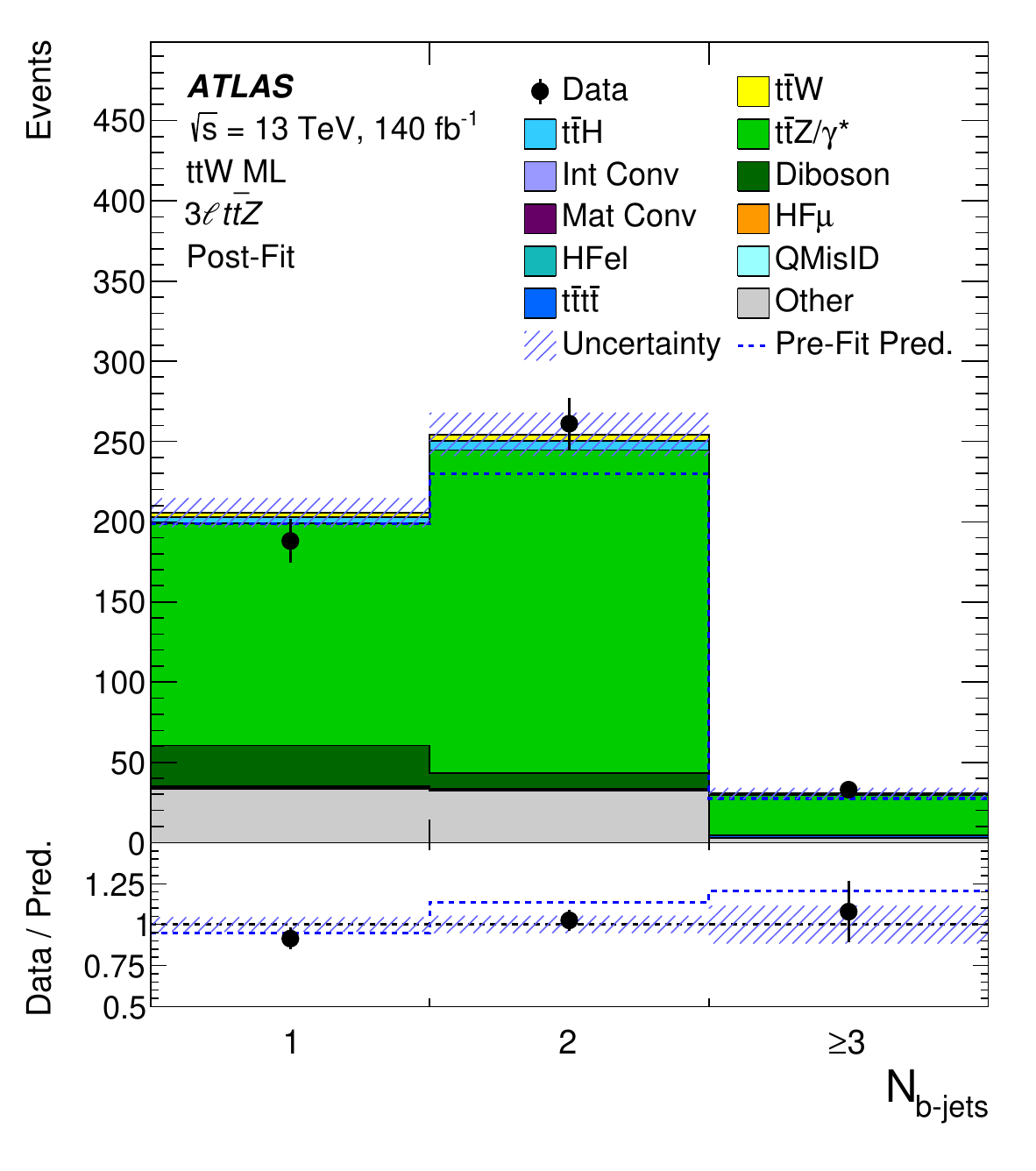}\label{fig:ttZCR_Nbjpost}}
\caption{
Comparison between data and the signal-plus-background prediction for (a) the total yield in the \lllVV CR and (b) the distribution of the $b$-jet multiplicity in the \lllttZ CR after the \VV jet multiplicity correction.
The signal and background contributions after the likelihood fit to data (\enquote{Post-Fit}) under the signal-plus-background hypothesis are shown as filled histograms.
The total signal-plus-background prediction before the likelihood fit to data (\enquote{Pre-Fit Bkg.}) is shown as a dashed blue histogram in the upper panel.
The ratio of the data to the total prediction (\enquote{Pred.}) is shown in the lower panel, separately for the post-fit prediction (black points) and pre-fit background (dashed blue line).
The combined statistical and systematic uncertainty in the prediction is indicated by the blue hatched band.
The \enquote{Other} background contribution is dominated by $tZ$ and $WtZ$ production.
}
\end{figure}
 
\subsubsection{Other irreducible backgrounds}
\label{sec:other_irreducible_backgrounds}
 
The rate of background events from internal conversions (denoted as ``Int Conv'') with $m(e^+e^-)<1~\gev$ is estimated using two dedicated CRs: \lllintc and \lllmatc.
The main contributions to these background events originate from $t\bar{t}\gamma^\ast$ with $m(e^{+}e^{-}) < 1$~GeV and $Z \to \ell^{+}\ell^{-}\gamma^*(\to e^{+}e^{-})$.
The total yield in each category is used in the likelihood fit to determine the normalisation factor $\lambda^\textrm{IntC}_e = 1.07 \pm 0.24$, where the uncertainty is mainly statistical.

\subsection{Reducible backgrounds}
\label{sec:bkg_reducible}
 
\subsubsection{Non-prompt leptons}
\label{sec:bkg_reducible_nonprompt}
 
Non-prompt leptons originate from material conversions, heavy-flavour
hadron decays, or the improper reconstruction of other particles, with
a composition strongly depending on the lepton quality requirements
and varying across event categories. These backgrounds are in general
small in all \ll and \lll SRs and are estimated from simulation,
with the normalisation determined by the likelihood fit. The main
contribution to the non-prompt-lepton background is from $\ttbar$
production, followed by much smaller contributions from $V$+jets and
single-top-quark processes. The non-prompt leptons in the simulated event
samples are labelled according to whether they originate from HF or
light-flavour (LF) hadron decays, or from a material conversion
candidate. The HF category includes leptons from bottom and charm
decays. The LF category is negligible in this analysis, and thus the
non-prompt leptons from hadronic decays are dominated by HF non-prompt
leptons.
 
Two corrections are applied to the \ttbar and overall non-prompt-lepton background
simulations before the fit. First, the \ttbar$+$~$\geq${}1~$b$-jet contribution
is multiplied by a factor of 1.3 as previously measured by ATLAS~\cite{HIGG-2020-23}.
Second, the shape of the
$b$-jet multiplicity distribution in the non-prompt-lepton background simulation is
corrected to match data in an orthogonal \llSS validation region
enriched in non-prompt leptons, where one of the leptons must pass a
looser non-prompt-lepton BDT score but not pass the $M$ lepton WP.
 
Several of the event categories introduced in Section~\ref{sec:events}
were designed to be enriched in specific processes and are used to
derive normalisation factors to improve the modelling in the
simulation.  The \lllmatc CR is enriched in material conversions (denoted as ``Mat Conv'')
and only the total event yield is used. There are six \ll CRs enriched in contributions
from HF non-prompt leptons in \ttbar events,
i.e.\ \llttetm, \llttemt, \llttemm, \llttmutm, \llttmumt,
and \llttmumm. In these CRs, the distribution of the transverse momentum
$p_{\textrm{T}}^{\ell 1}$ of the sub-leading lepton is used.
Requiring at least one $M_{\textrm{ex}}$ lepton in these CR events
provides separation from the processes with prompt leptons, especially
$\ttW$, and thus optimises the sensitivity to the HF non-prompt
electron and muon contributions. The $\ttW$ signal contribution in
these control regions is at most 11\% of the total
expected yield. Normalisation factors for three non-prompt-lepton
background contributions are estimated from the simultaneous likelihood fit to data.
The normalisation factor for HF non-prompt leptons is estimated separately
for electrons and muons, and is denoted by $\lambda^\textrm{had}_e$ and
$\lambda^\textrm{had}_\mu$ respectively. An additional normalisation
factor $\lambda^\textrm{MatC}_e$ is determined for the background from
material conversions. The measured normalisation factors are
$\lambda^\textrm{had}_e = 0.83 \pm 0.31$,
$\lambda^\textrm{had}_\mu = 1.01 \pm 0.21$, and
$\lambda^\textrm{MatC}_e = 1.15 \pm 0.31$, where the
uncertainties are mainly statistical.
 
Figures~\ref{fig:HFe_TMCR} and~\ref{fig:HFmu_TMCR} display the $p_\textrm{T}^{\ell 1}$ distribution in the \llttetm and \llttmutm CRs after the likelihood fit to data. As shown in the figures, the purity of the HF non-prompt-lepton background is 43\% and 61\%, respectively, which was achieved when using the exclusive $M_{\textrm{ex}}$ lepton working point.
 
\begin{figure}[t]
\centering
\subfloat[]{\includegraphics[width=0.40\textwidth]{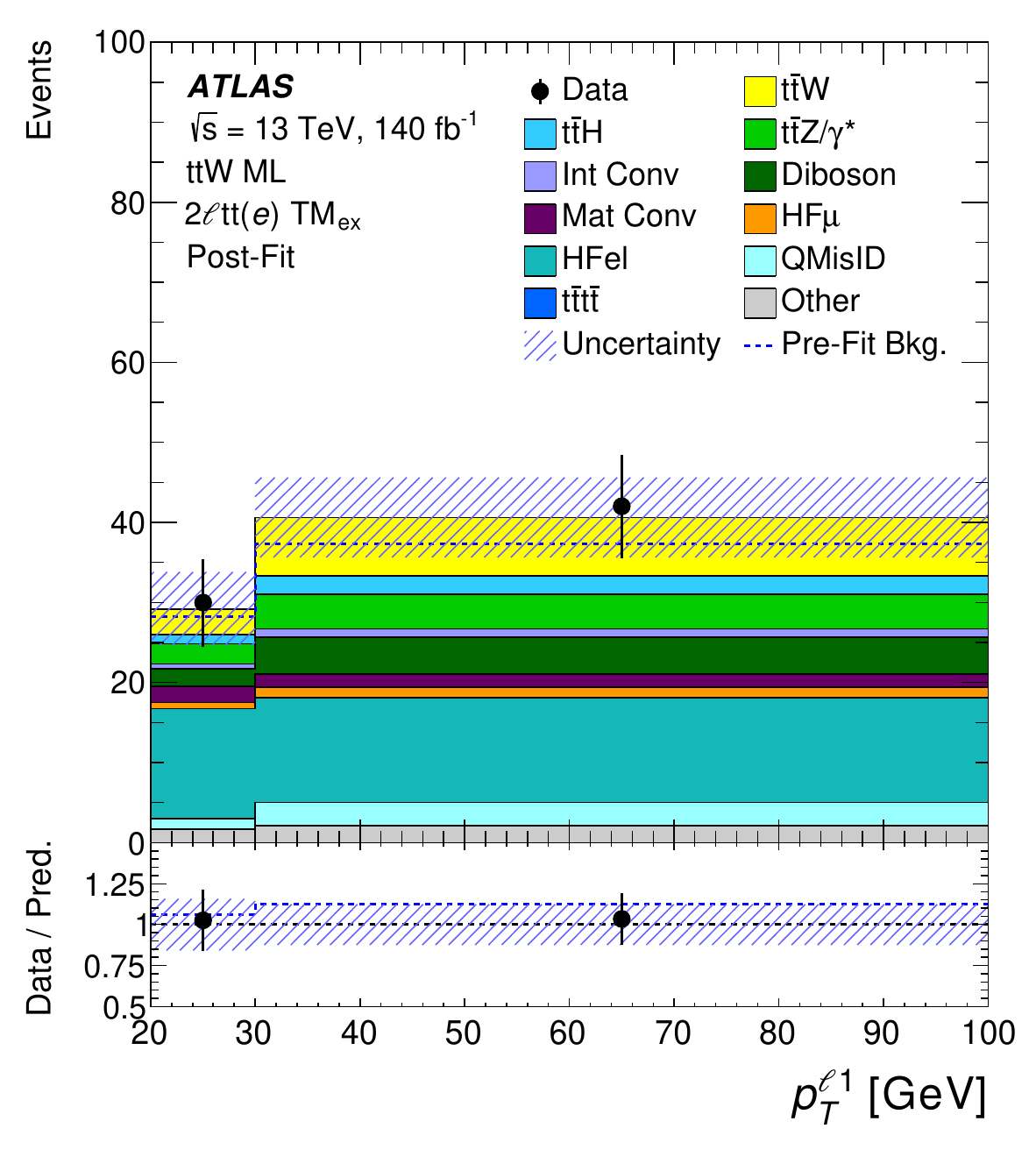}\label{fig:HFe_TMCR}}
\subfloat[]{\includegraphics[width=0.40\textwidth]{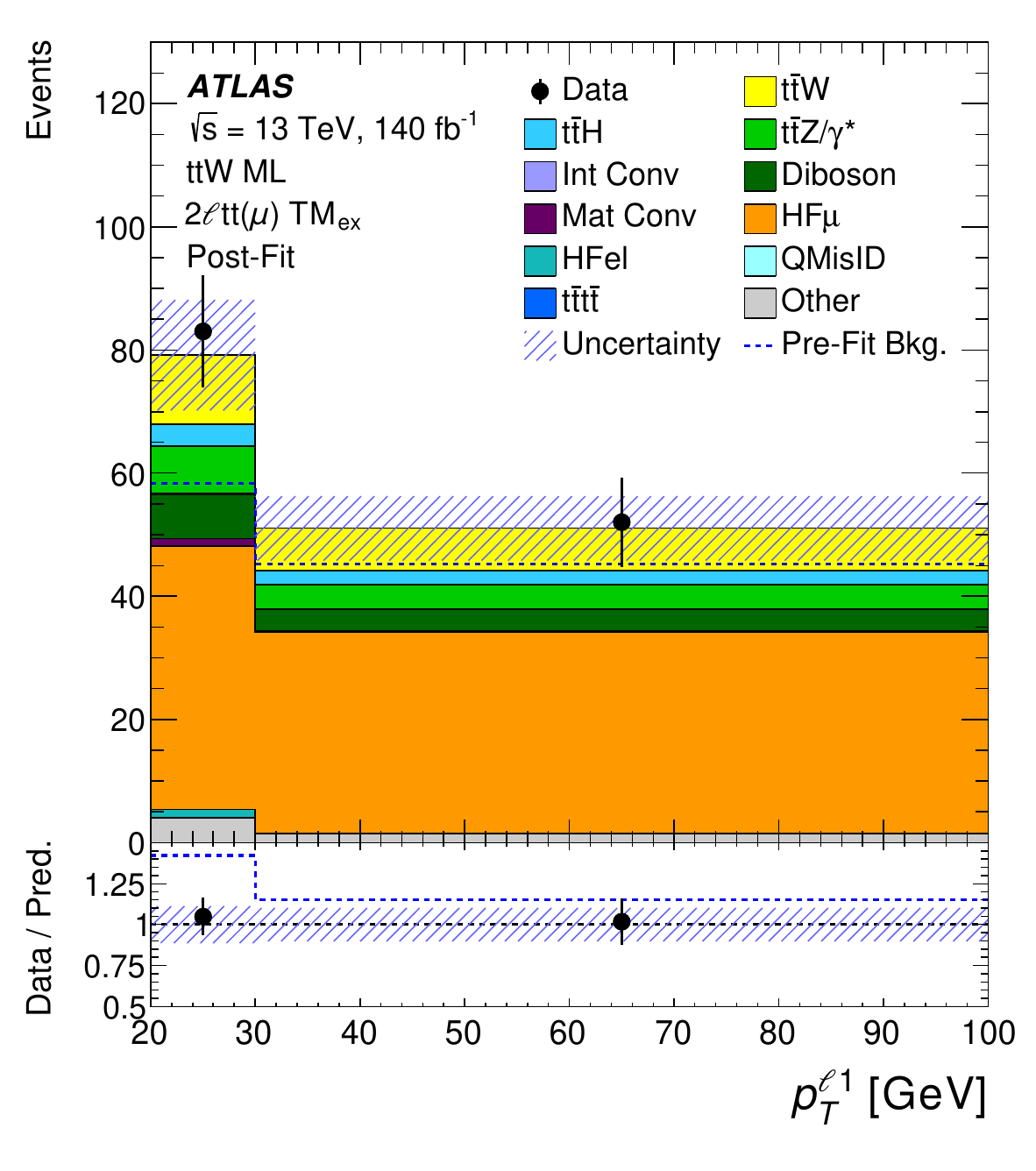}\label{fig:HFmu_TMCR}}
\caption{
Comparison between data and the signal-plus-background prediction for the distribution of the transverse momentum of the sub-leading lepton ($p_\textrm{T}^{\ell 1}$) in (a) the \llttetm CR and (b) the \llttmutm CR.
The signal and background contributions after the likelihood fit to data (\enquote{Post-Fit}) under the signal-plus-background hypothesis are shown as filled histograms.
The total signal-plus-background prediction before the likelihood fit to data (\enquote{Pre-Fit Bkg.}) is shown as a dashed blue histogram in the upper panel.
The ratio of the data to the total prediction (\enquote{Pred.}) is shown in the lower panel, separately for the post-fit prediction (black points) and pre-fit background (dashed blue line).
The combined statistical and systematic uncertainty in the prediction is indicated by the blue hatched band.
The last bin in each figure contains the overflow.}
\end{figure}
 
\subsubsection{Charge misassignment}
\label{sec:bkg_reducible_chg_miss}
 
Backgrounds due to leptons with an incorrectly assigned charge
primarily affect the $\ll$ channel and arise mainly from $t\bar{t}$
production, with one electron either undergoing hard bremsstrahlung
followed by an asymmetric conversion ($e^\pm \to e^\pm \gamma^* \to
e^\pm e^+ e^-$) or having mismeasured track curvature.  The muon charge
misassignment rate is negligible in the $\pt$ range relevant to this
analysis. The electron charge misassignment rate is measured in data
by using samples of $Z \to e^+e^-$ events reconstructed as either same-charge
pairs or opposite-charge pairs, with the non-$Z$ background
subtracted via a sideband method utilising events outside the $Z$ boson mass
window. The sideband region is defined by fitting a Gaussian
function to the $Z$ boson mass distribution and selecting events with
$m(\ell\ell)$ more than one and less than four standard deviations
from the fitted mean.
 
The charge misassignment rate is extracted from the ratio of the
numbers of same-charge and opposite-charge events close to the $Z$-boson mass
through a likelihood approach taking into account the possibility that
both electron charges are misassigned. The rates are parameterised
as a function of electron $\pt$ and $|\eta|$, and vary from about
$10^{-5}$ for low-$\pt$ electrons ($17 \leq \pt \leq 50~\gev$) with
$|\eta| \leq 1.37$, to about $4 \times 10^{-3}$ for high-$\pt$
electrons ($\pt \geq 100~\gev$) with $2 \leq |\eta| \leq 2.47$.  To
estimate the QMisID background in each of the corresponding event
categories, the measured charge misassignment rate is applied to
data events satisfying the requirements of the $\ll$ channels, except
that the two leptons must have opposite charges.


\section{Systematic uncertainties}
\label{sec:systematics}

 
Many sources of systematic uncertainty are considered for the
\xs measurements. These are grouped
into three categories: instrumental uncertainties that affect physics
object reconstruction; modelling uncertainties relevant for MC
simulation-based estimations; and uncertainties affecting
the estimation of misreconstructed-lepton backgrounds.
 
\subsection{Instrumental uncertainties}
\label{sec:inst_syst}
Instrumental systematic uncertainties related to the trigger
efficiency, lepton reconstruction and identification, jet calibration,
$b$-tagging, and non-prompt-lepton BDT working point
calibrations are considered in the analysis.
Physics object reconstruction and calibration are provided together with a systematic uncertainty model for propagating the corresponding uncertainties to the analysis. The calibrations are applied either as an overall event reweighting, or scale factor, to correct the simulation to the data, or as a rescaling (smearing) of the object's energy or momentum scale (resolution) and come with associated uncertainties.
The combined effect of these uncertainties is less than 10\% in the inclusive measurement,
but reaches approximately 18\% in the higher jet-multiplicity bins of the differential measurement.
 
Uncertainties associated with the lepton selection arise from the
trigger, reconstruction, identification and isolation efficiencies,
and the lepton momentum scale and
resolution~\cite{PERF-2015-10,EGAM-2018-01,MUON-2018-03,MUON-2022-01}.
Uncertainties in the non-prompt-lepton BDT calibration are estimated through a
$Z\to\ell\ell$ tag-and-probe method and cover uncertainties related to the
$Z(\to\ell\ell)$+jets MC modelling, the template cut/shape,
the $m_{\ell\ell}$ window, the tag-and-probe lepton selections, the multijet background, the non-prompt
lepton background, the luminosity, the \xs{}s of the considered
processes, and the limited number of events in simulation and data. The combined effect of
these uncertainties is approximately 2\% in the inclusive measurement
and typically less than 5\% in the differential measurements.
 
Uncertainties associated with jet reconstruction and calibration
arise from the jet energy scale~(JES), the JVT requirement and the jet
energy resolution~(JER). The JES and its uncertainties are derived by
combining information from test-beam data, collision data and
simulation~\cite{JETM-2018-05}. The JES and JER have 30 and 13
components, respectively, in the fit. The uncertainties in the JES, JER
and JVT increase at lower jet \pt.  The combined effect of these
uncertainties is approximately 1.5\% in the inclusive measurement,
while in the differential measurements it is approximately 5\% for
intermediate jet multiplicities and rises to 12\%--18\% for low and high
jet multiplicities.
 
The efficiency of the $b$-tagging algorithm is measured for each
jet flavour using control samples in data and in simulation. From
these measurements, correction factors are derived to correct the
tagging rates in the simulation. For \bjet{}s, the
correction factors and their uncertainties are estimated from data
using dileptonic \ttbar events~\cite{FTAG-2018-01,FTAG-2019-07}.
For $c$-jets, they are derived from jets arising from \Wboson boson decays
in \ttbar events~\cite{FTAG-2020-08}.
For light-flavour jets, the correction factors are derived from dijet
events~\cite{FTAG-2019-02}. Uncertainties from sources affecting
the $b$- and $c$-tagging efficiencies are evaluated as a function of
jet \pt, including bin-to-bin correlations. The uncertainties in the
efficiency for tagging light-flavour jets depend on the jet \pt and
$|\eta|$.  An additional uncertainty is assigned to account for the
extrapolation of the $b$-tagging efficiency measurement from the \pt
region used to determine the correction factors to regions with higher
\pt. The combined effect of these uncertainties is approximately 0.2\%
in the inclusive measurement, whereas in the differential measurements, it is
approximately 3\% in the intermediate \bjet multiplicity bins and
rises to 10\%--18\% for lower and higher \bjet multiplicities.
 
The uncertainty in the reweighting of the MC pile-up distribution to
match the distribution in data is evaluated by varying the reweighting
factors and has a very small impact on either the inclusive or
differential results.
 
The uncertainty in the combined 2015--2018 integrated luminosity is 0.83\% \cite{DAPR-2021-01},
obtained using the LUCID-2 detector \cite{LUCID2} for the primary luminosity measurements,
complemented by measurements using the inner detector and calorimeters.

\subsection{MC modelling uncertainties}\label{sec:systematics_theory}
 
Systematic uncertainties in the \ttW MC predictions due to missing
higher-order QCD corrections in the modelling are estimated by doubling and
halving the factorisation and renormalisation scales chosen for the nominal sample.
To estimate the uncertainty due to ambiguities in the ME and
PS algorithm and parameter choices, the nominal \Sherpa
prediction is compared with the prediction of the \MGPY[8] FxFx sample
described in Section~\ref{sec:datamc}. In addition, a dedicated parton
shower model uncertainty is estimated as the relative
difference between the \POWPY[8] and \POWHER[7] predictions and
applied to the nominal \ttW prediction. Uncertainties in the PDF
modelling are included by varying the value of \alphas and using
different PDF sets.  The \ttW modelling uncertainty is one of the largest
uncertainties in the analysis and leads to a combined impact (summing
each contribution in quadrature) of 6\% in the inclusive measurement and,
depending on the bin, 10\%--20\% in the differential measurement.
 
The uncertainty in \ttH modelling due to the initial-state radiation
(ISR) is estimated using weights in the matrix element calculation and
the parton shower. The factorisation and renormalisation scales are
varied, and ISR \alphas variations are taken from the A14 tune.
Additionally, uncertainties in the ME (PS algorithm -- parton shower, hadronisation and underlying-event modelling)
are estimated by comparing the nominal \POWPY[8] MC simulation
with \MGPY[8] (\POWHER[7]).
 
Uncertainties due to missing higher orders in QCD are estimated for
\ttZ production by varying the nominal factorisation and
renormalisation scales by factors of $0.5$ and $2.0$. Uncertainties in
additional-jet modelling are estimated with ISR \alphas
variations taken from the A14 tune. Parton shower,
hadronisation and underlying-event modelling uncertainties are estimated
by comparing the nominal \MGPY[8] MC simulation with
\MGHER[7].
 
Uncertainties in $VV$ modelling due to missing higher orders
in QCD are estimated by varying the nominal factorisation and
renormalisation scales by factors of $0.5$ and $2.0$.
In addition, systematic uncertainties associated with
the data-driven $N_\mathrm{jets}$-dependent correction
factors applied to $VV$ production are derived from
the fit function parameter uncertainties.

Alternative $t\bar{t}$ MC samples are used for shower uncertainties in
the HF non-prompt-lepton background estimate and for uncertainties in
the extrapolation from the CRs to the SRs. For this, the \POWHER[7.1.3]
and \SHERPA[2.2.10] samples are used.

The uncertainties in higher-order \xs{}s are taken into
account by using the values given in
Section~\ref{sec:datamc} for all background processes except
those whose normalisation is measured in the fit.
In the $t\bar{t}Z$ MC sample, the component
containing additional heavy-quark-initiated jets is normalised to the theoretical prediction and assigned a 50\%
uncertainty.

\subsection{Non-prompt-lepton background modelling uncertainties}\label{sec:systematics_HF_conv_QmisID}
 
For the estimation of non-prompt-lepton backgrounds from HF decays and
conversions, the following uncertainties are taken into account.
 
Uncertainties are estimated for the modelling of the non-prompt-lepton BDT input variables,
namely the muon's energy deposit in the calorimeter relative to the expected value
($E_\mathrm{cluster}/E_\mathrm{expected}$), the electron track \pt
divided by the jet track \pt, and the secondary vertex's longitudinal significance, using tracks with $\pt > 500$~MeV for non-prompt muons.
These uncertainties are included as variations that can affect the shape of distributions in each region, but not their normalisation.
 
Extrapolation uncertainties in the non-prompt-lepton background yields are
derived to account for non-prompt-lepton rate differences between the
\textit{Tight--not-VeryTight} and \textit{VeryTight} non-prompt-lepton BDT working points.
These uncertainties are obtained by comparing the non-prompt-lepton
efficiencies in the nominal \ttbar simulation with those in an alternative
\ttbar simulation as a function of lepton \pt and separately for
electrons and muons.
A constant uncertainty of 20\% is estimated for electrons and muons.
 
Uncertainties in the $t\bar{t}$ parton shower are assessed by using the
\POWHER[7] $t\bar{t}$ MC simulation as a variation of the nominal
\POWPY[8] prediction, separately for the electron and muon channels.
 
Uncertainties in the modelling of \ttbar production in association
with heavy-flavour jets are accounted for by assigning an uncorrelated
50\% uncertainty to the $\ttbar+b$ and $\ttbar+c$ background
processes.
 
Uncertainties in the electron conversion background extrapolation from
$Z$-enriched to \ttbar-enriched regions are derived from the
residual data/simulation mismodelling in
dedicated \ll validation regions with two tight same-sign leptons,
requiring one of them to be a conversion candidate.
Internal and material conversion extrapolation
uncertainties of 50\% and 10\%, respectively, are applied.
 
A systematic uncertainty of 10\%--60\% is assigned to the background
from electrons with misidentified charge. The uncertainty increases
with electron \pt\ and decreases with $|\eta|$. The uncertainty is
assessed by combining the uncertainties in the measurement of the
charge misassignment rate, the changes in the rate when varying the
$m_Z$ window selection, and the different rates measured in data and
$Z\rightarrow e^+e^-$ simulation.


\section{Results}
\label{sec:results}

 
A maximum-likelihood fit to all bins in the SRs and the 10 CRs (see Tables~\ref{tab:selectionSR} and~\ref{tab:selectionCR})
is performed to simultaneously determine the background and the \ttW signal yields that are most consistent with the data.
In the six HF non-prompt electron and muon CRs
(\llttetm, \llttemt, \llttemm, \llttmutm, \llttmumt, \llttmumm),
the sub-leading lepton's transverse momentum ($\pt^{\ell1}$) distribution is used, whereas in the two conversion CRs
(\lllintc, \lllmatc) and the \lllVV CR, the total event yield (i.e.~a single bin) is
used. In the \lllttZ CR, the $b$-jet multiplicity is fitted.
 
For the inclusive \ttW \xs measurement, the event yields in 56 SR bins (48 for \llSS and 8 for \lll) are fitted to data together with the CRs, and the results are shown in Section~\ref{sec:inclxsec}.
For the differential \ttW \xs measurement, various distributions in eight SRs (two \llSS and six \lll signal region categories) are fitted to data along with the same CRs used in the inclusive measurement. The corresponding results are shown in Section~\ref{sec:diffxsec}.

\subsection{Inclusive \xs measurement}
\label{sec:inclxsec}

 
The \ttW \xs is measured inclusively and for the fiducial regions
defined in Section~\ref{sec:events}.
The total event yields in the signal categories described in
Table~\ref{tab:selectionSR} are fitted simultaneously with the
control regions to data.
The likelihood function ${\cal L}(\mu,\vec{\lambda}, \vec{\theta})$ is
constructed as a product of Poisson probability terms over all bins
considered in the measurement, and depends on the following: the signal-strength
parameter, $\mu$, defined as a multiplicative factor applied to the
predicted yield for the \ttW signal; $\vec{\lambda}$, the
normalisation factors for several backgrounds (see
Section~\ref{sec:background}); and $\vec{\theta}$, a set of nuisance
parameters (NP) encoding systematic uncertainties in the signal and
background expectations~\cite{Cranmer:2012sba}. Systematic
uncertainties can impact the estimated signal and background rates,
the migration of events between categories, and the shape of the
fitted distributions; they are summarised in
Section~\ref{sec:systematics}.  Both $\mu$ and $\vec{\lambda}$ are
treated as free parameters in the likelihood fit. The NPs
$\vec{\theta}$ allow variations of the expectations for signal and
background according to the systematic uncertainties, subject to
Gaussian constraints in the likelihood fit.  Their fitted values
represent the deviations from the nominal expectations that are needed
to provide the best fit to the data.  Statistical uncertainties in each
bin due to the limited size of the simulated event samples are taken into
account with dedicated parameters, using the Beeston--Barlow \enquote{lite}
technique~\cite{Barlow:1993dm}.
 
The observed and expected yields in
the 10 CRs and the 56 SRs after the combined likelihood fit under the
signal-plus-background hypothesis are compared in
Figures~\ref{fig:summary_incl_CR} and~\ref{fig:summary_incl_SR}.  The
corresponding post-fit yields for the SRs and CRs are shown in
Tables~\ref{tab:SRspostfityields},~\ref{tab:CRspostfityields2L},
and~\ref{tab:CRspostfityields3L}.
There are no significant deviations from the predictions after the fit to data.
 
\begin{figure}[t!]
\centering
\includegraphics[height=0.5\textwidth]{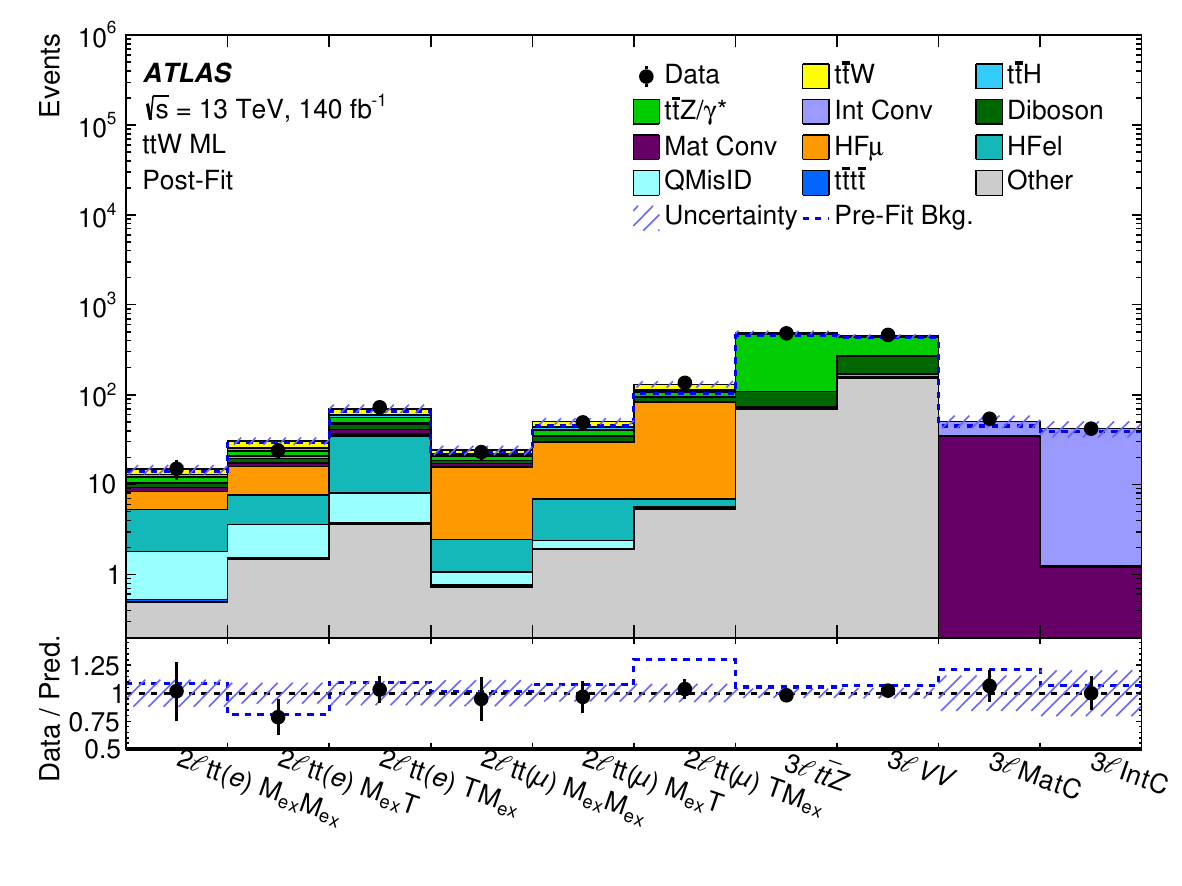}
\caption{Comparison between data and the signal-plus-background prediction for the event yields in the ten control region categories.
The signal and background contributions after the likelihood fit to data (\enquote{Post-Fit}) under the signal-plus-background hypothesis are shown as filled histograms.
The total signal-plus-background prediction before the likelihood fit to data (\enquote{Pre-Fit Bkg.}) is shown as a dashed blue histogram in the upper panel.
The ratio of the data to the total prediction (\enquote{Pred.}) is shown in the lower panel, separately for the post-fit prediction (black points) and pre-fit background (dashed blue line).
The combined statistical and systematic uncertainty in the prediction is indicated by the blue hatched band. }
\label{fig:summary_incl_CR}
\end{figure}
 
\begin{figure}[t!]
\centering
\subfloat[]{\includegraphics[height=0.4\textwidth]{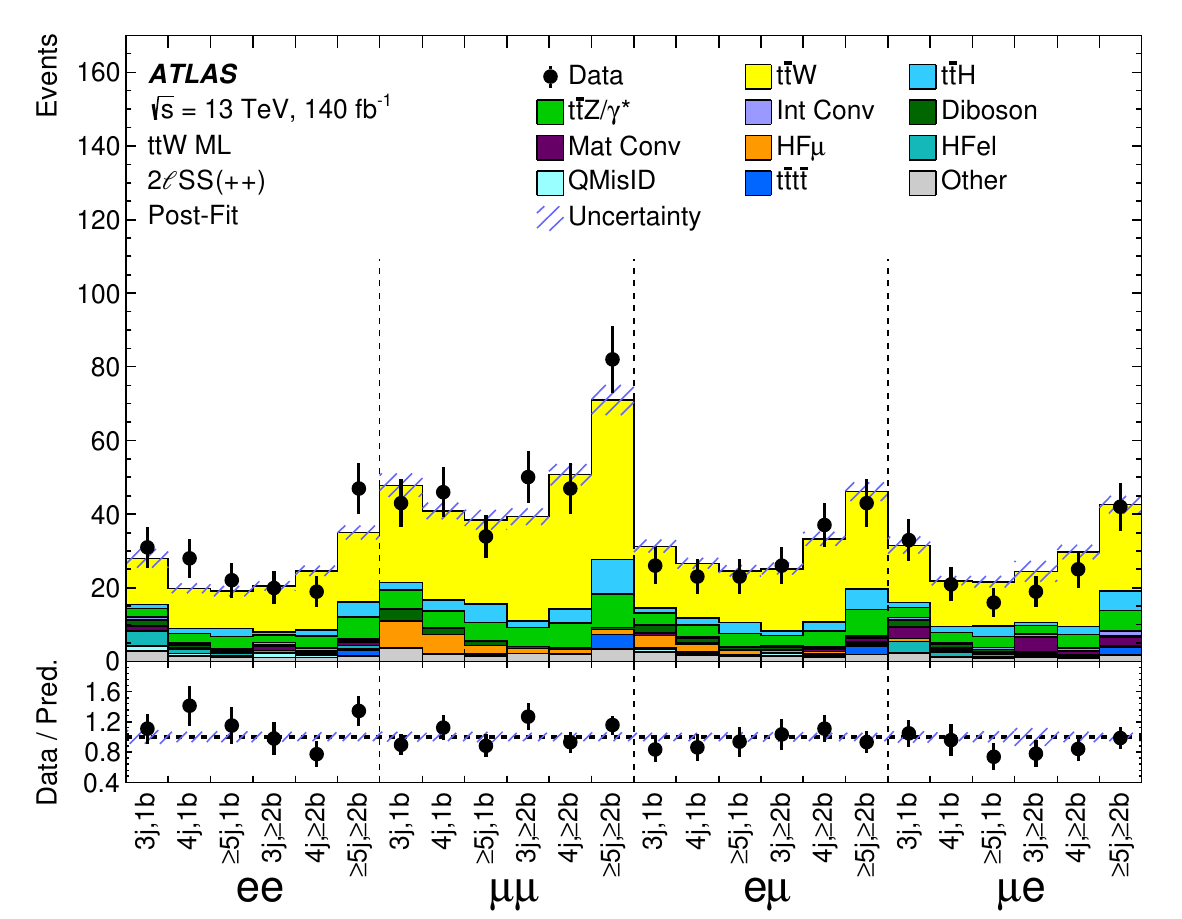}}
\subfloat[]{\includegraphics[height=0.4\textwidth]{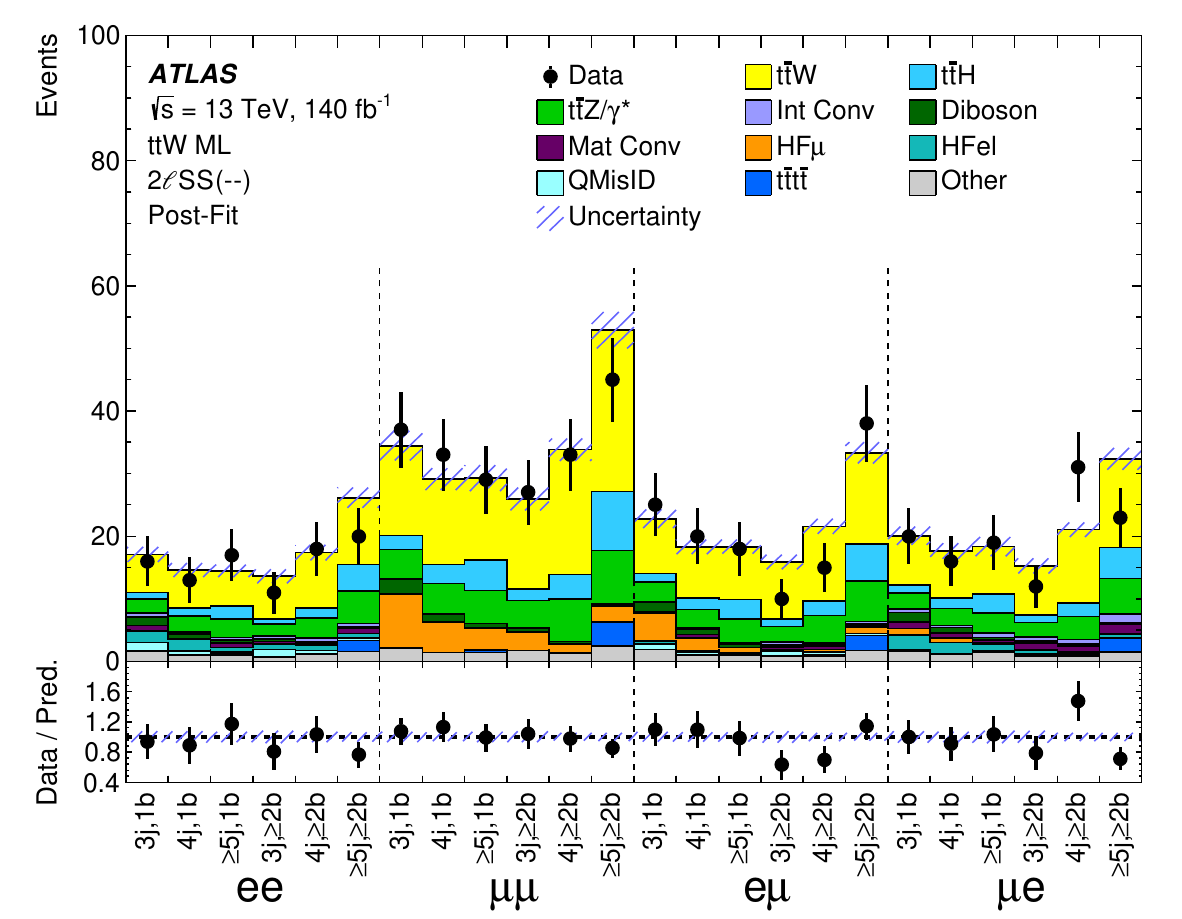}}\\
\subfloat[]{\includegraphics[height=0.4\textwidth]{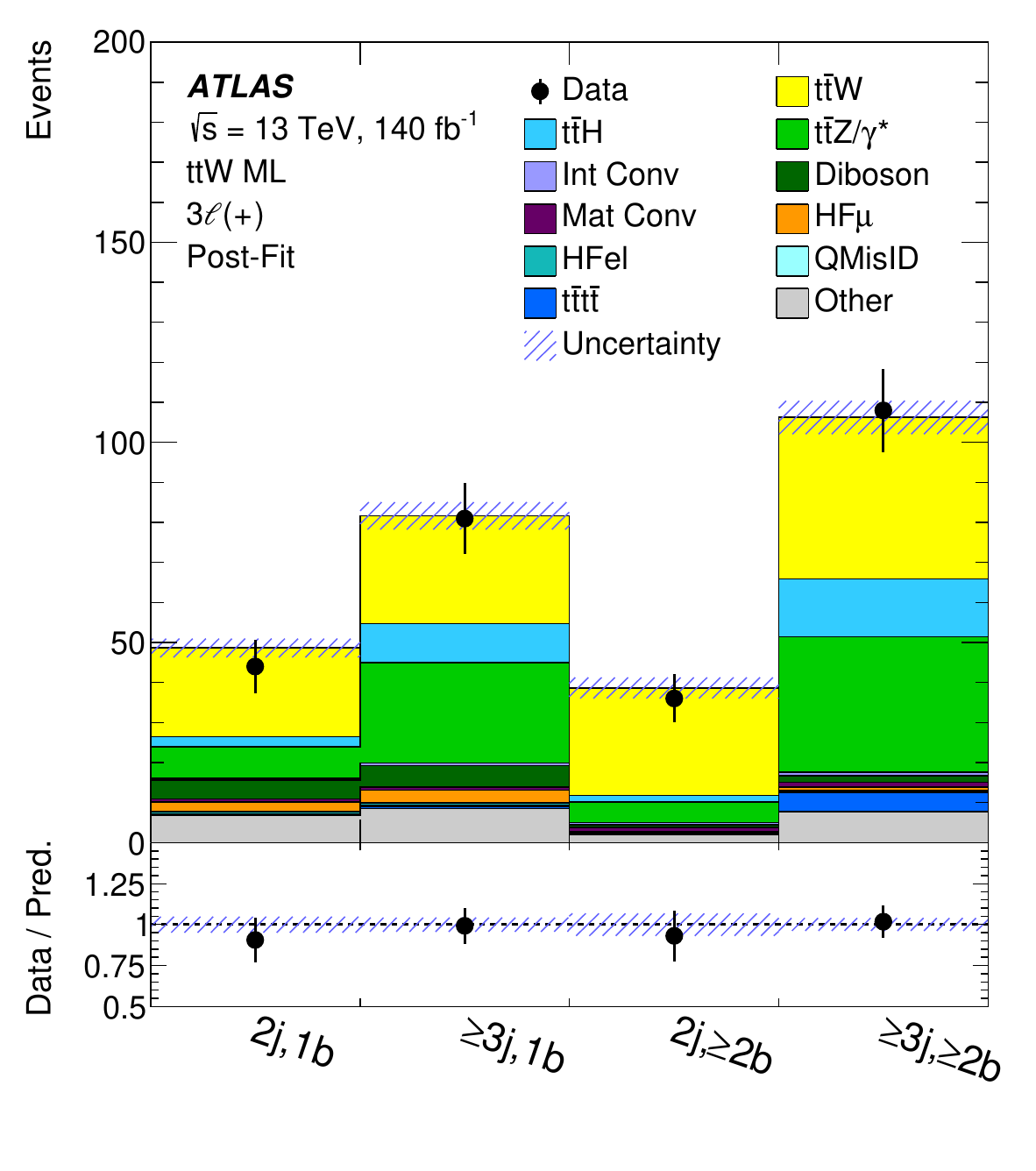}}
\subfloat[]{\includegraphics[height=0.4\textwidth]{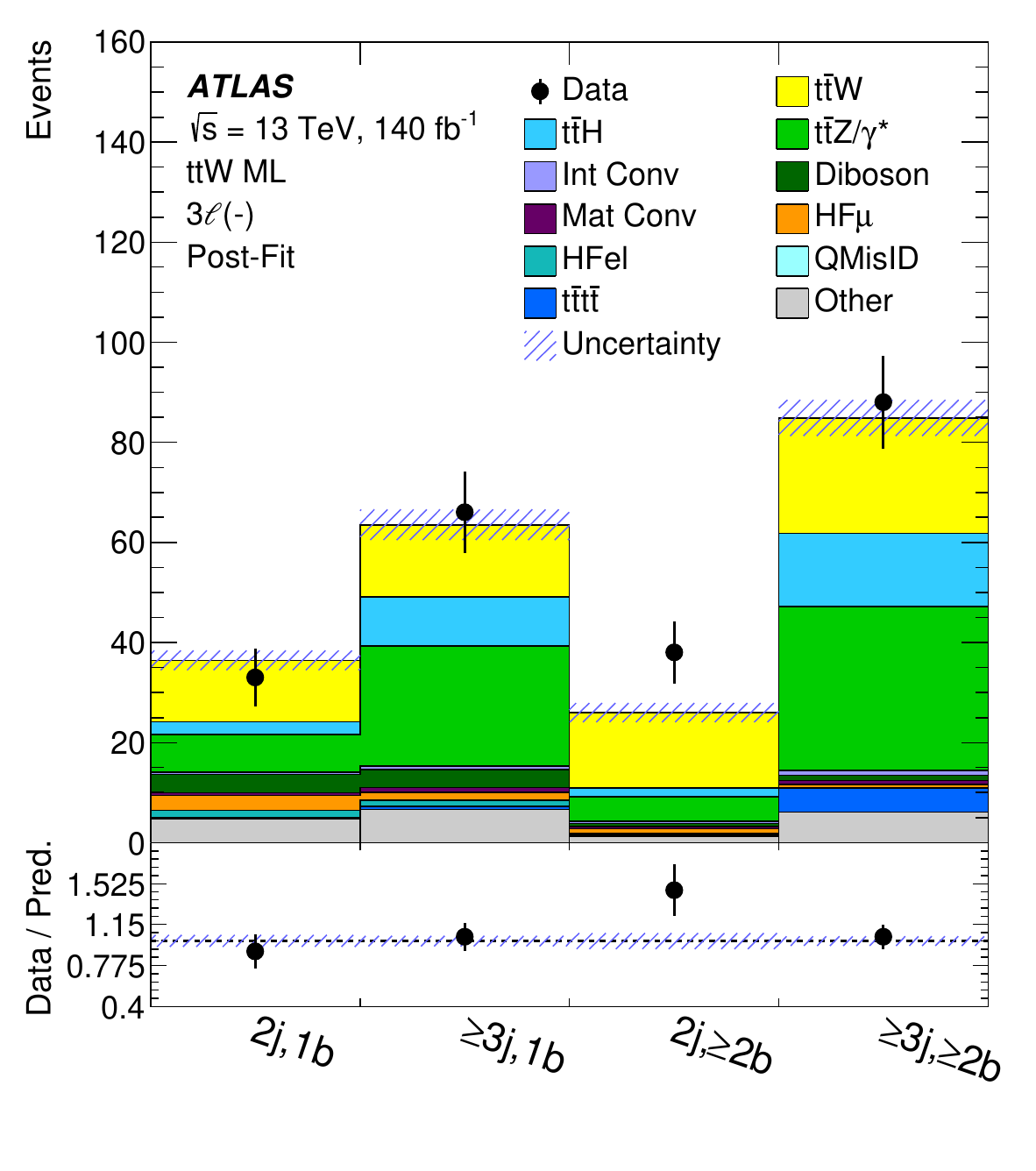}}
\caption{Comparison between data and the signal-plus-background prediction for the event yields in the (a) \twopp, (b) \twomm, (c) \threep~and (d) \threem~signal region categories.
The signal and background contributions after the likelihood fit to data (\enquote{Post-Fit}) under the signal-plus-background hypothesis are shown as filled histograms.
The ratio of the data to the total post-fit prediction (\enquote{Pred.}) is shown in the lower panel.
The combined statistical and systematic uncertainty in the prediction is indicated by the blue hatched band.
}
\label{fig:summary_incl_SR}
\end{figure}
 
\begin{table}[htbp]
\begin{center}
\caption{Summary of observed and predicted yields in the four signal region categories. The background prediction is shown after the combined likelihood fit to data under the signal-plus-background hypothesis across all control region and signal region categories. The quoted uncertainties consider all statistical and systematic uncertainties after the fit to data. The uncertainty on the total yield can be smaller than the quadratic sum of uncertainties of the individual contributions due to the anti-correlations resulting from the likelihood fit. \label{tab:SRspostfityields}}
\begin{tabular}{l|c|c|c|c}
\hline
& \textbf{$\ell^-\ell^-$ SR}    & \textbf{$\ell^+\ell^+$ SR}   & \textbf{$\ell^+\ell^-\ell^-$ SR}  & \textbf{$\ell^-\ell^+\ell^+$ SR} \\
\hline \hline
$t\bar{t}W$                     & $261\pm18$                   & $468\pm30$                     & $64\pm6$                            & $115\pm10$\\
$t\bar{t}H$                     & $66\pm9$                        & $65\pm9$                         & $29\pm4$                            & $28\pm4$  \\
$t\bar{t}Z/\gamma^*$     		& $95\pm9$                        & $100\pm10$                    & $69\pm8$                             & $71\pm8$  \\
Int Conv 						& \phantom{0}$9\pm5$       & \phantom{0}$9\pm5$      & \phantom{0}$2.5\pm1.4$      & \phantom{0}$2.5\pm1.4$   \\
Diboson                          & $16\pm5$                        & $22\pm7$                        & \phantom{0}$9.1\pm2.7$      & $12\pm4$  \\
Mat Conv                        & $11.2\pm3.3$                  & $19\pm5$                        & \phantom{0}$2.5\pm0.8$      & \phantom{0}$3.9\pm1.1$   \\
HF($\mu$)                      & \phantom{0}$35\pm11$   & \phantom{0}$30\pm10$   & \phantom{0}$6.1\pm2.5$     & \phantom{0}$6.6\pm2.6$   \\
HF(e)                              & $14\pm6$                        & $15\pm6$                        & \phantom{0}$2.9\pm1.3$      & \phantom{0}$2.0\pm0.8$   \\
QMisID                           & \phantom{0}$8.2\pm2.7$ & \phantom{0}$8.2\pm2.7$ & \phantom{0}$0.69\pm0.15$ & \phantom{0}$0.66\pm0.13$ \\
$t\bar{t}t\bar{t}$              & $12\pm9$                         & $12\pm9$                        & \phantom{0}$5\pm4$           & \phantom{0}$5\pm4$   \\
Other                              & $32\pm5$                         & $42\pm6$                        & $18.5\pm3.1$                       & $25.0\pm3.3$  \\
\hline \hline
Total                               & $559\pm14$                      & $788\pm23$                    & $209\pm7$\phantom{0}       & $273\pm9$\phantom{0} \\
\hline
Data                               & 546                                    & 803                                  & 225                                      & 269           \\
\hline
\end{tabular}
\end{center}
\end{table}

\begin{table}[htbp]
\begin{center}
\caption{Summary of observed and predicted yields in the \llSS control region categories. The background prediction is shown after the combined likelihood fit to data under the signal-plus-background hypothesis across all control region and signal region categories. The quoted uncertainties consider all statistical and systematic uncertainties after the fit to data. The uncertainty on the total yield can be smaller than the quadratic sum of uncertainties of the individual contributions due to the anti-correlations resulting from the likelihood fit. \label{tab:CRspostfityields2L}}
\resizebox{1.0\textwidth}{!}{
\begin{tabular}{l|c|c|c|c|c|c}\hline
& \textbf{\llttemm} & \textbf{\llttemt} & \textbf{\llttetm} & \textbf{\llttmumm}& \textbf{\llttmumt} &  \textbf{\llttmutm} \\
\hline\hline
$t\bar{t}W$                     & $1.8\pm0.5$       & $5.1\pm0.5$       & $10.4\pm1.4$\phantom{0}      & $2.4\pm0.7$       & $7.3\pm1.2$        &  $17.9\pm2.5$\phantom{0}       \\
$t\bar{t}H$                     & $0.79\pm0.19$     & $1.73\pm0.32$     & $3.5\pm0.6$       & $1.11\pm0.21$     & $2.8\pm0.5$        &  $5.8\pm1.0$        \\
$t\bar{t}Z/\gamma^*$            & $1.49\pm0.19$      & $3.2\pm0.4$       & $6.8\pm0.7$       & $2.07\pm0.29$     & $5.2\pm0.6$        &  $11.4\pm1.4$\phantom{0}       \\
Int Conv 						  & $0.18\pm0.12$     & $0.9\pm0.5$       & $1.6\pm0.9$       & $0.11\pm0.07$     & $0.54\pm0.32$      &  $0.27\pm0.16$      \\
Diboson                         & $1.2\pm0.3$      & $2.3\pm0.6$       & $6.8\pm1.4$       & $1.4\pm0.4$       & $5\pm1$            &  $10.6\pm2.2$\phantom{0}       \\
Mat Conv                        & $0.95\pm0.31$     & $1.5\pm1.3$       & $3.7\pm1.7$       & $1.5\pm0.5$       & $1.0\pm0.4$        &  $1.3\pm0.8$        \\
HF($\mu$)                         & $3.0\pm0.8$       & $8.1\pm1.6$       & $1.9\pm0.9$       & $13.1\pm2.5$\phantom{0}      & $22\pm4$\phantom{0}           &  $75\pm11$          \\
HF(e)                            & $3.5\pm1.0$       & $4.1\pm1.2$       & $27\pm9$\phantom{0}          & $1.4\pm0.5$       & $4.5\pm1.5$        &  $1.3\pm0.9$        \\
QMisID                          & $1.3\pm0.8$       & $2.0\pm1.4$       & $4.2\pm2.1$       & $0.31\pm0.2$\phantom{0}      & $0.43\pm0.35$      &  $0.21\pm0.09$      \\
$t\bar{t}t\bar{t}$              & $0.033\pm0.027$   & $0.035\pm0.029$   & $0.07\pm0.06$   & $0.039\pm0.032$   & $0.07\pm0.05$    &  $0.11\pm0.09$      \\
Other                           & $0.49\pm0.07$     & $1.49\pm0.17$     & $3.7\pm0.4$       & $0.72\pm0.09$     & $1.88\pm0.17$      &  $5.3\pm0.8$        \\
\hline\hline
Total                           & $14.6\pm1.8$\phantom{0}  & $30.4\pm2.8$\phantom{0}  & $69\pm7$\phantom{0}  & $24.0\pm2.8$\phantom{0}  & $50\pm4$\phantom{0}  & $129\pm11$\phantom{0}         \\
\hline
Data                            & 15                & 24                & 72                & 23                & 49                 &  135                \\
\hline
\end{tabular}
}
\end{center}
\end{table}
 
\begin{table}[htbp]
\begin{center}
\caption{Summary of observed and predicted yields in the \lll control region categories. The background prediction is shown after the combined likelihood fit to data under the signal-plus-background hypothesis across all control region and signal region categories.\label{tab:CRspostfityields3L}}
\begin{tabular}{l|c|c|c|c}\hline
& \textbf{\lllintc} & \textbf{\lllmatc} & \textbf{\lllttZ} & \textbf{\lllVV}  \\
\hline\hline
$t\bar{t}W$                   & -- & -- & $7.2\pm0.8$ & $13.0\pm0.9$\phantom{0}\\
$t\bar{t}H$                   & -- & -- & $10.0\pm1.5$\phantom{0} & $4.3\pm0.6$\\
$t\bar{t}Z/\gamma^*$         & -- & -- & $360\pm40$\phantom{0} & $160\pm23$\phantom{0}\\
Int Conv						 & $41\pm9$\phantom{0} & $15\pm4$\phantom{0} & $0.055\pm0.032$ & $0.41\pm0.25$\\
Diboson                       & $0.035\pm0.020$ & $0.88\pm0.20$ & $35\pm11$ & $100\pm28$\phantom{0}\\
Mat Conv                      & $1.2\pm0.9$ & $34\pm9$\phantom{0} & $0.48\pm0.30$ & $2.1\pm0.7$\\
HF($\mu$)                       & -- & -- & $1.19\pm0.35$ & $8.4\pm2.7$\\
HF(e)                          & -- & -- & $0.40\pm0.22$ & $7.1\pm2.4$\\
QMisID                        & -- & -- & $0.19\pm0.13$ & $0.8\pm0.4$\\
$t\bar{t}t\bar{t}$            & -- & -- & $1.7\pm1.3$ & $0.05\pm0.04$\\
Other                         & -- & -- & $68\pm19$ & $151\pm20$\phantom{0}\\
\hline\hline
Total                         & $42\pm9$\phantom{0} & $50\pm8$\phantom{0} & $487\pm21$\phantom{0} & $446\pm20$\phantom{0}\\
\hline
Data                          & 42 & 54 & 482 & 460\\
\hline
\end{tabular}
\end{center}
\end{table}

The best-fit value of the $\ttW$ \xs, $\sigma(\ttW)$, is:
\begin{equation*}
\sigma(\ttW) = 880\pm50\,\mathrm{(stat.)}\pm70\,\mathrm{(syst.)} = 880\pm80~\mathrm{fb}.
\label{eq:ttwxs}
\end{equation*}
 
It is compatible with the SM NNLO \xs of $745\pm 50\,\textrm{(scale)}\pm 13\,\textrm{(2-loop approx.)}\pm 19\;\textrm{(PDF, \alphas)}$~fb from Ref.~\cite{Buonocore_2023} at the level of $1.4\sigma$. It is also compatible with the SM \xs of $722\,^{+70}_{-78}\textrm{(scale)}\pm 7\,\textrm{(PDF)}$~fb from Ref.~\cite{Frederix:2021agh} at the level of $1.5\sigma$.
The systematic uncertainty, 7.9\%, is somewhat larger than the 5.8\% reported by CMS
in Ref.~\cite{CMS-TOP-21-011}. However, the systematic uncertainty would decrease to 5.6\%
if uncertainties related to algorithmic choices in the \ttW ME generator and PS,
which are not considered in the CMS measurement, were removed.
This modified version of the measurement does not include
ISR/FSR and colour reconnection \ttW uncertainties, which are considered in
the CMS measurement and amount to 0.8\% and 1\%, respectively.

The fiducial \ttW \xs, $\sigma_{\textrm{fid}}(\ttW)$, based on the selection
defined in Section~\ref{sec:events}, is also measured. The fit to data is set
up identically to the inclusive \xs measurement but with modified
systematic uncertainties to remove fiducial acceptance effects.
The measured value is:
\begin{equation*}
\sigma_{\textrm{fid}}(\ttW) = 20.9\pm1.1\,\mathrm{(stat.)}\pm1.9\,\mathrm{(syst.)} = 20.9\pm2.2~\mathrm{fb}.
\label{eq:ttwxsfid}
\end{equation*}
 
The relative uncertainty in the fiducial \xs measurement is slightly larger
than that of the inclusive measurement, which is contrary to what might
usually be expected. This is due to significant differences
in detector-level event reconstruction efficiency between the \Sherpa and \MGPY[8] FxFx \ttW generator models.
There are also differences in fiducial acceptance between the two generators,
but with opposite impact. Therefore, the two effects partially cancel out in the inclusive fit,
but the efficiency difference shows its full effect in the fiducial fit.

Figure~\ref{fig:results_ttWxs}(a) compares the
measured \ttW \xs with theory predictions from \SHERPA (as described in
Section~\ref{sec:datamc}), the \MGPY[8] FxFx prescription including
EWK corrections from Ref.~\cite{Frederix:2021agh}, the NLO+NNLL
calculation in Ref.~\cite{Kulesza:2020nfh}, and the NNLO calculation in
Ref.~\cite{Buonocore_2023}. The measurement from
CMS~\cite{CMS-TOP-21-011} is also shown.  The measured inclusive \xs
is larger than the theoretical predictions, and is compatible with the
measurement from CMS.

\begin{figure}[htbp]
\centering
\subfloat[]{\includegraphics[width=0.49\textwidth]{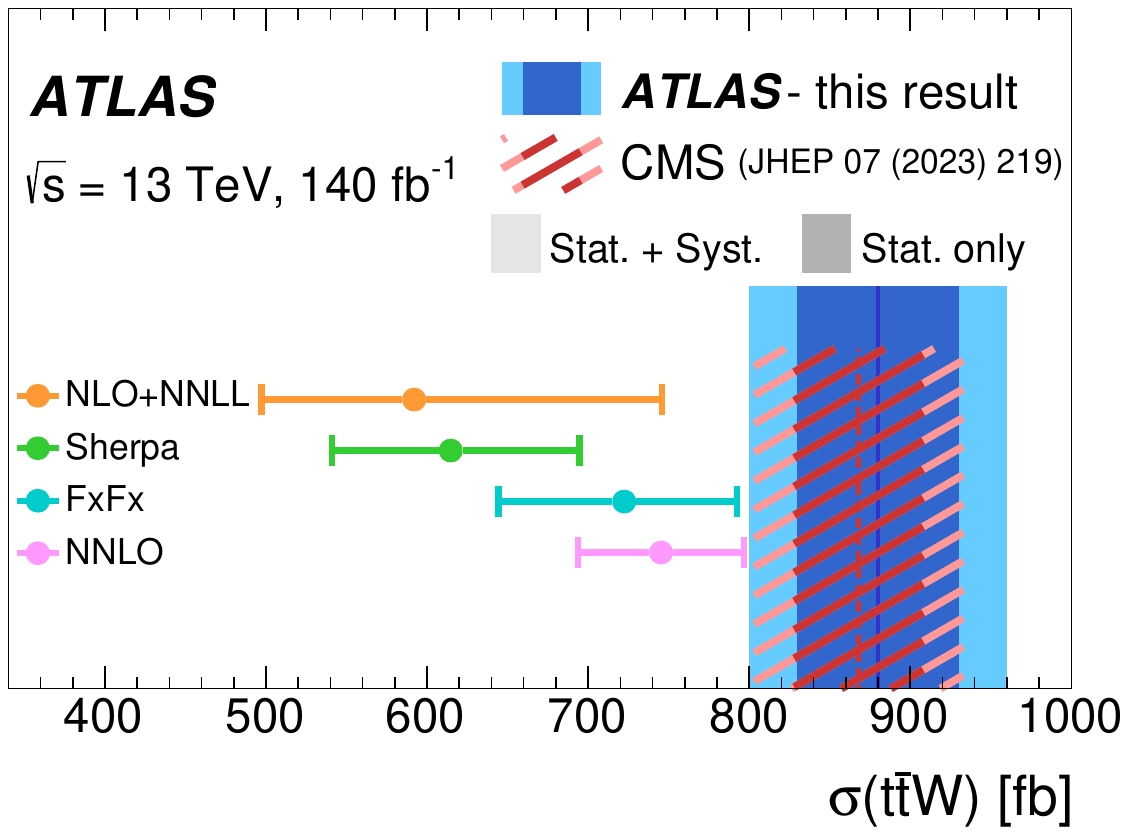}}
\subfloat[]{\includegraphics[width=0.49\textwidth]{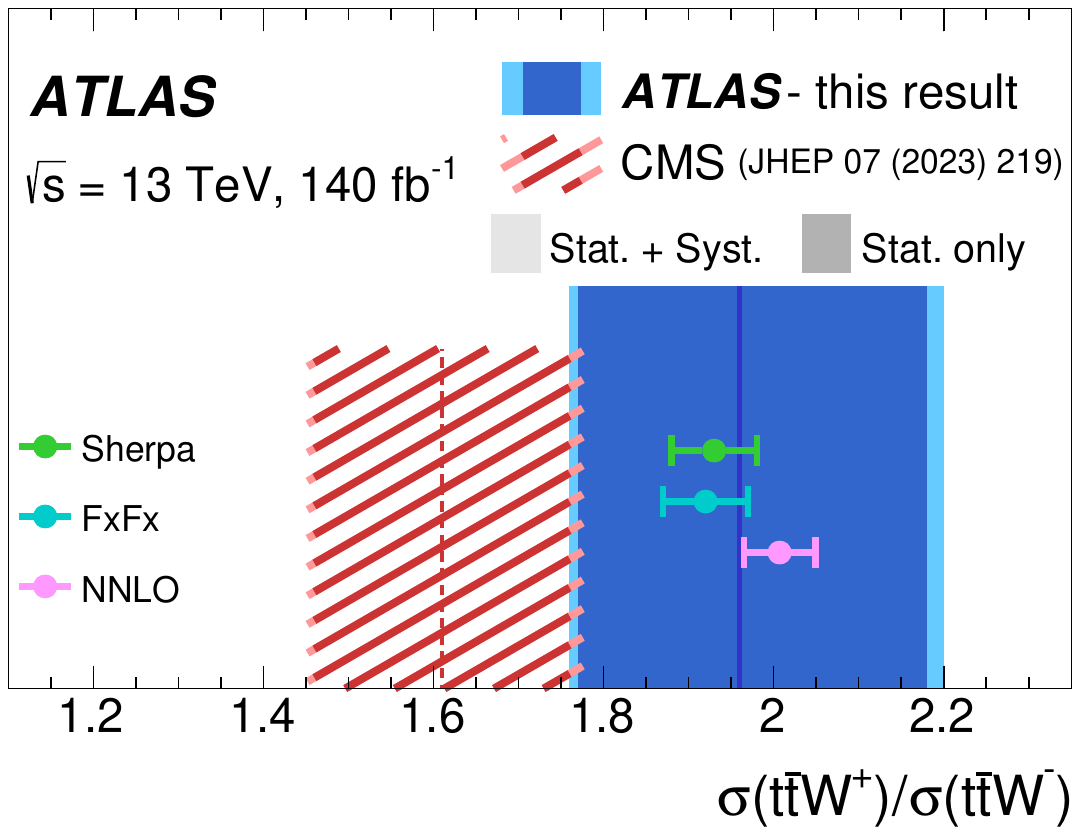}}
\caption{Comparison of (a) the measured inclusive \ttW \xs and (b) the $\ttW^{+}/\ttW^{-}$ \xs ratio with theoretical predictions from \SHERPA, the \MGPY[8] FxFx prescription including EWK corrections from Ref.~\cite{Frederix:2021agh}, the NLO+NNLL calculation from Ref.~\cite{Kulesza:2020nfh}, and the NNLO calculation from Ref.~\cite{Buonocore_2023}, and also the measurement from CMS~\cite{CMS-TOP-21-011}. The theoretical uncertainty in $\ttW^{+}/\ttW^{-}$ is calculated assuming scale variations are fully correlated between $\ttW^{+}$ and $\ttW^{-}$.}  \label{fig:results_ttWxs}
\end{figure}

Additional measurements are performed using the same analysis model,
but measuring the \xs{}s of $\sigma({\ttW}^{+})$ and $\sigma({\ttW}^{-})$
simultaneously. Splitting the measurement by $W$ boson charge is
interesting because a strong charge asymmetry is expected due to the
associated asymmetry in the valence quark PDF, and this feature is
often used to separate \ttW production from other processes with similar final states,
such as \ttH and \tttt production. The following values were measured:
\begin{equation}
\begin{split}
\sigma({\ttW}^{+}) = 583\pm34\,\mathrm{(stat.)}\pm47\,\mathrm{(syst.)} = 583\pm58~\mathrm{fb}, \\
\sigma({\ttW}^{-}) = 296\pm28\,\mathrm{(stat.)}\pm29\,\mathrm{(syst.)} = 296\pm40~\mathrm{fb}. \\
\end{split}
\label{eq:muttwplusminus}
\end{equation}
 
The ratio of the \xs{}s is extracted with another fit configuration that depends on $\mu(R)$, defined as a multiplicative factor applied to the ratio $R(\ttW) = \sigma({\ttW}^{+})/\sigma({\ttW}^{-})$ evaluated using Eq.~(\ref{eq:muttwplusminus}). The resulting measurement of $R(\ttW)$ is:
 
\begin{equation*}
R(\ttW) = 1.96\pm0.21\,\mathrm{(stat.)}\pm0.09\,\mathrm{(syst.)} = 1.96\pm0.22.
\label{eq:chargeasymm_best}
\end{equation*}
 
Figure~\ref{fig:results_ttWxs}(b) compares the measured charge
ratio with theoretical predictions and the measurement
from CMS. The measured value is in good agreement with the theoretical
prediction of
$R(\ttW)=2.007\pm2.1\%\,\mathrm{(scale)}\pm 1.7\%~\textrm{(PDF)}$
from Ref.~\cite{Buonocore_2023},
as well as those of $R(\ttW)=1.93\pm 0.02\,\mathrm{(scale)}\pm 0.04~\mathrm{(PDF)}$
from \SHERPA (as described in Section~\ref{sec:datamc}) and
$R(\ttW)=1.92\pm 0.02\,\mathrm{(scale)}\pm 0.04~\mathrm{(PDF)}$ from the
\MGPY[8] FxFx prescription including EWK corrections from Ref.~\cite{Frederix:2021agh}.
This measurement yields a value higher than the one measured by CMS, but
each measurement's uncertainty has a large statistical component.

\begin{figure}[htbp]
\centering
\includegraphics[width=0.6\textwidth]{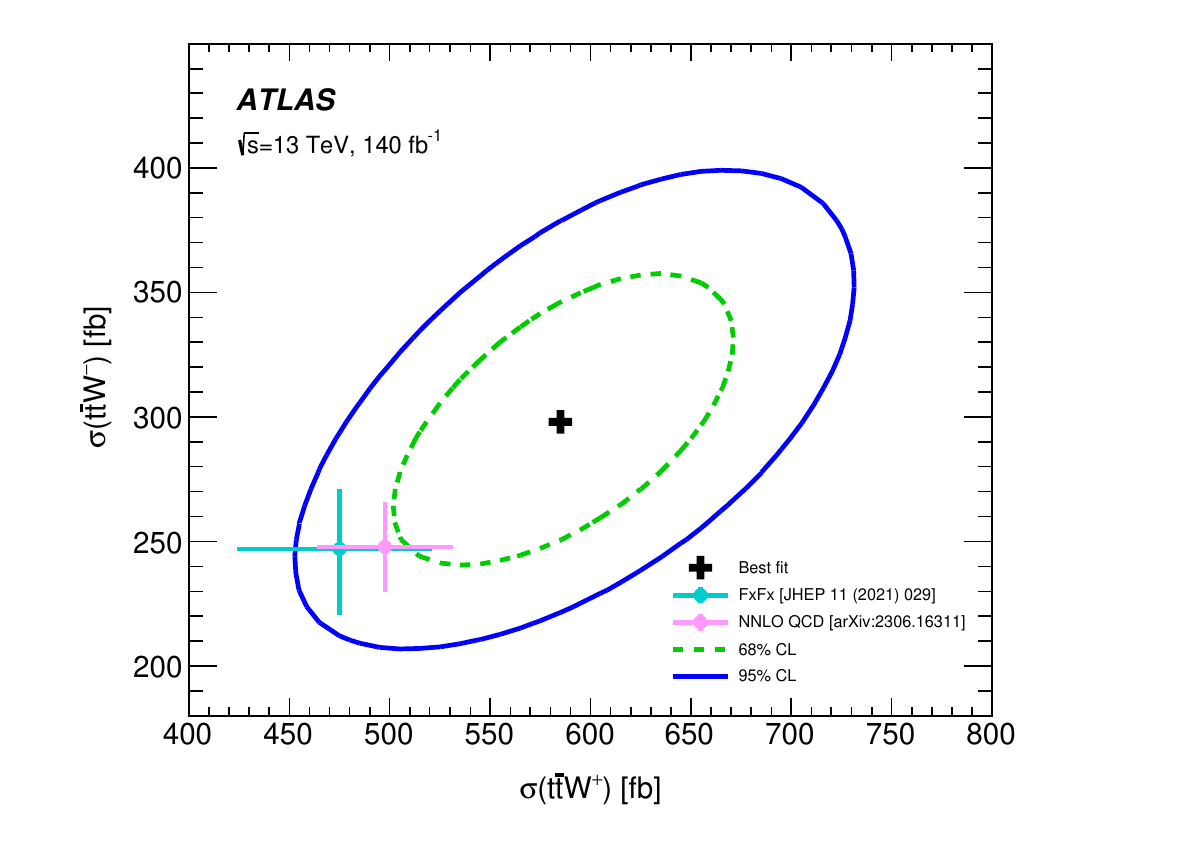}
\caption{Two-dimensional likelihood scan of the $\ttW^{+}$ and $\ttW^{-}$ cross-sections. The black cross indicates the best-fit point and the green and lilac markers show the SM predictions from Refs.~\cite{Frederix:2021agh} and~\cite{Buonocore_2023}, respectively. The blue dashed and solid lines correspond to the 68\% and 95\% confidence interval contours, respectively.}
\label{fig:results_pmcorr}
\end{figure}
 
Figure~\ref{fig:results_pmcorr} shows a two-dimensional likelihood
scan of the fitted values of $\ttW^{+}$ and $\ttW^{-}$ compared to the
SM prediction. There is a 58\% correlation between the $\ttW^{+}$ and
$\ttW^{-}$ measurements.

The parameter of interest can also be defined to be the relative
charge asymmetry, \rqasym, defined as:
 
\begin{equation*}
\rqasym = \frac{\sigma({\ttW}^{+}) - \sigma({\ttW}^{-})}{\sigma({\ttW}^{+}) + \sigma({\ttW}^{-})}.
\label{eq:relchargeasymm_def}
\end{equation*}
 
The measured value resulting from the best fit to data is:
\begin{equation*}
\rqasym = 0.33\pm0.05\,\mathrm{(stat.)}\pm0.02\,\mathrm{(syst.)} = 0.33\pm0.05.
\label{eq:relchargeasymm_best}
\end{equation*}
 
This value is in good agreement with the theoretical prediction
of $0.322\pm 0.003\,\mathrm{(scale)}\pm 0.007\,\mathrm{(PDF)}$ from \SHERPA.
 
The contributions from the different sources of uncertainty affecting the measured \xs are shown grouped by categories in Table~\ref{tab:sys_impact}, and with the breakdown for each individual nuisance parameter in Figure~\ref{fig:ranking}.
The most important systematic uncertainties arising from theoretical
predictions are in the modelling of the \ttW signal, followed by the
uncertainty in the prompt-lepton background normalisation, where the
theoretical uncertainties in the \ttH and \tttt production
normalisations are dominant.  The leading instrumental systematic
uncertainties originate from the $b$-tagging efficiency and the
non-prompt-lepton BDT's prompt-lepton efficiency.

\begin{table}[b]
\centering
\caption{The impact of the main sources of uncertainty on the measured quantities. The impact of each uncertainty on $\sigma(\ttW)$ is assessed by fixing the relevant fit parameters for a given uncertainty and re-fitting to data. The reduction in the total uncertainty in this modified fit gives the quoted impact. Different individual components used in the
fit are combined into categories. The leading contributions to the prompt-lepton background normalisation uncertainty are the \ttH and \tttt normalisation uncertainties. In the fourth row, \enquote{Fakes} stands for non-prompt leptons.}
\begin{tabular}{l|c|c|c|c}
\hline
& \textbf{$\frac{\Delta\sigma(\ttW)}{\sigma(\ttW)}$[\%]} & \textbf{$\frac{\Delta\sigma_{\textrm{fid}}(\ttW)}{\sigma_{\textrm{fid}}(\ttW)}$[\%]} & \textbf{$\frac{\Delta R(\ttW)}{R(\ttW)}$[\%]} & \textbf{$\frac{\Delta \rqasym}{\rqasym}$[\%]}  \\  \hline\hline
\ttW ME modelling                 & 6.0    & 7.0   & 0.9   & 1.3  \\
Prompt-lepton bkg.~norm.          & 3.1    & 3.1   & 1.6   & 2.1  \\
Lepton isolation BDT              & 2.3    & 2.3   & 0.7   & 1.0  \\
Fakes/$VV$/\ttZ norm.~(free to vary) & 2.3    & 2.5   & 1.7   & 2.4  \\
Non-prompt-lepton bkg.~modelling  & 2.0    & 2.0   & 2.4   & 3.2  \\
Trigger                           & 1.9    & 1.9   & 0.5   & 0.6  \\
MC statistics                     & 1.5    & 1.6   & 1.9   & 2.6  \\
\ttW PDF                          & 1.5    & 1.5   & 2.1   & 2.8  \\
Jet energy scale                  & 1.3    & 1.4   & 1.2   & 1.7  \\
Prompt-lepton bkg.~modelling      & 1.3    & 1.3   & 1.2   & 1.7  \\
Luminosity                        & 0.9    & 0.9   & 0.1  & \phantom{0}$<$0.10\phantom{$<$} \\
Charge Mis-ID                     & 0.7    & 0.7   & 0.4   & 0.6  \\
Jet energy resolution             & 0.5    & 0.5   & 0.6   & 0.9 \\
Flavour tagging                   & 0.4   & 0.5  & 0.5   & 0.6  \\
\ttW PS modelling                 & 0.4    & 0.4   & \phantom{0}0.29   & 0.4  \\
\ttW scale                        & \phantom{0}0.24   & \phantom{0}$<$0.10\phantom{$<$}   & \phantom{0}0.29   & \phantom{0}0.29  \\
Electron/photon reconstruction  & \phantom{0}0.21   & \phantom{0}0.22   & \phantom{0}0.11  & \phantom{0}0.13  \\
Muon                              & \phantom{0}0.15  & \phantom{0}0.17 & \phantom{0}$<$0.10\phantom{$<$} & 0.2  \\
\MET                               & \phantom{0}$<$0.10\phantom{$<$}  & \phantom{0}$<$0.10\phantom{$<$} & \phantom{0}0.15  & \phantom{0}0.21  \\
Pile-up                           & \phantom{0}$<$0.10\phantom{$<$}  & \phantom{0}0.11  & \phantom{0}$<$0.10\phantom{$<$} & \phantom{0}$<$0.10\phantom{$<$}  \\
\hline
Total systematic uncertainty  & 8\phantom{.0}    & 9\phantom{.0}  & 5\phantom{.0}   & 7\phantom{.0}   \\
Statistical uncertainty           & 5\phantom{.0}    & 5\phantom{.0}   & 10\phantom{.00}  & 14\phantom{.00}   \\ \hline
\hline
\textbf{Total}                    & 9\phantom{.0}   & 11\phantom{.00}  & 11\phantom{.00}  & 15\phantom{.00}    \\ \hline
\end{tabular}
\label{tab:sys_impact}
\end{table}

\begin{figure}[htbp]
\centering
\subfloat[]{\includegraphics[width=0.65\textwidth]{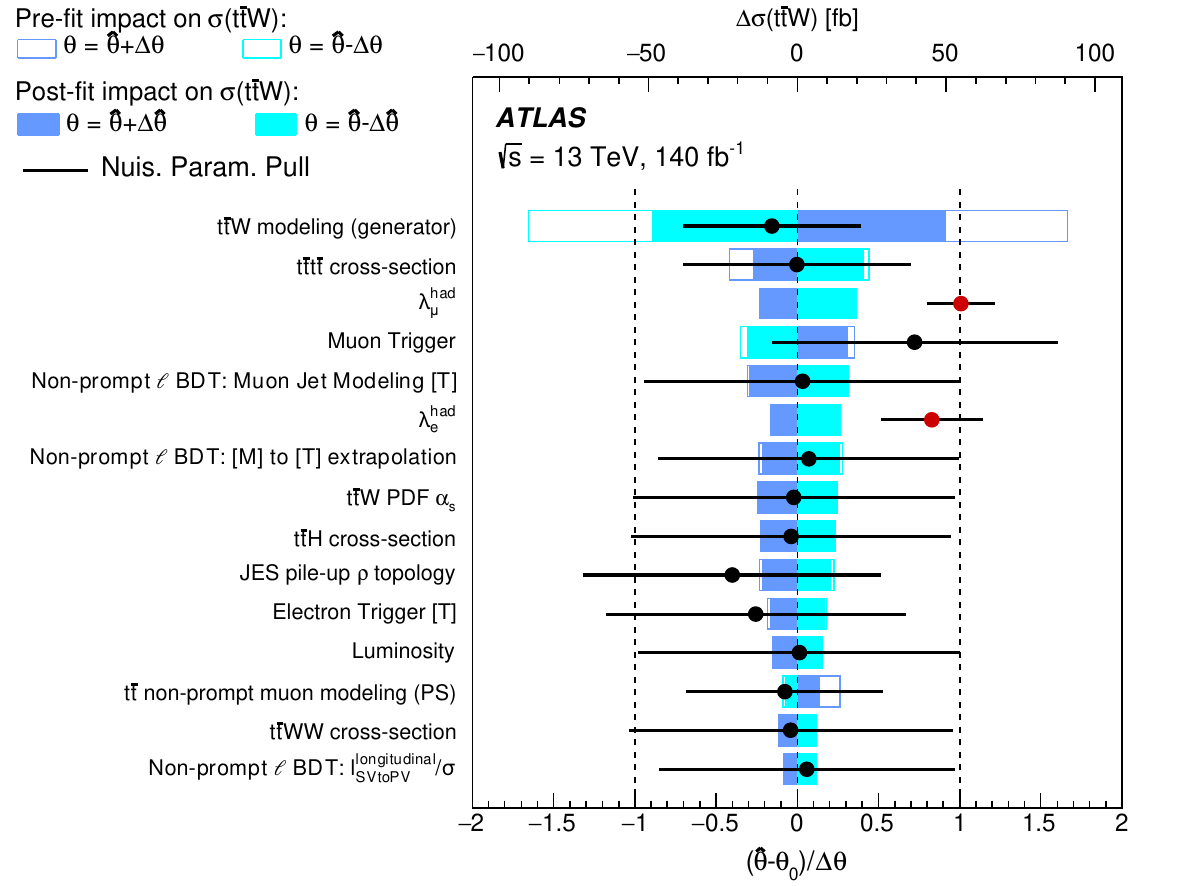}}\\
\subfloat[]{\includegraphics[width=0.65\textwidth]{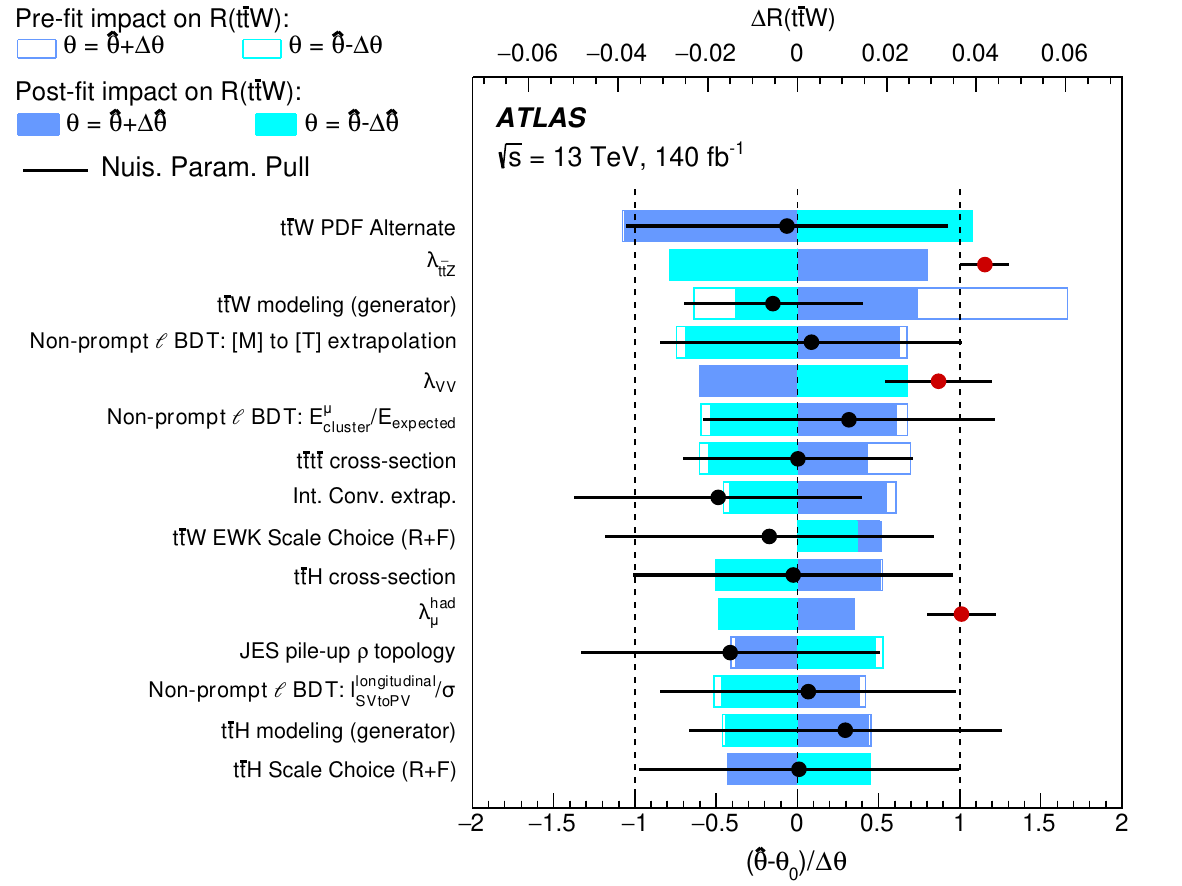}}
\caption{Ranking of the nuisance parameters and background normalisation factors included in the fit according to their impact on (a) the \ttW \xs, $\sigma(\ttW)$, and (b) $R(\ttW)$. In both (a) and (b), only the 15 nuisance parameters with the largest impacts are shown.
The empty (solid) blue rectangles illustrate the pre-fit (post-fit) impacts on the parameter of interest, $\sigma(\ttW)$ in (a) and $R(\ttW)$ in (b), corresponding to the top axis. The pre-fit (post-fit) impact of each nuisance parameter, $\Delta\sigma(\ttW)$ in (a) and $\Delta R(\ttW)$ in (b), is calculated as the difference in the fitted value of (a) $\sigma(\ttW)$ or (b) $R(\ttW)$ between the nominal fit and the fit when fixing the corresponding nuisance parameter to $\boldsymbol{\hat{\theta}} \pm \boldsymbol{\Delta\theta}$ $(\boldsymbol{\hat{\theta}} \pm \boldsymbol{\Delta\hat{\theta}})$, where $\boldsymbol{\hat{\theta}}$ is the best-fit value of the nuisance parameter and  $\boldsymbol{\Delta\theta}$  $(\boldsymbol{\Delta\hat{\theta}})$ is its pre-fit (post-fit) uncertainty.
The black points show the pulls of the nuisance parameters relative to zero, $\boldsymbol{\theta}_{0}=0$.
These pulls and their relative post-fit errors, $\boldsymbol{\Delta\hat{\theta}}/\boldsymbol{\Delta\theta}$, refer to the bottom axis.
The background normalisation factors $\lambda$ (red points) are pulled relative to one, $\boldsymbol{\theta}_{0}=1$. The fitted values and their post-fit errors refer to the bottom axis.
The systematic uncertainties with \enquote{[T]} affect only the \enquote{tight} leptons.}
\label{fig:ranking}
\end{figure}


\clearpage
 
\subsection{Differential \xs measurement}
\label{sec:diffxsec}

 
The differential \ttW measurements were made as a function of
six different observables.  The chosen observables were selected
because they have either an experimental motivation (e.g.\ variables
where modelling discrepancies were observed before) or a theoretical
motivation (e.g.\ variables that have significant distribution shape changes due
to NLO corrections).
The full set of measured observables is given in
Table~\ref{tab:unfold_vars}.
 
\begin{table}[h]
\caption{Definition of observables measured. \label{tab:unfold_vars}}
\begin{tabular}{cl}
\hline
Variable & Definition \\
\hline
\njets    & Number of selected jets with $\pT>25$~GeV and $|\eta|<2.5$\\
\HTjet    & Scalar sum of the transverse momenta of selected jets with $\pT>25$~GeV and $|\eta|<2.5$\\
\HTlep    & Scalar sum of the transverse momenta of selected leptons\\
\DRlb     & Angular separation between the leading lepton and the leading $b$-tagged jet\\
\DPhillSS & Absolute azimuthal separation between the two leptons of the same-sign pair \\
\DEtallSS & Absolute pseudorapidity separation between the two leptons of the same-sign pair \\
\hline
\end{tabular}
\end{table}
 
A profile likelihood unfolding procedure~\cite{blobel2002unfolding} (PLU) is used to
measure particle-level differential \xs{}s for one observable at a time in the fiducial
phase space defined in Section~\ref{sec:events}. A likelihood model
is built for a detector-level distribution by treating the
contribution from each generator-level bin as a subcomponent of the
detector-level signal.  The likelihood takes the following form:
 
\begin{equation*}{\cal L}(\sigma,\vec{\theta},\vec{\lambda}) = \prod_{i}{P\left(N_i | L_{\textrm{int}} \displaystyle\sum_{j}{\mathcal{R}_{ij}(\vec{\theta})\sigma_j(\vec{\theta})+ B_i(\vec{\theta},\vec{\lambda})}\right)} \times \prod_{k}{G(\theta_{k})} \times \prod_{l}R(\sigma_l,\tau_l),
\end{equation*}

where $P$ and $G$ are probability density functions of the Poisson and Gaussian distributions, respectively. The index $i$ indicates the detector-level bin, $j$ indicates the particle-level bin index and $k$ the systematic uncertainty
index. The variable $N_{i}$ is the number of observed data events in detector-level bin
$i$, $L_{\textrm{int}}$ is the integrated luminosity,
$\mathcal{R}_{ij}$ is the $(i,j)$ bin of the response matrix, $\sigma_j$ is the unfolded
\xs in particle-level bin $j$, and $B_{i}$ is the number of
background events in detector-level bin $i$, which depends on both
$\vec{\lambda}$, the ratios of the measured background yields to the
predicted values, and $\vec{\theta}$, the systematic-uncertainty nuisance parameters.
The regularisation term, $R$, is controlled via the $\tau_l$ parameter that determines the level of
regularisation used in the unfolding of generator-level bin $l$, where $l=2,\ldots,N_{\textrm{bins}}-1$.
In practice, the value of $\sigma_j$ is extracted by fitting for the number of events in
particle-level bin $j$ in each fiducial region, given by $N_j$ = $L_{\textrm{int}}\sigma_{j}$ with
$\sigma_j = \mu_{j} \sigma^{\textrm{MC}}_{j}$, and it is the signal
strengths $\mu_{j}$ that vary in the fit. For a given observable and fiducial region,
the value of $\tau_l$ is common to all $l$ bins and is shown in Table~\ref{tab:tau_values}.

\begin{table}[h]
\centering
\caption{Choice of regularisation parameter $\tau$ used for each observable and region. \label{tab:tau_values}}
\begin{tabular}{c |c c c c}
\hline
Observable & \twopp & \threep & \twomm & \threem \\
\hline
\njets & 0.7 & 0.3 & 0.4 & 0.2 \\
\HTjet & 0.6 & 0.3 & 0.4 & 0.2  \\
\HTlep & 0.5 & 0.1 &  0.2 & 0.1  \\
\DRlb & 0.3  & 0.2  & 0.3  & 0.3  \\
\DPhillSS & 0.1 &  0\phantom{.0} &  0\phantom{.0} & 0\phantom{.0} \\
\DEtallSS & 0.1 & 0.1 &  0.1 & 0\phantom{.0}  \\
\hline
\end{tabular}
\end{table}

The response matrix is defined as:
\begin{equation*}
\mathcal{R}_{ij} = \frac{1}{\alpha_i}\epsilon_{j}\mathcal{M}_{ij}.
\end{equation*}
The migration matrix, $\mathcal{M}_{ij}$, which quantifies the
bin-to-bin migrations of events owing to resolution effects when
going from particle level to detector level, is defined as:
\begin{equation*}
\mathcal{M}_{ij} = \frac{N_{ij}^{\text{det.}\,\cap\, \text{fid.}}}{N_j^{\text{det.}\,\cap\, \text{fid.}}}.
\end{equation*}
The superscript $\textrm{det.}\,\cap\, \text{fid.}$ indicates events that pass both the detector-level event selection and the fiducial selection. The numerator $N^{\textrm{det.}\,\cap\,\textrm{fid.}}_{ij}$ is the expected number of detector-level events in detector-level bin $i$ and particle-level bin $j$, while
the denominator $N_j^{\text{det.}\,\cap\, \text{fid.}}$ is the expected number of detector-level events in particle-level bin $j$ summed over all detector-level bins.
 
The migration matrix is then corrected with acceptance
factors, $\alpha_i$, which define the fraction of events
that satisfy the detector-level selection and originate from
configurations outside the particle-level fiducial selection:
\begin{equation*}
\alpha_i= \frac{N_i^{\text{det.}\,\cap\, \text{fid.}}}{N_i^{\text{det.}}}.
\end{equation*}
 
Finally, efficiency corrections, $\epsilon_j$, are applied
to account for events that satisfy the fiducial phase-space
selection criteria but are not reconstructed in the detector:
\begin{equation*}
\epsilon_j= \frac{N_j^{\text{det.}\,\cap\, \text{fid.}}}{N_j^\text{fid.}}.
\end{equation*}
Here, $N_i^\text{det.}$ ($N_j^\text{fid.}$) represents the expected number of events in the $i^{\mathrm{th}}$ ($j^{\mathrm{th}}$) bin of the detector-level (fiducial particle-level) histogram and $N_i^{\text{det.}\,\cap\, \text{fid.}}$ is the expected number of detector-level events in detector-level bin $i$ summed over all particle-level bins.

Systematic uncertainties, evaluated at detector level, impact the
unfolded \xs measurements through modifications of the response matrix.
 
The binning scheme is optimised for each observable, except \njets,
so that at least $60\%$ of the total
number of events in a given particle-level bin are in the diagonal element
of the migration matrix and there is low statistical
uncertainty for the yields in a given reconstruction-level bin
(i.e.\ taking into account the signal and background MC samples). The
statistical uncertainty is required to be below 10\% (16\%) for high-purity
(low-purity) regions to obtain similar bin granularity.
 
Figure~\ref{fig:Main:Unfolding:migEffAcc:nJets} shows the migration
matrix, normalised to the total number of events in each truth bin,
and the acceptance and efficiency corrections for \njets.  The \lll
regions are split according to the number of OS-SF lepton pairs.  The
fraction of events in the diagonal cells of the migration matrix shows
the resolution of each observable.  In the case of \njets, the
fractions along the diagonal are quite large, 60\%--90\%, for the \ll
regions, whereas for the \lll regions the migrations between bins are
split among their subregions in a roughly 1:2:1 ratio.  The migrations
between bins are generally smaller for the lepton-based observables,
with angular variables showing the smallest migrations, and larger for
the jet-based observables. The simultaneous unfolding of all fiducial
regions shows a 10\% migration of events from \lll regions into \llSS
regions.

\begin{figure}[htp]
\centering
\subfloat[]{\includegraphics[width=0.48\textwidth]{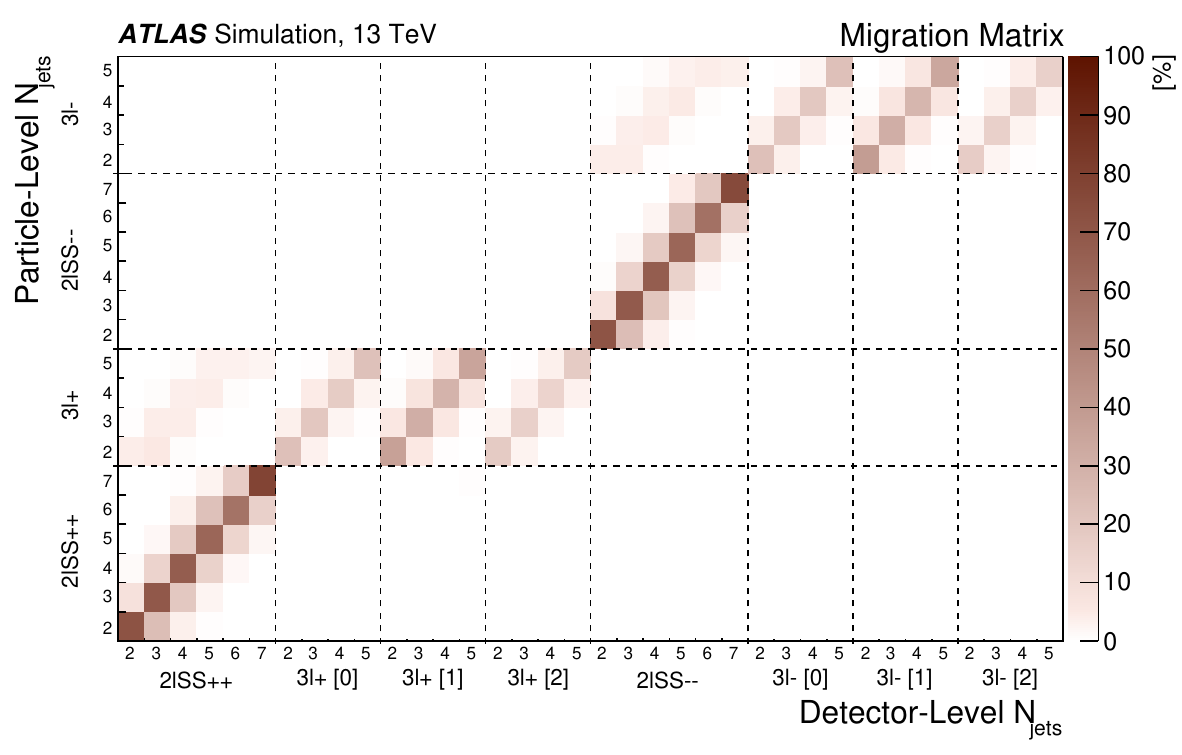}}
\subfloat[]{\includegraphics[width=0.48\textwidth]{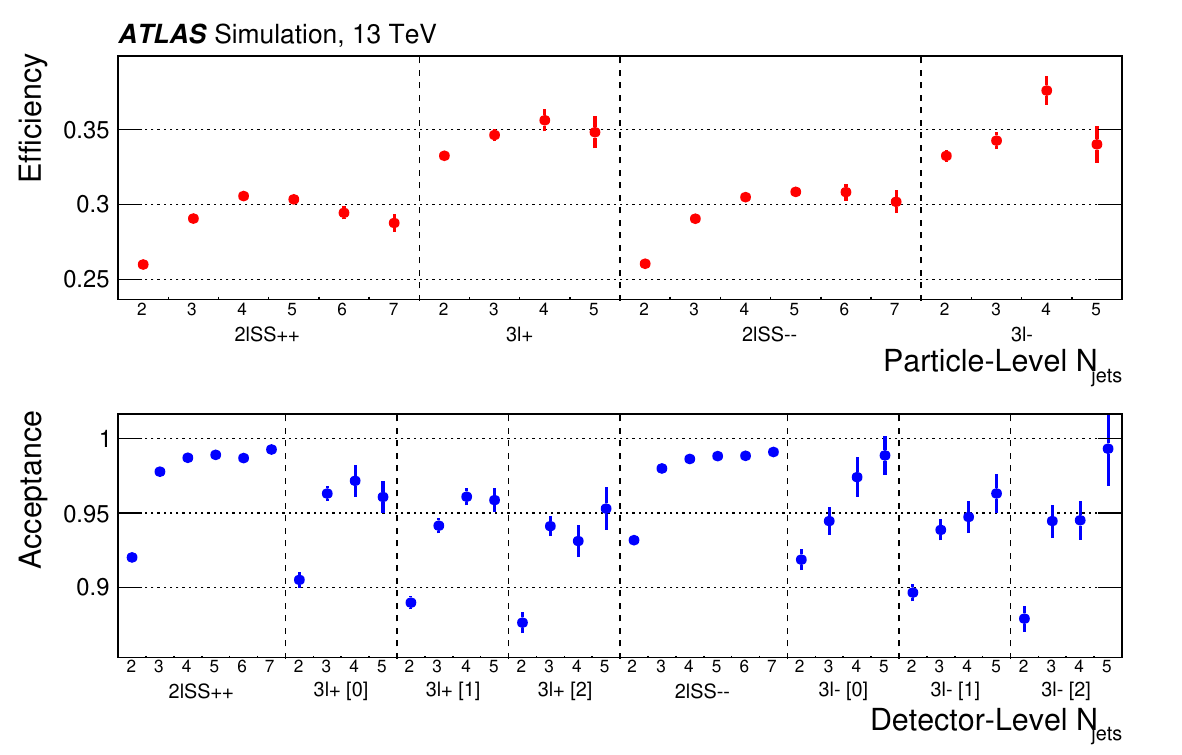}}
\caption{The (a) normalised migration matrix and (b) efficiency/acceptance corrections for $N_\text{jets}$ calculated in \Sherpa[2.2.10] with EWK corrections. The \lll regions are split according to the number of OS-SF lepton pairs, shown in brackets. The highest jet-multiplicity bin in each subregion is inclusive.}  \label{fig:Main:Unfolding:migEffAcc:nJets}
\end{figure}
 
The detector-level distributions of \njets are shown before and
after the fit to data in Figures~\ref{fig:obs_Njets_SRtemplate}(a)
and~\ref{fig:obs_Njets_SRtemplate}(b) respectively.
The \lll regions are split according to the number of OS-SF lepton pairs,
improving the discrimination between the signal \ttW and the \ttZ and \VV background processes.
 
\begin{figure}[htp]
\centering
\subfloat[]{\includegraphics[width=0.48\textwidth]{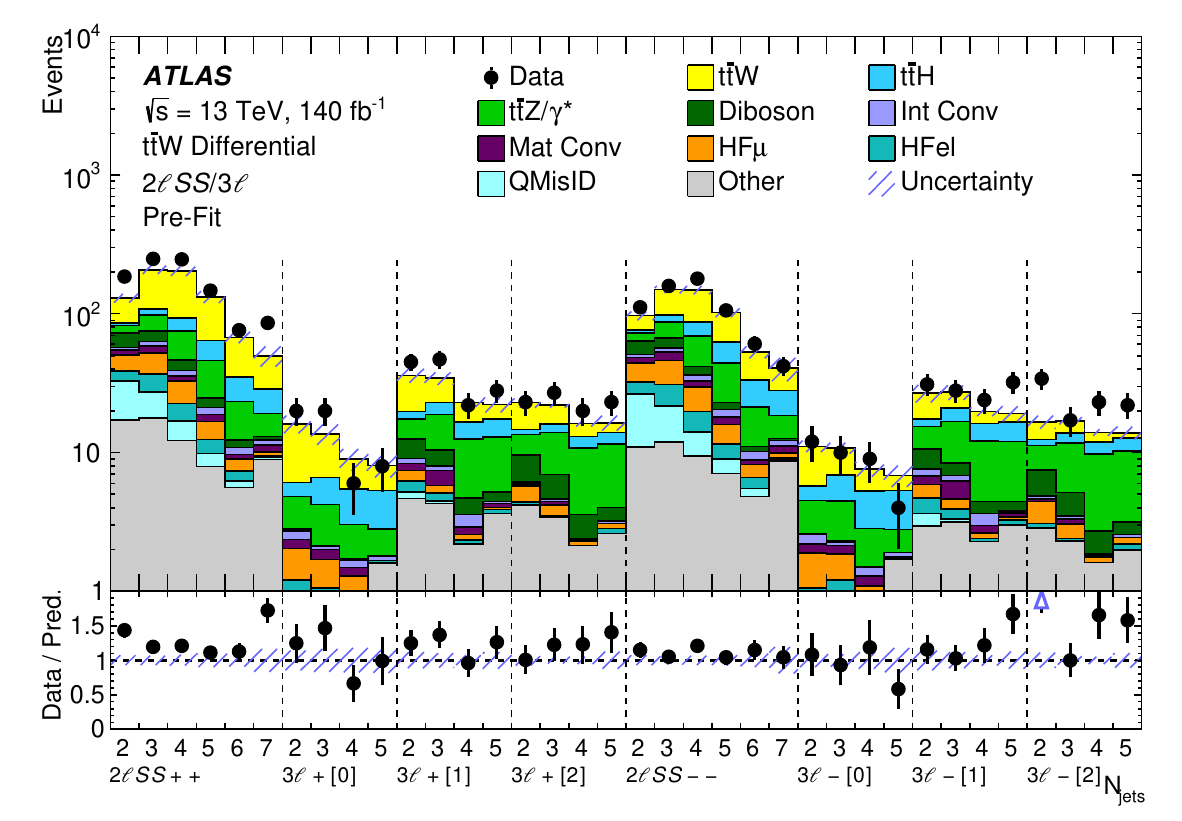}}
\subfloat[]{\includegraphics[width=0.48\textwidth]{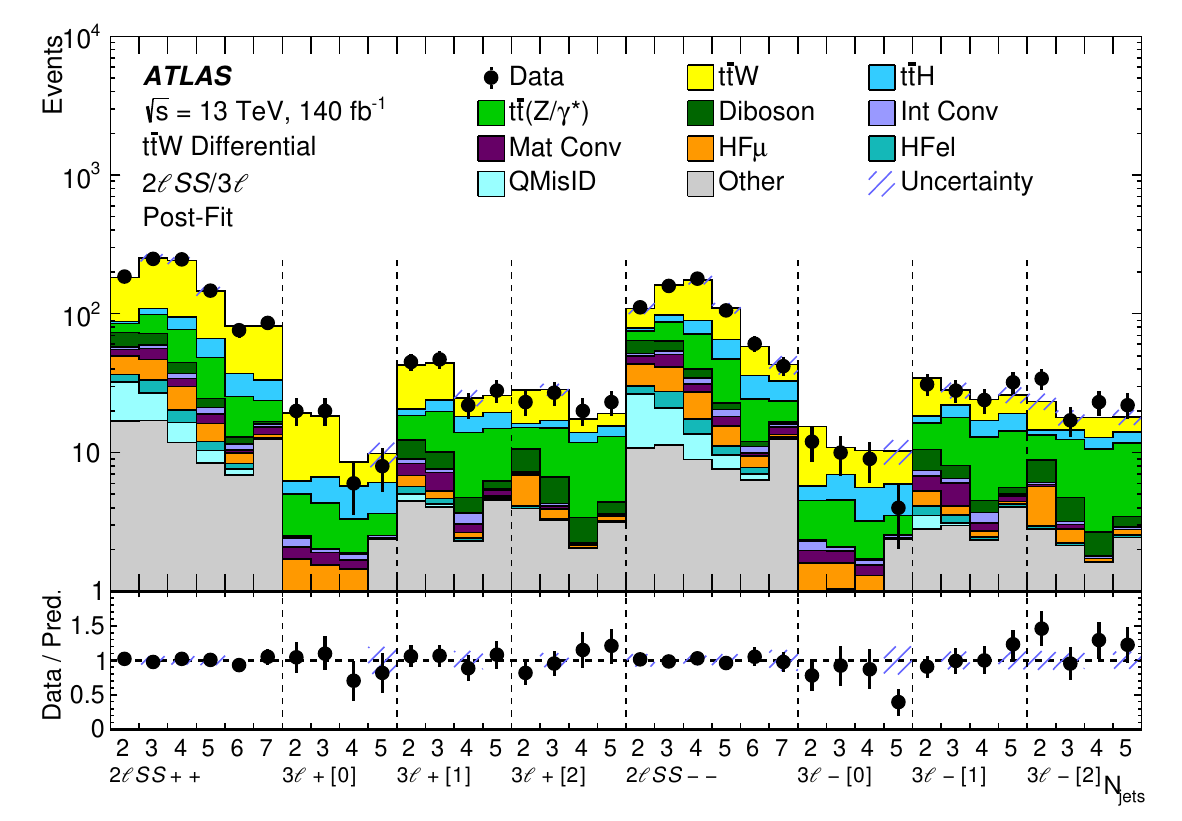}}
\caption{Signal region datasets for the \njets unfolding in (a) the pre-fit case and (b) the post-fit case. The \lll regions are split according to the number of OS-SF lepton pairs, shown in brackets.  The highest jet-multiplicity bin in each subregion is inclusive. The signal and background contributions before and after the likelihood fit to data under the signal-plus-background hypothesis are shown as filled histograms. The ratio of the data to the total pre-fit and post-fit predictions (\enquote{Pred.}) is shown in the lower panel. The combined statistical and systematic uncertainty in the prediction is indicated by the blue hatched band. The blue triangles indicate points that are outside the vertical range of the figure.}\label{fig:obs_Njets_SRtemplate}
\end{figure}

Tikhonov regularisation~\cite{Phillips:1962ofa,Tikhonov63,Tikhonov77,Cowan:1998ji}
in the PLU technique is implemented by an additional term, $R(\sigma_j,\tau_j)$, in the
likelihood function that constrains the discrete second derivative of
the generator-level bin normalisation factors to be close to zero. The
regularisation strengths, $\tau_j$, are fixed parameters that assume a common
value across bins in each signal region and observable. The chosen $\tau_j$ values are optimised
independently for each signal region and observable.  The optimisation
procedure is based on statistical bootstrapping to estimate the value of $\tau$ that minimises
the global correlation coefficient $\rho$ of the unfolded covariance matrix~\cite{Schmitt:2017}.
It is given by the following expression:
\begin{equation*}
\rho = \left<\sqrt{1-\left(V_{ii}V^{-1}_{ii}\right)^{-1}}\right>_{i\in{1,\dots,N_\mathrm{bins}}},
\end{equation*}
where $V_{ij}$ is the covariance matrix of the bin normalisation factors.
Pseudo-experiments are sampled from detector-level templates obtained
from the background predictions and alternative \ttW sample.
Another optimisation procedure is studied, which selects the maximum value of $\tau$ that does not
increase the $\chi^2$ between the unfolded observable's distribution and the alternative \ttW generator-level
template. The $\chi^2$ is defined as:
\begin{equation*}
\chi^2 = \sum_{i,j}^n \Delta\sigma_i C_{i,j}^{-1} \Delta\sigma_j,
\end{equation*}
where $\Delta \sigma_i$ is the difference between the unfolded data and the MC prediction and $C_{i,j}^{-1}$ is the inverse of the covariance matrix of the unfolded cross sections.
The $\chi^2$-based procedure yields larger values for $\tau$ than the $\rho$-based procedure.
The $\rho$-based procedure is ultimately chosen because it is not sensitive to the choice of
alternative \ttW generator-level template, while the $\chi^2$-based procedure will give a large bias when
the shape of the observed data distribution differs from the nominal \ttW prediction more than the alternative \ttW prediction does.
 
Various closure and stress tests of the unfolding procedure are
performed.  Full closure is achieved in fits to an
Asimov~\cite{Cowan:2010js} dataset.~\footnote{The Asimov dataset is an artificial data set
defined so as to make estimators for all parameters equal to their true values.}
Additional validation tests are
performed, for example by changing normalisations in certain
regions or applying linear slopes in the particle-level
distributions. In both cases, perfect retrieval of the injected
particle-level distribution is observed, as expected from the PLU
technique and the chosen regularisation technique.  A stress test
evaluating the bias in the unfolding procedure is performed by
unfolding an Asimov dataset obtained using an alternative signal
model.  A similar stress test evaluating the sensitivity of the
unfolding procedure to the limited number of simulated events is
performed by splitting the nominal signal sample into two parts of
equal size, where one part is used for the unfolding and the other is
used to generate the Asimov dataset.
Good closure is obtained in all cases.  A comprehensive
stress test is performed to characterise the level of bias as a
function of $\tau$ by unfolding Asimov datasets generated by injecting
quadratic shapes into the signal templates, either with an enhancement
in the core of the distribution or in the upper tail. In the case
where shapes have 200\% enhancement, the median bias in any bin
relative to its total uncertainty is between 1\% and 30\%.  An
additional data-driven test is performed to evaluate the bias in
the unfolded observed data. Each observable is first unfolded with no
regularisation to obtain a particle-level distribution for the data
with no bias. The distributions are fitted with a second-order polynomial
to capture the lowest-order shape that is removed by regularisation.
These fits are used to generate smooth Asimov pseudo-data templates
that are unfolded in fits with regularisation, but only including
statistical uncertainties. The correct pseudo-data shapes are in
general recovered well, with deviations that are typically very small relative
to statistical uncertainties across all bins and observables, and with
$\chi^2$ $p$-values near 1.0 for the observables listed in Table~\ref{tab:unfold_vars}.

As expected, given the larger-than-predicted measured inclusive \xs, the absolute
differential \xs{}s measured in data are larger than the theoretical
predictions in most regions of phase space. To better understand normalisation and shape effects in the
comparison of the measurements with predictions, the absolute and
normalised \xs{}s are shown for each of the measured observables.
Several different representations of the differential \xs
measurements are shown for each observable. Firstly, the measured
absolute differential \xs{}s for combined lepton-charge regions, the
measured normalised differential \xs{}s for combined lepton-charge regions,
and the differential \rqasym measurement are shown.
Secondly, measurements of both the absolute and normalised \xs{}s and
the differential \xs{}s are shown for each lepton-charge channel.
Although the parameterisation of the yields in the different lepton-charge regions
differs between the two cases, the underlying data are the same.
 
The results are shown in
Figures~\ref{fig:main:Unfolding:nJets}--\ref{fig:main:Unfolding:DEtallSS}.
The measurements are compared with the theoretical predictions
previously described in Section~\ref{sec:datamc}. Only the results in
the \lll channel are compared with a dedicated theoretical prediction
based on the procedure outlined in Ref.~\cite{Bevilacqua:2020srb} that
includes off-shell effects. This prediction is calculated at
parton level and corrected to particle level by applying bin-by-bin
scaling factors that account for non-perturbative effects, such as multiparton
interactions and hadronisation.
 
The compatibility of the measured differential \xs distributions and the
predictions is assessed by calculating $\chi^2$ values and probabilities,
for the relevant number of degrees of freedom, properly taking into
account uncertainty correlations. These values are shown in
Tables~\ref{tab:chi22lSSnorm} and~\ref{tab:chi23lnorm} for the
normalised \xs{}s in the \llSS and \lll regions respectively.
They are shown for the normalised \xs{}s rather than the absolute \xs{}s
in order to give a better indication of how well the theoretical predictions
agree with the data, without having to take account of the known
difference between the predicted and measured inclusive absolute \xs{}s.
The $\chi^2$ $p$-values for compatibility of the absolute \xs measurements
and the \SHERPA predictions are in the range of 6\%--70\%, and typically
less than 10\% in the \llSS region.
Tables~\ref{tab:unfolded_chi2_2lSS_asym} and~\ref{tab:unfolded_chi2_3l_asym}
show the $\chi^2$ values and $p$-values for compatibility of the predicted and
measured \rqasym distributions in the \llSS and \lll regions respectively.

\begin{table}[h]
\footnotesize
\caption{$\chi^2$ and $p$-values calculated for unfolded normalised \xs distributions in the \llSS region.\label{tab:chi22lSSnorm}}
\resizebox{\textwidth}{!}{
\begin{tabular}{ c | c | c | c | c | c | c | c | c | c | c | c}
\hline
Observable & NDF &\multicolumn{2}{c|}{Sherpa 2.2.10} & \multicolumn{2}{c|}{MG5aMC+Py8 FxFx} & \multicolumn{2}{c|}{MG5aMC+Py8 Incl.} & \multicolumn{2}{c|}{Powheg+Py8} & \multicolumn{2}{c}{Powheg+H7}\\
\cline{3-12}
& & $\chi^2$ & $p$-value & $\chi^2$ & $p$-value & $\chi^2$ & $p$-value & $\chi^2$ & $p$-value & $\chi^2$ & $p$-value\\
\hline\hline
\njets & 5 & 6.5 & 0.26 & 7.8 & 0.17 & 6.9 & 0.23 & 7.2 & 0.21 & 6.8 & 0.24\\
\HTjet & 5 & 2.0 & 0.85 & 2.3 & 0.81 & 1.9 & 0.86 & 2.5 & 0.78 & 2.7 & 0.75\\
\HTlep & 7 & 6.1 & 0.53 & 6.3 & 0.51 & 5.9 & 0.55 & 5.9 & 0.56 & 5.9 & 0.55\\
\DRlb & 7 & 6.8 & 0.45 & 7.2 & 0.41 & 7.2 & 0.41 & 7.2 & 0.40 & 7.1 & 0.41\\
\DPhillSS & 7 & 1.7 & 0.97 & 1.7 & 0.98 & 1.9 & 0.96 & 1.8 & 0.97 & 1.9 & 0.97\\
\DEtallSS & 6 & 7.0 & 0.32 & 8.1 & 0.23 & 12.1\phantom{0} & 0.06 & 10.2\phantom{0} & 0.12 & 10.1\phantom{0} & 0.12\\
\hline
\end{tabular}
}
\end{table}

\begin{table}[h]
\footnotesize
\caption{$\chi^2$ and $p$-values calculated for unfolded normalised \xs distributions in the \lll region.\label{tab:chi23lnorm}}
\resizebox{\textwidth}{!}{
\begin{tabular}{ c | c | c | c | c | c | c | c | c | c | c | c | c | c}
\hline
Observable & NDF &\multicolumn{2}{c|}{Sherpa 2.2.10} & \multicolumn{2}{c|}{Off-Shell} & \multicolumn{2}{c|}{MG5aMC+Py8 FxFx} & \multicolumn{2}{c|}{MG5aMC+Py8 Incl.} & \multicolumn{2}{c|}{Powheg+Py8} & \multicolumn{2}{c}{Powheg+H7}\\
\cline{3-14}
& & $\chi^2$ & $p$-value & $\chi^2$ & $p$-value & $\chi^2$ & $p$-value & $\chi^2$ & $p$-value & $\chi^2$ & $p$-value & $\chi^2$ & $p$-value\\
\hline
\hline
\njets & 3 & 0.4 & 0.95 &-&-& 0.3 & 0.96 & 0.5 & 0.92 & 1.0 & 0.81 & 1.1 & 0.77\\
\HTjet & 4 & 1.5 & 0.82 &-&-& 1.2 & 0.88 & 1.9 & 0.76 & 2.2 & 0.70 & 2.8 & 0.59\\
\HTlep & 5 & 1.6 & 0.90 & 2.7 & 0.75 & 1.9 & 0.86 & 2.2 & 0.81 & 2.1 & 0.84 & 1.9 & 0.86\\
\DRlb & 5 & 9.1 & 0.10 & 8.5 & 0.13 & 9.3 & 0.10 & 8.9 & 0.11 & 8.8 & 0.12 & 8.8 & 0.12\\
\DPhillSS & 5 & 4.5 & 0.48 & 4.5 & 0.48 & 4.5 & 0.48 & 4.3 & 0.51 & 4.4 & 0.49 & 4.3 & 0.50\\
\DEtallSS & 5 & 6.0 & 0.31 & 5.6 & 0.35 & 5.8 & 0.32 & 5.5 & 0.36 & 5.4 & 0.37 & 5.6 & 0.35\\
\hline
\end{tabular}
}
\end{table}

\begin{table}[h]
\footnotesize
\caption{$\chi^2$ and $p$-values calculated for unfolded asymmetry distributions in the \llSS region.\label{tab:unfolded_chi2_2lSS_asym}}
\resizebox{\textwidth}{!}{
\begin{tabular}{ c | c | c | c | c | c | c | c | c | c | c | c}
\hline
Observable & NDF &\multicolumn{2}{c|}{Sherpa 2.2.10} &  \multicolumn{2}{c|}{MG5aMC+Py8 FxFx } & \multicolumn{2}{c|}{MG5aMC+Py8 Incl.} & \multicolumn{2}{c|}{Powheg+P8} & \multicolumn{2}{c}{Powheg+H7}\\
\cline{3-12}
& & $\chi^2$ & $p$-value & $\chi^2$ & $p$-value & $\chi^2$ & $p$-value & $\chi^2$ & $p$-value & $\chi^2$ & $p$-value\\
\hline    \hline
\njets & 6 & 3.8 & 0.70 & 4.0 & 0.68 & 3.5 & 0.75 & 3.3 & 0.78 & 3.2 & 0.78\\
\HTjet & 6 & 2.7 & 0.84 & 3.0 & 0.81 & 1.8 & 0.93 & 1.7 & 0.95 & 1.6 & 0.95\\
\HTlep & 8 & 5.7 & 0.68 & 5.6 & 0.69 & 3.0 & 0.94 & 3.4 & 0.91 & 3.3 & 0.91\\
\DRlb & 8 & 5.1 & 0.75 & 5.3 & 0.72 & 4.5 & 0.81 & 4.4 & 0.82 & 4.4 & 0.81\\
\DPhillSS & 8 & 8.2 & 0.41 & 9.0 & 0.34 & 7.4 & 0.49 & 7.5 & 0.48 & 7.5 & 0.48\\
\DEtallSS & 7 & 8.9 & 0.26 & 8.8 & 0.26 & 7.8 & 0.35 & 8.1 & 0.32 & 8.0 & 0.33\\
\hline
\end{tabular}
}
\end{table}
 
\begin{table}[h]
\scriptsize
\caption{$\chi^2$ and $p$-values calculated for unfolded asymmetry distributions in the \lll region.\label{tab:unfolded_chi2_3l_asym}}
\resizebox{\textwidth}{!}{
\begin{tabular}{ c | c | c | c | c | c | c | c | c | c | c | c | c | c}
\hline
Observable & NDF &\multicolumn{2}{c|}{Sherpa 2.2.10} & \multicolumn{2}{c|}{Off-Shell} & \multicolumn{2}{c|}{MG5aMC+Py8 FxFx } & \multicolumn{2}{c|}{MG5aMC+Py8 Incl.} & \multicolumn{2}{c|}{Powheg+Py8} & \multicolumn{2}{c}{Powheg+H7}\\
\cline{3-14}
& & $\chi^2$ & $p$-value & $\chi^2$ & $p$-value & $\chi^2$ & $p$-value & $\chi^2$ & $p$-value & $\chi^2$ & $p$-value & $\chi^2$ & $p$-value\\
\hline\hline
\njets & 4 & 4.3 & 0.37 &-&-& 4.7 & 0.32 & 4.5 & 0.34 & 4.8 & 0.31 & 4.6 & 0.33\\
\HTjet & 5 & 4.1 & 0.53 &-&-& 4.4 & 0.50 & 4.5 & 0.48 & 4.6 & 0.47 & 4.5 & 0.48\\
\HTlep & 6 & 3.6 & 0.73 & 3.6 & 0.73 & 3.6 & 0.73 & 3.7 & 0.72 & 3.6 & 0.73 & 3.6 & 0.73\\
\DRlb & 6 & 3.3 & 0.77 & 3.3 & 0.77 & 3.9 & 0.69 & 4.1 & 0.66 & 3.9 & 0.69 & 3.9 & 0.69\\
\DPhillSS & 6 & 7.2 & 0.30 & 7.2 & 0.30 & 7.7 & 0.26 & 7.6 & 0.27 & 7.8 & 0.26 & 7.6 & 0.27\\
\DEtallSS & 6 & 4.9 & 0.56 & 5.1 & 0.53 & 5.4 & 0.50 & 5.4 & 0.49 & 5.3 & 0.50 & 5.1 & 0.53\\
\hline
\end{tabular}
}
\end{table}

Figure~\ref{fig:main:Unfolding:nJets}(a) shows the measured
absolute \xs, normalised \xs and \rqasym as a function
of \njets. The corresponding uncertainty breakdown is shown in
Figure~\ref{fig:main:Unfolding:nJets}(b). In most bins the largest
uncertainties come from the limited amount of data; these dominate
across all bins of the \lll channel and at low and high jet
multiplicities in the \llSS channel. The dominant systematic
uncertainty stems from the \ttW model and is due to the difference between
the unfolded results from different choices of \ttW reference
sample. At large jet multiplicities the theoretical uncertainty of
the background model and non-prompt-lepton BDT are the largest.  The total uncertainty in the normalised \xs
measurements is somewhat reduced; a major contribution to this is the
smaller \ttW model uncertainty after removing normalisation effects.
The \rqasym measurement uncertainty is dominated by the statistical
uncertainty across the range of jet multiplicities because many systematic
uncertainties cancel each other out in the ratio.
The absolute \xs measured in data is higher than all the theoretical predictions,
the largest discrepancy being at low jet multiplicities in the \llSS channel.
At high multiplicities the predictions with additional partons using NLO
merging are in better agreement with the data than the NLO+PS
prediction from \POWHEG, as expected, for both the \llSS and \lll
channels. Taking the scale uncertainties into account, these are
compatible with the data for jet multiplicities greater than four.
Uncertainties due to the scales are approximately 10\% for low jet
multiplicities and increase to about 40\% for higher
multiplicities.  The main \ttW contribution to the \llSS selection is
the final state with two leptonically decaying $W$ bosons, one from
the decay of a top quark and one from the associated $W$ boson. This
gives rise to a final state with four partons from the hadronic decay
of the remaining $W$ boson and the $b$-quarks from the top quark
decays. Therefore, the disagreement between theory and data at lower
jet multiplicities is independent of the modelling of additional
partons and is observed for both the NLO+PS generators and those with
additional partons using NLO merging.  The shape of the \rqasym
distribution as a function of the number of jets shows some tension
between data and the predictions. The measured asymmetry is
larger at low and high jet multiplicities while being in agreement at
intermediate numbers of four or five jets in the \llSS channel. This is
driven by a large mismodelling in the \twopp~region, while
the \twomm~jet multiplicity distribution is in better agreement in
both shape and overall \xs. This can be seen in right panel of
Figure~\ref{fig:main:Unfolding:nJets}(c) for the
normalised \xs measurements, where, especially for the \llSS channel,
the agreement in shape is rather good for the predictions with
additional partons using NLO merging.
 
\begin{figure}[htbp]
\centering
\subfloat[]{\includegraphics[width=0.90\textwidth]{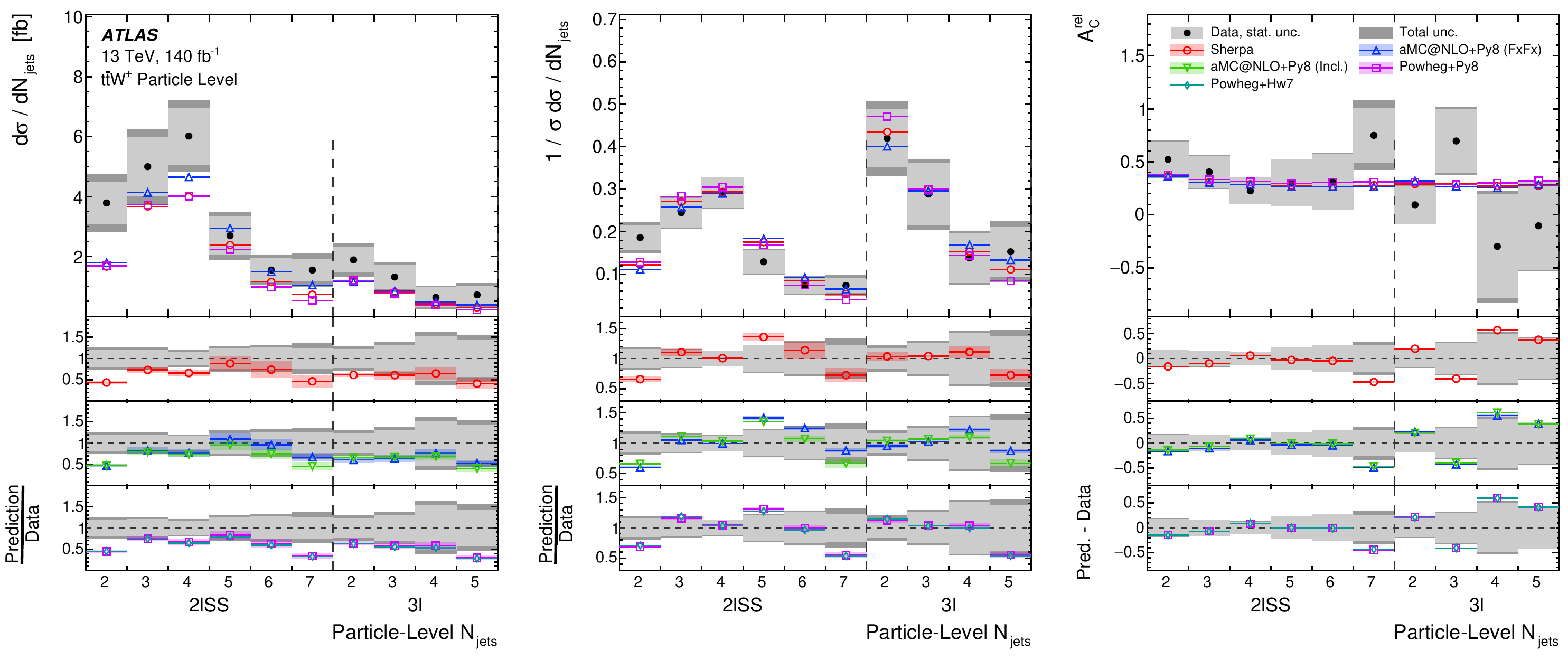}}\\
\subfloat[]{\includegraphics[width=0.90\textwidth]{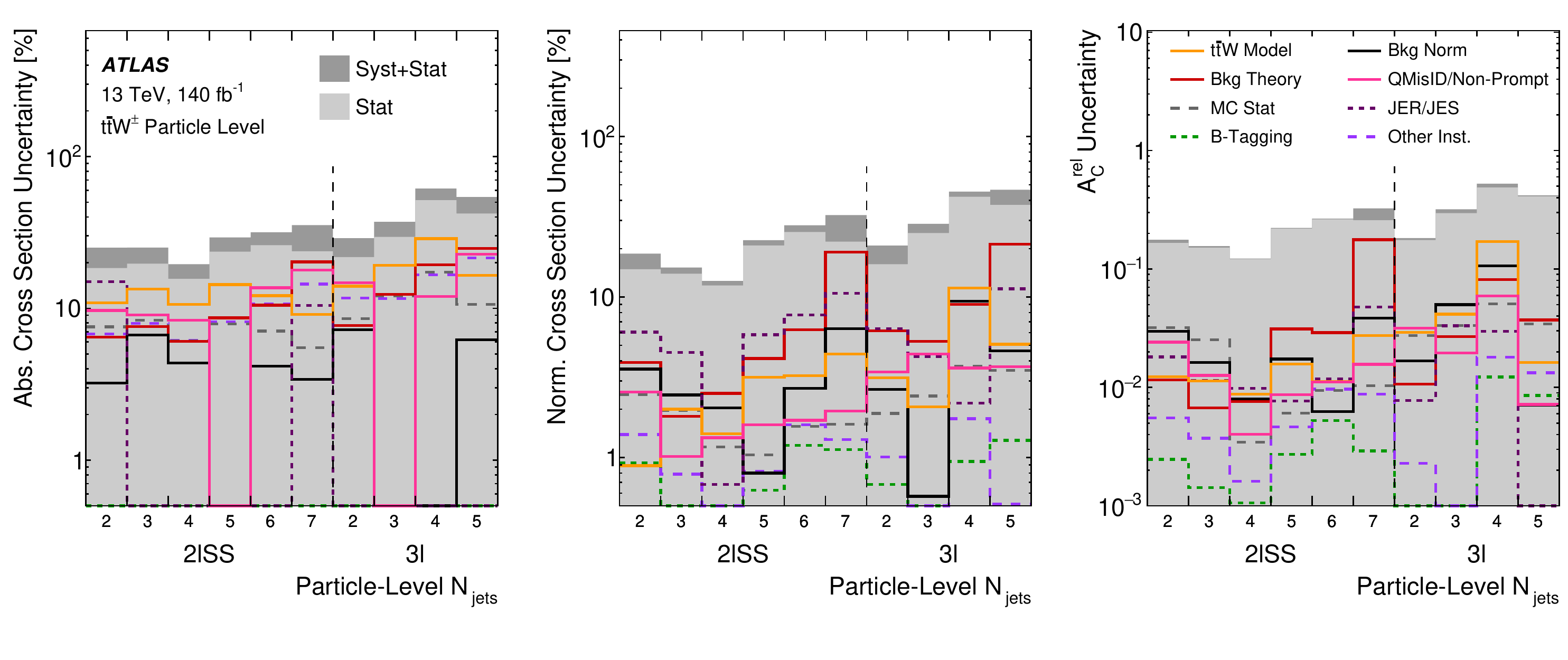}}\\
\subfloat[]{\includegraphics[width=0.90\textwidth]{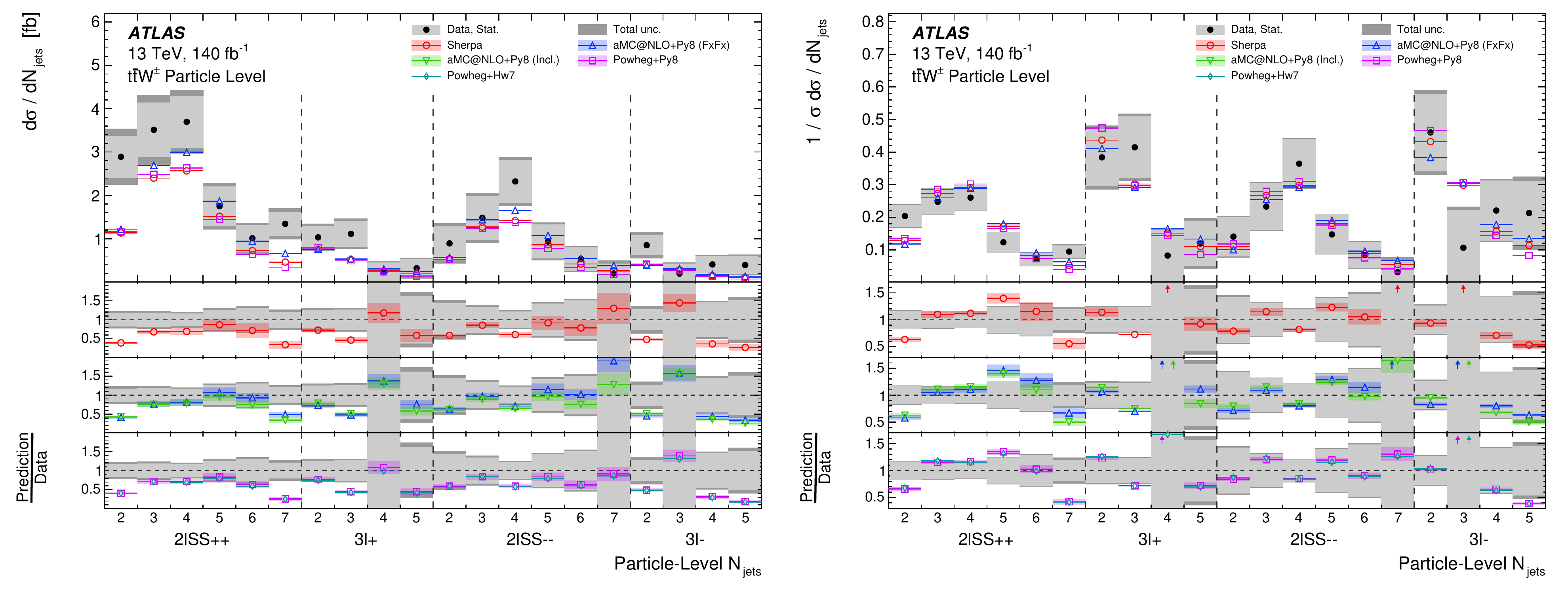}}
\caption{Unfolded distributions of (a) the lepton-charge-combined absolute \xs, the normalised \xs and \rqasym, (b) the corresponding uncertainty breakdowns, and (c) unfolded distributions of the lepton-charge-split absolute and normalised \xs{}s, as a function of jet multiplicity. The coloured markers overlaid upon the \xs measurements show different theoretical predictions. The breakdown of statistical-only uncertainties is shown in light grey in the bottom panels, and combined statistical and systematic uncertainties are shown in dark grey. The dotted/dashed/full lines in (b) show the relative contributions of grouped sources of systematic uncertainty. The triangles indicate points that are outside the vertical range of the figure.}
\label{fig:main:Unfolding:nJets}
\end{figure}

\begin{figure}[htbp]
\centering
\subfloat[]{\includegraphics[width=0.90\textwidth]{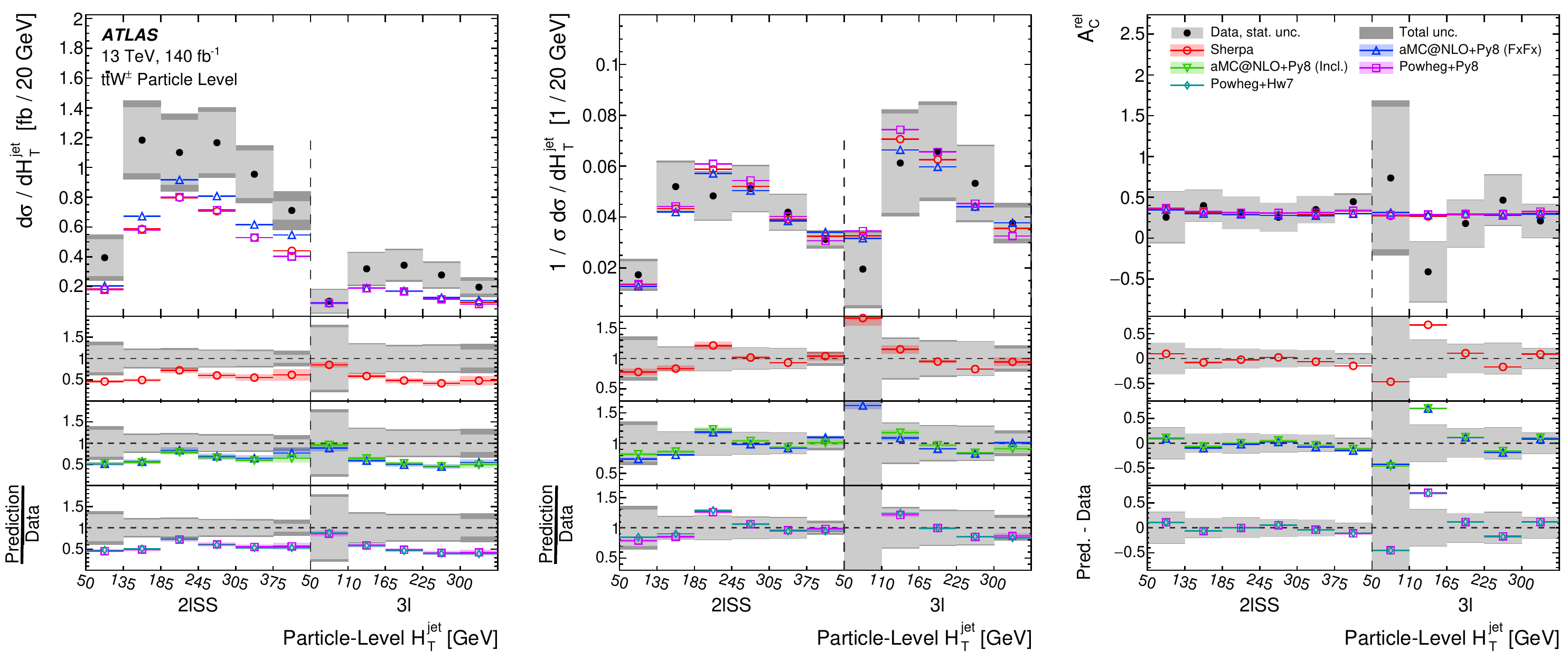}}\\
\subfloat[]{\includegraphics[width=0.90\textwidth]{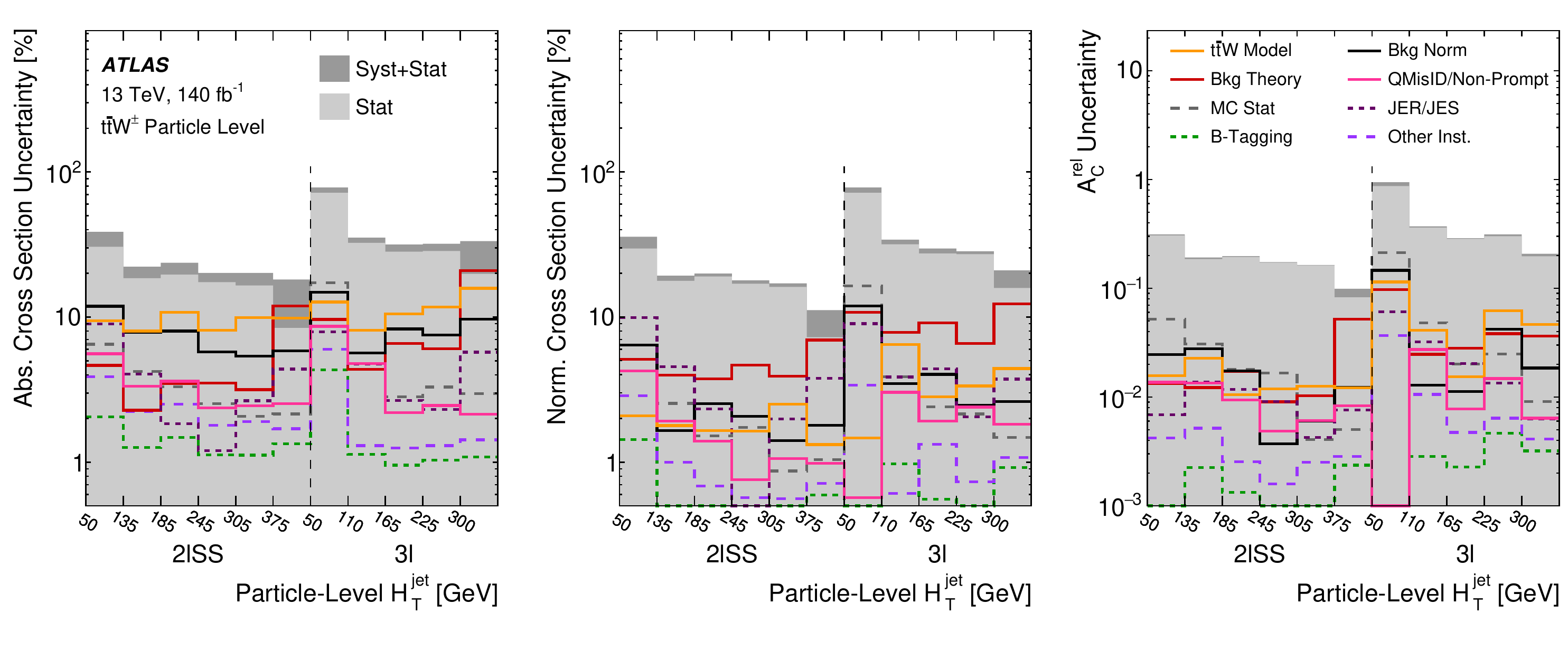}}\\
\subfloat[]{\includegraphics[width=0.90\textwidth]{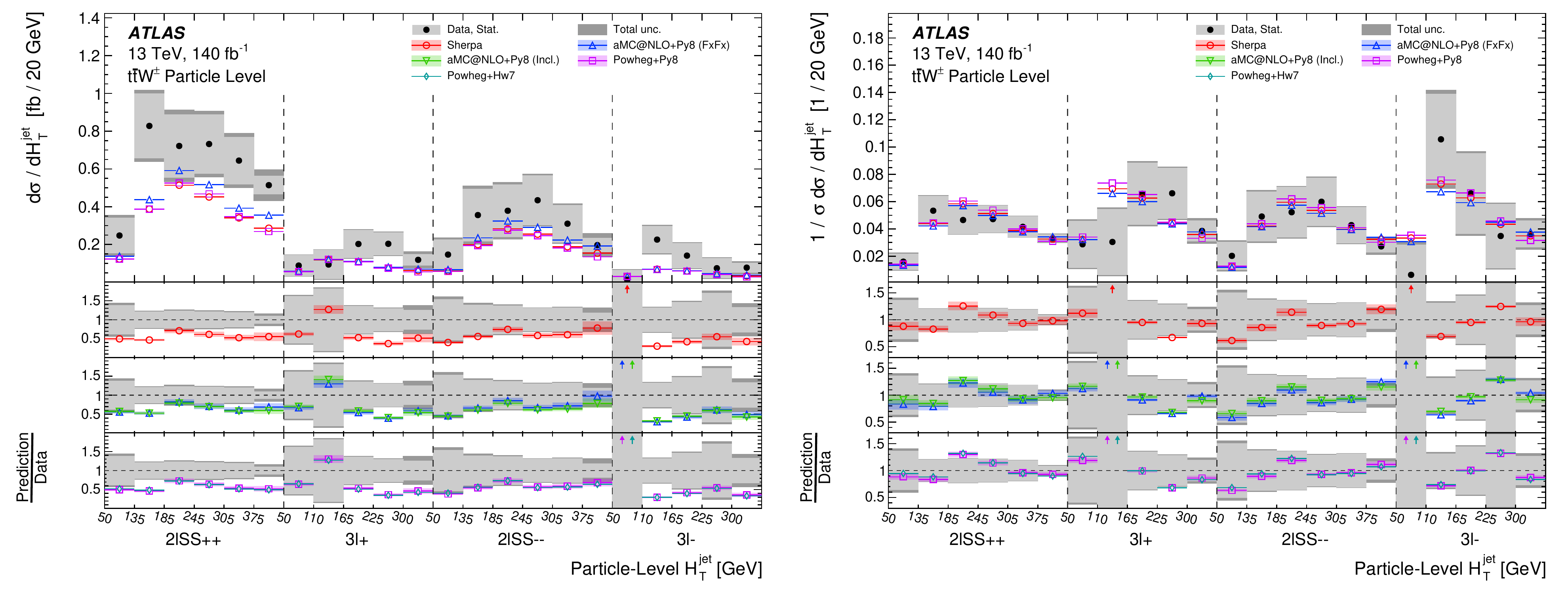}}
\caption{Unfolded distributions of (a) the lepton-charge-combined absolute \xs, the normalised \xs and \rqasym, (b) the corresponding uncertainty breakdowns, and (c) unfolded distributions of the lepton-charge-split absolute and normalised \xs{}s, as a function of \HTjet. The coloured markers overlaid upon the \xs measurements show different theoretical predictions. The breakdown of statistical-only uncertainties is shown in light grey in the bottom panels, and combined statistical and systematic uncertainties are shown in dark grey. The dotted/dashed/full lines in (b) show the relative contributions of grouped sources of systematic uncertainty. The triangles indicate points that are outside the vertical range of the figure.}
\label{fig:main:Unfolding:HT_jet}
\end{figure}

Figure~\ref{fig:main:Unfolding:HT_jet}(a) shows the
measured absolute \xs, normalised \xs and \rqasym as a function of \HTjet.
The corresponding uncertainty breakdowns are shown in
Figure~\ref{fig:main:Unfolding:HT_jet}(b). Again, the limited amount of
data is the largest source of uncertainty across most bins of the
measurement. The main systematic uncertainties come from the modelling
of the \ttW signal process, with an important additional contribution
from the jet energy scale and resolution, particularly in the low \HTjet
region.
The absolute \xs measurements indicate a slight improvement in
modelling at high \HTjet from the predictions with additional partons using NLO merging, similarly to
what was seen for \njets.  The normalised \xs measurements, shown in the
middle panel of Figure~\ref{fig:main:Unfolding:HT_jet}(a), show that the modelling of
the distribution's shape is generally very good in the \llSS
channel, while the predictions favour lower values than the data in
the \lll channel. The \rqasym distribution predictions are also in agreement with data in
the \llSS region, but not in the \lll region. This seems to be driven
by a significant difference between data and predictions in
the \threep region, as shown in
Figure~\ref{fig:main:Unfolding:HT_jet}(c).
 
\begin{figure}[htbp]
\centering
\subfloat[]{\includegraphics[width=0.90\textwidth]{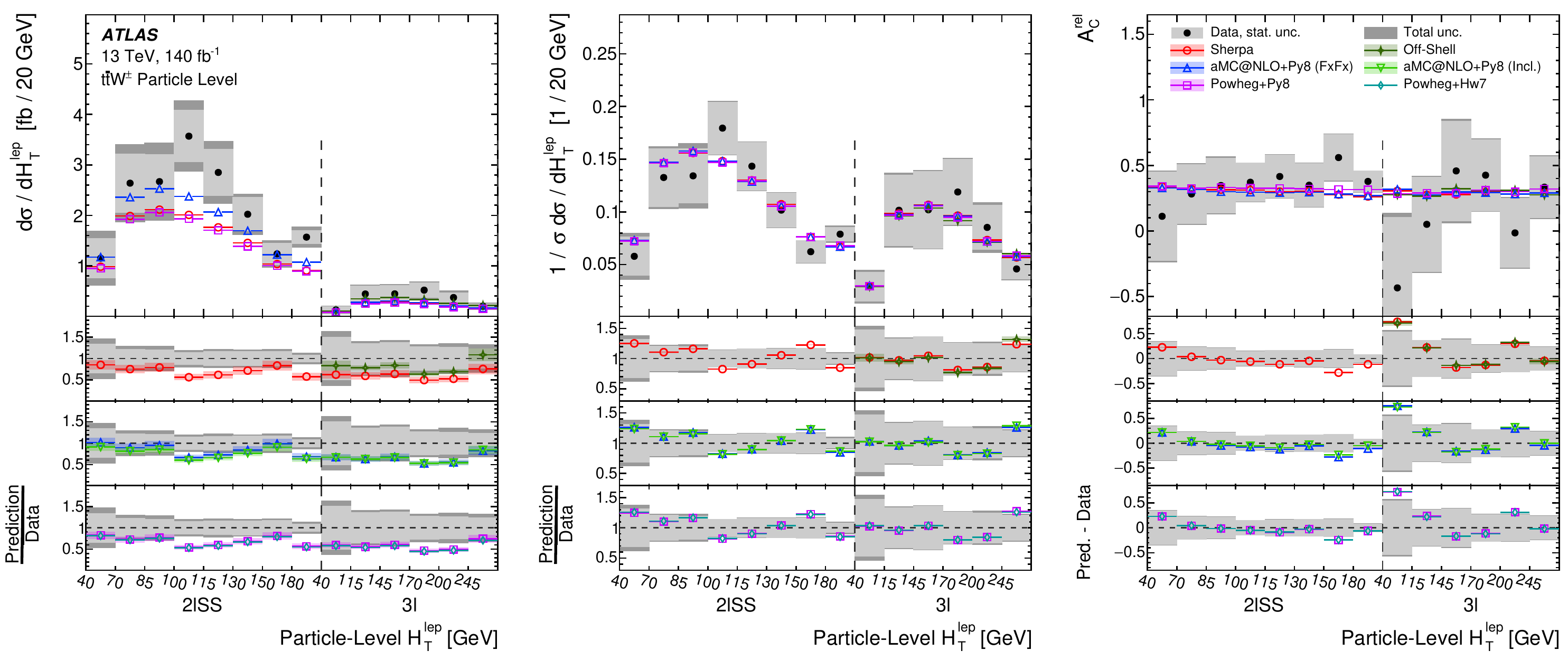}}\\
\subfloat[]{\includegraphics[width=0.90\textwidth]{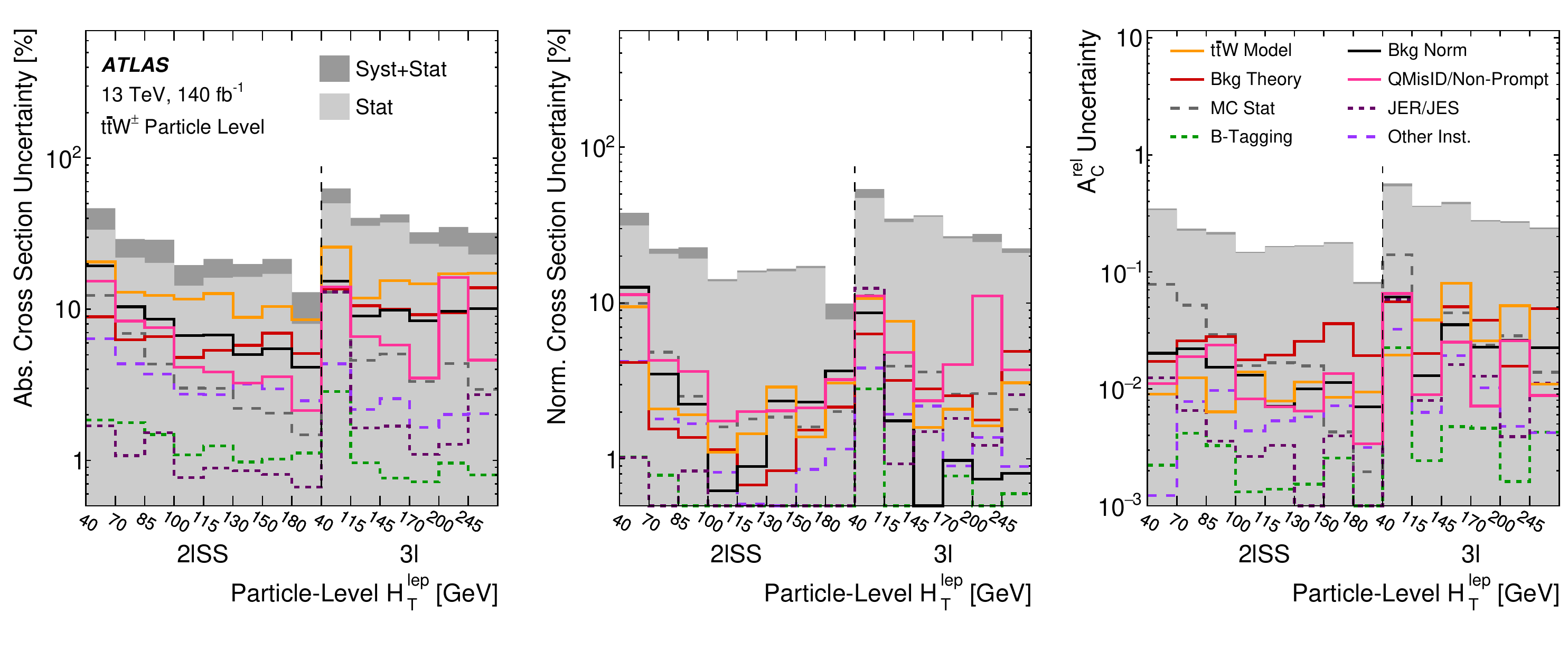}}\\
\subfloat[]{\includegraphics[width=0.90\textwidth]{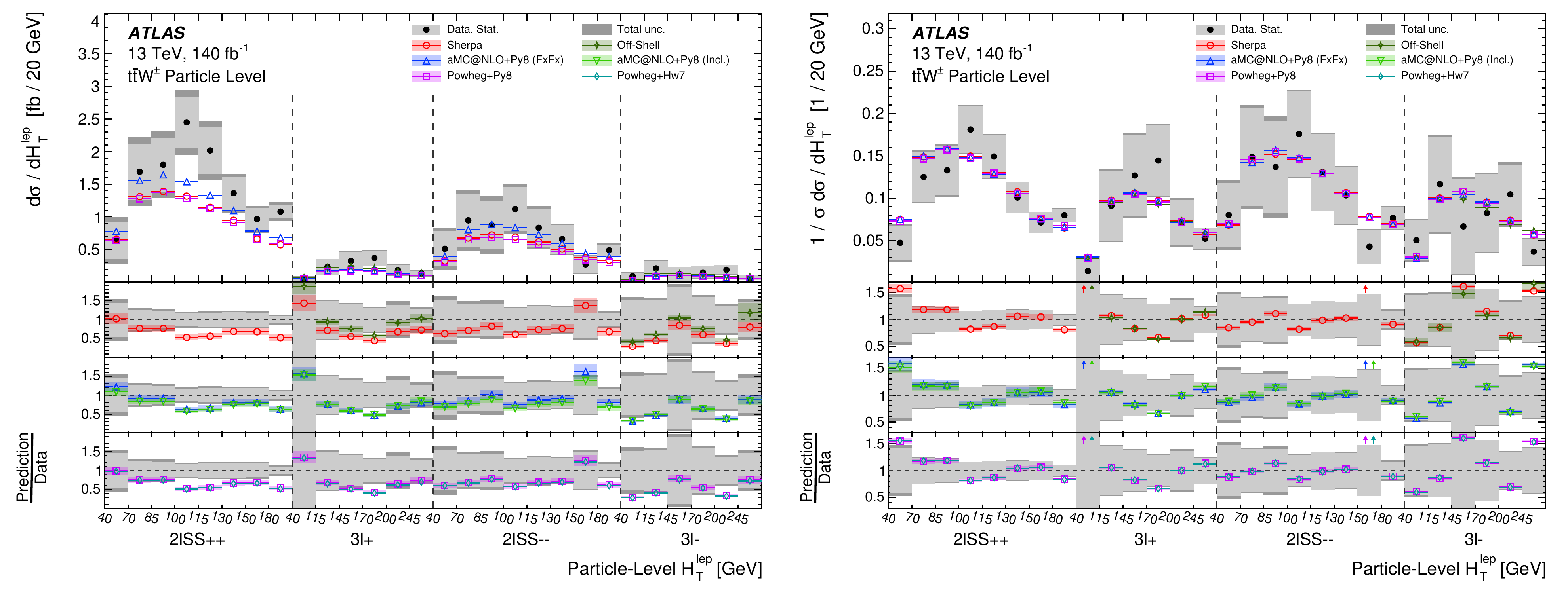}}
\caption{Unfolded distributions of (a) the lepton-charge-combined absolute \xs, the normalised \xs and \rqasym, (b) the corresponding uncertainty breakdowns, and (c) unfolded distributions of the lepton-charge-split absolute and normalised \xs{}s, as a function of \HTlep. The coloured markers overlaid upon the \xs measurements show different theoretical predictions. The breakdown of statistical-only uncertainties is shown in light grey in the bottom panels, and combined statistical and systematic uncertainties are shown in dark grey. The dotted/dashed/full lines in (b) show the relative contributions of grouped sources of systematic uncertainty. The triangles indicate points that are outside the vertical range of the figure.}
\label{fig:main:Unfolding:HT_lep}
\end{figure}
 
Figure~\ref{fig:main:Unfolding:HT_lep}(a) shows the
measured absolute \xs, normalised \xs and \rqasym as a function of \HTlep.
The corresponding uncertainties are shown in
Figure~\ref{fig:main:Unfolding:HT_lep}(b). The dominant systematic
uncertainty across the \HTlep range comes from the \ttW model, with
additional contributions from the background normalisation and lepton
charge misidentification and non-prompt-lepton BDT at lower \HTlep values.
This observable is less sensitive to the modelling of additional
jet activity in the event, so the NLO+PS predictions and the
predictions with additional partons using NLO merging give similar
results. In the \lll channel, the prediction including off-shell effects reports a slightly higher absolute \xs than
the predictions with additional partons using NLO merging and
therefore has overall better agreement with the data. The \enquote{off-shell}
$p$-value for the absolute cross-section distribution is 0.85 compared to a value of 0.61 for the \POWPY[8]
prediction.  All predictions are generally in good agreement with each
other in the normalised \xs results for \HTlep. Although the predicted
\HTlep distribution shape differs slightly from the data, this is not very
significant.  It is interesting to note that a significant decrease
in the systematic uncertainty is observed, especially at intermediate
values of \HTlep, for the normalised \xs measurement shown in
Figure~\ref{fig:main:Unfolding:HT_lep}(b). This is due to the large
decrease in several key uncertainties, including uncertainties in the \ttW model
and background normalisation and theoretical uncertainties in the
background modelling.  The values of \rqasym in
data tend to be lower than predicted in both the \llSS and \lll channels for
lower values of \HTlep, the origin of which can be seen in
Figure~\ref{fig:main:Unfolding:HT_lep}(c).

\begin{figure}[htbp]
\centering
\subfloat[]{\includegraphics[width=0.90\textwidth]{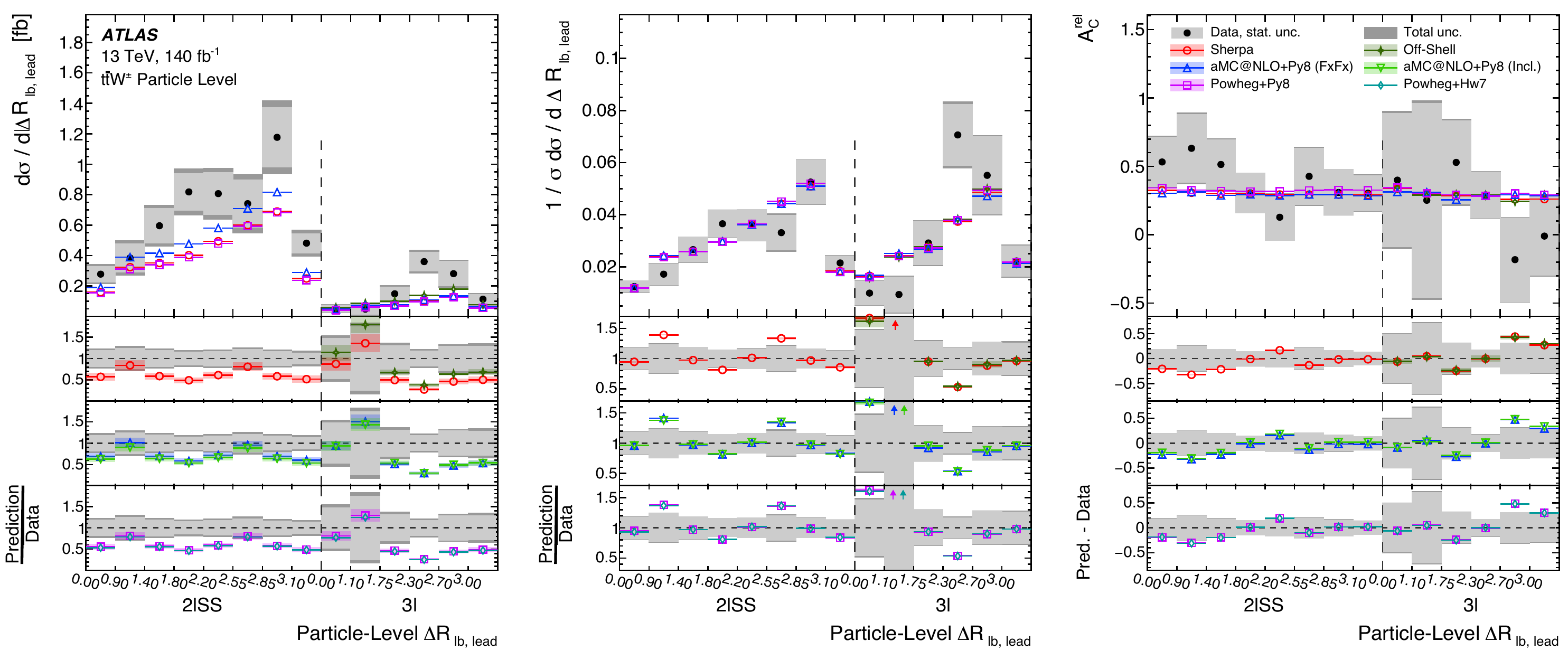}}\\
\subfloat[]{\includegraphics[width=0.90\textwidth]{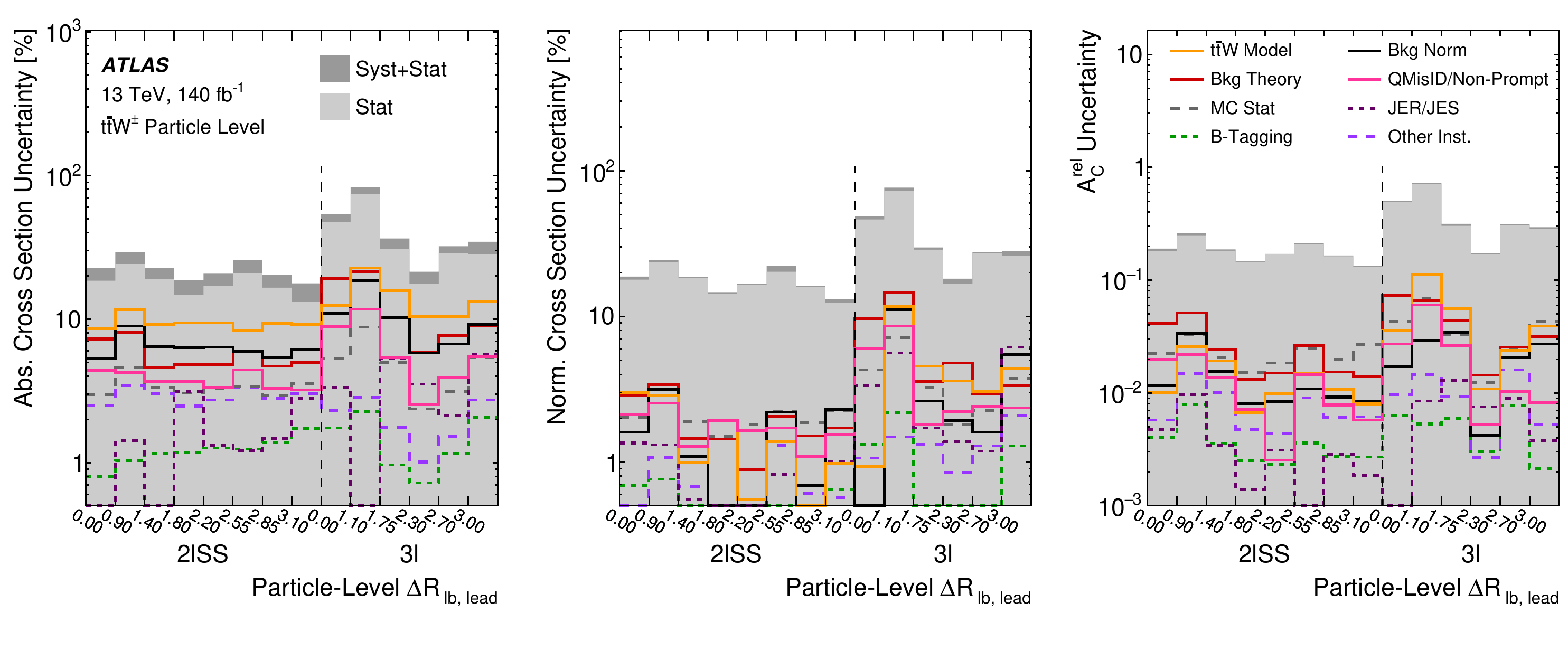}}\\
\subfloat[]{\includegraphics[width=0.90\textwidth]{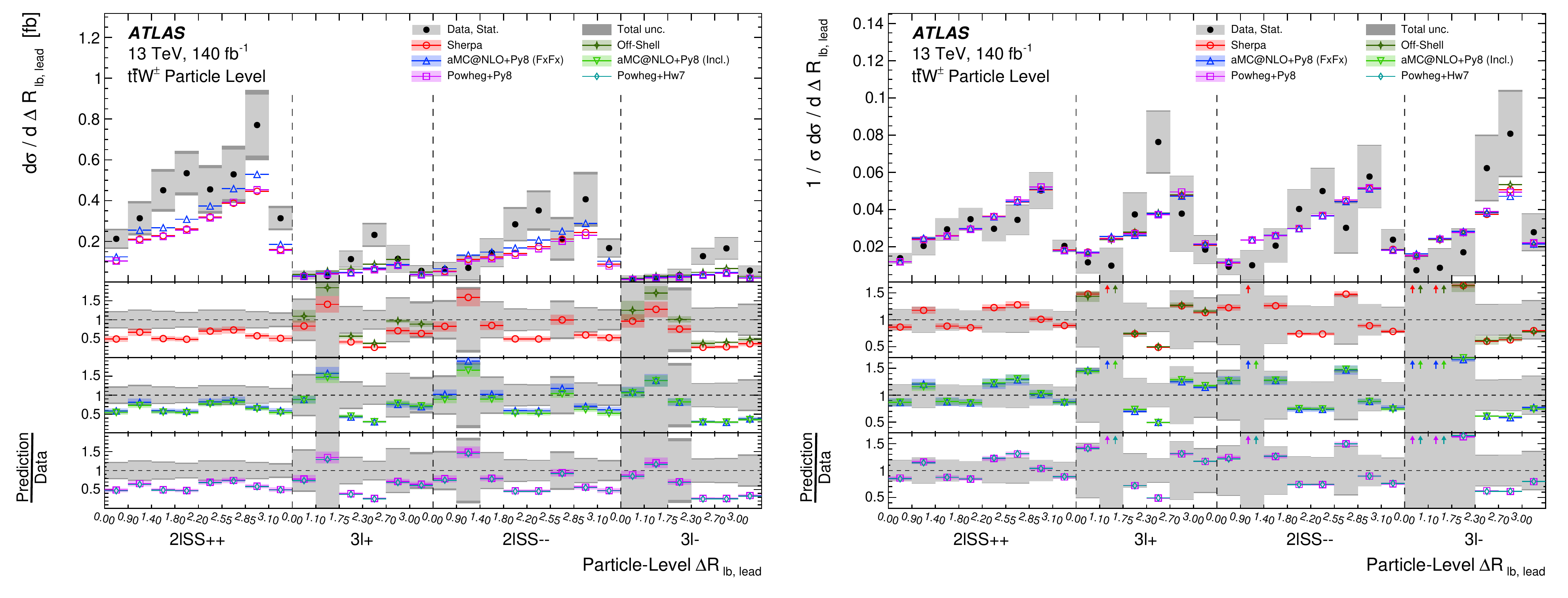}}
\caption{Unfolded distributions of (a) the lepton-charge-combined absolute \xs, the normalised \xs and \rqasym, (b) the corresponding uncertainty breakdowns, and (c) unfolded distributions of the lepton-charge-split absolute and normalised \xs{}s, as a function of \DRlb. The coloured markers overlaid upon the \xs measurements show different theoretical predictions. The breakdown of statistical-only uncertainties is shown in light grey in the bottom panels, and combined statistical and systematic uncertainties are shown in dark grey. The dotted/dashed/full lines in (b) show the relative contributions of grouped sources of systematic uncertainty. The triangles indicate points that are outside the vertical range of the figure.}
\label{fig:main:Unfolding:DRlb}
\end{figure}

Figure~\ref{fig:main:Unfolding:DRlb}(a) shows the
measured absolute \xs, normalised \xs and \rqasym as a function of \DRlb.
A significant decrease in the systematic uncertainty is observed for
the normalised \xs measurement across the range of \DRlb values, as shown in
Figure~\ref{fig:main:Unfolding:DRlb}(b).
A mild discrepancy between data and predictions is observed at
low \DRlb for \rqasym in the \llSS channel. This is because at
low \DRlb values in the \twomm channel the \xs{}s measured in data are
lower than the predictions, as shown in
Figure~\ref{fig:main:Unfolding:DRlb}(c), yielding the different
\rqasym values.

\begin{figure}[htbp]
\centering
\subfloat[]{\includegraphics[width=0.90\textwidth]{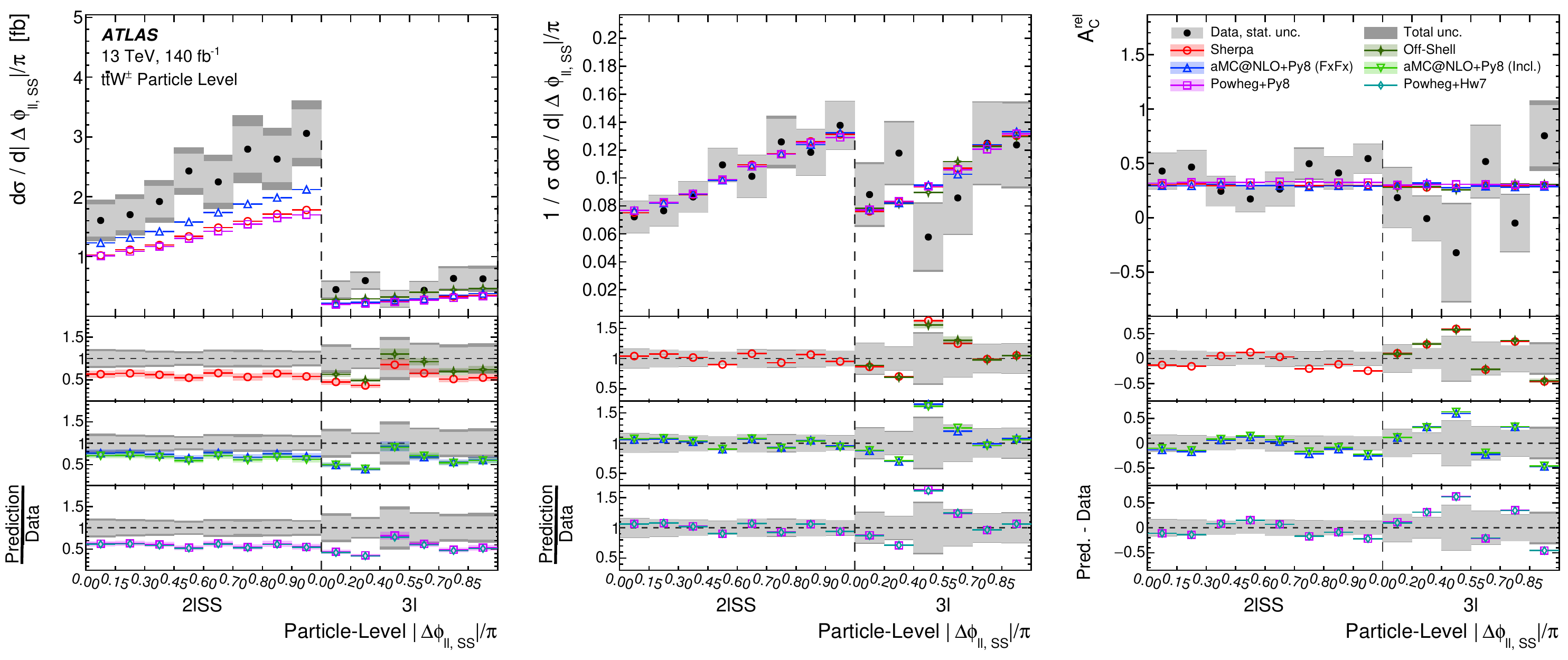}}\\
\subfloat[]{\includegraphics[width=0.90\textwidth]{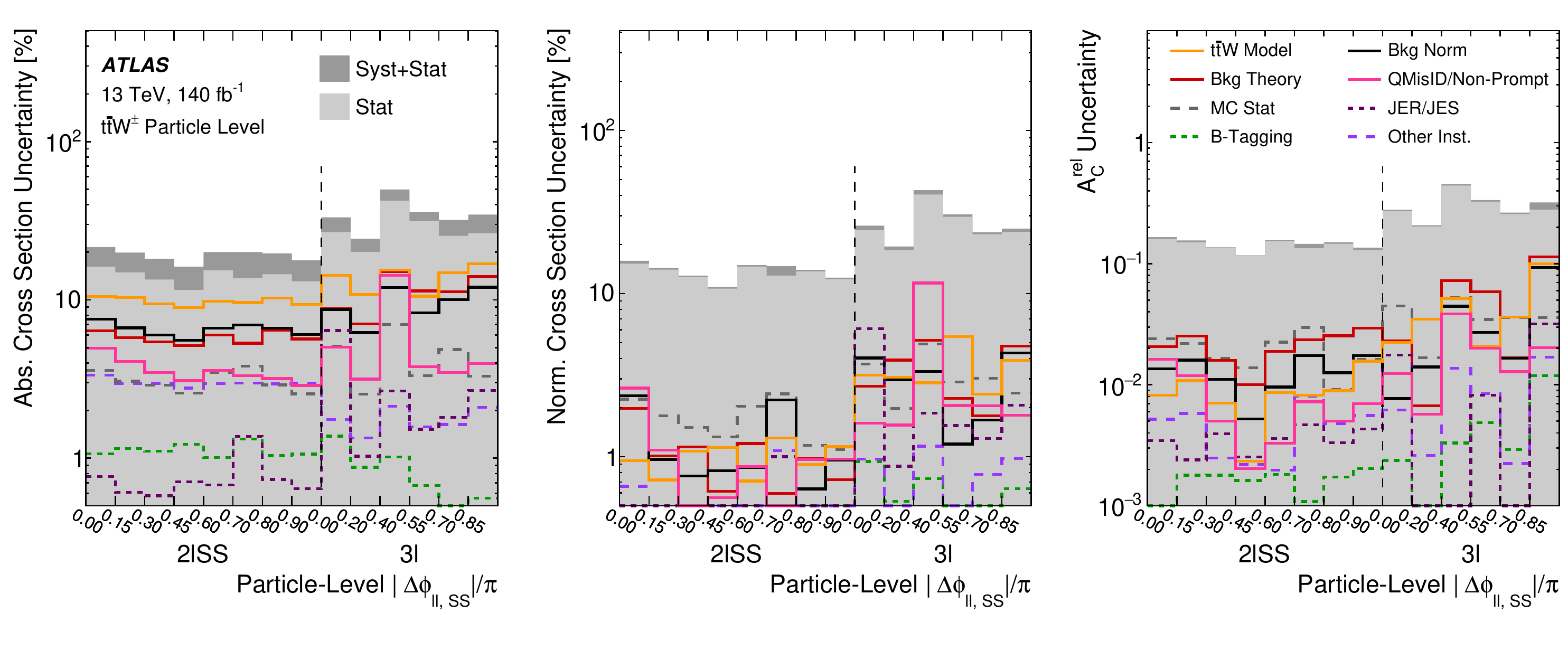}}\\
\subfloat[]{\includegraphics[width=0.90\textwidth]{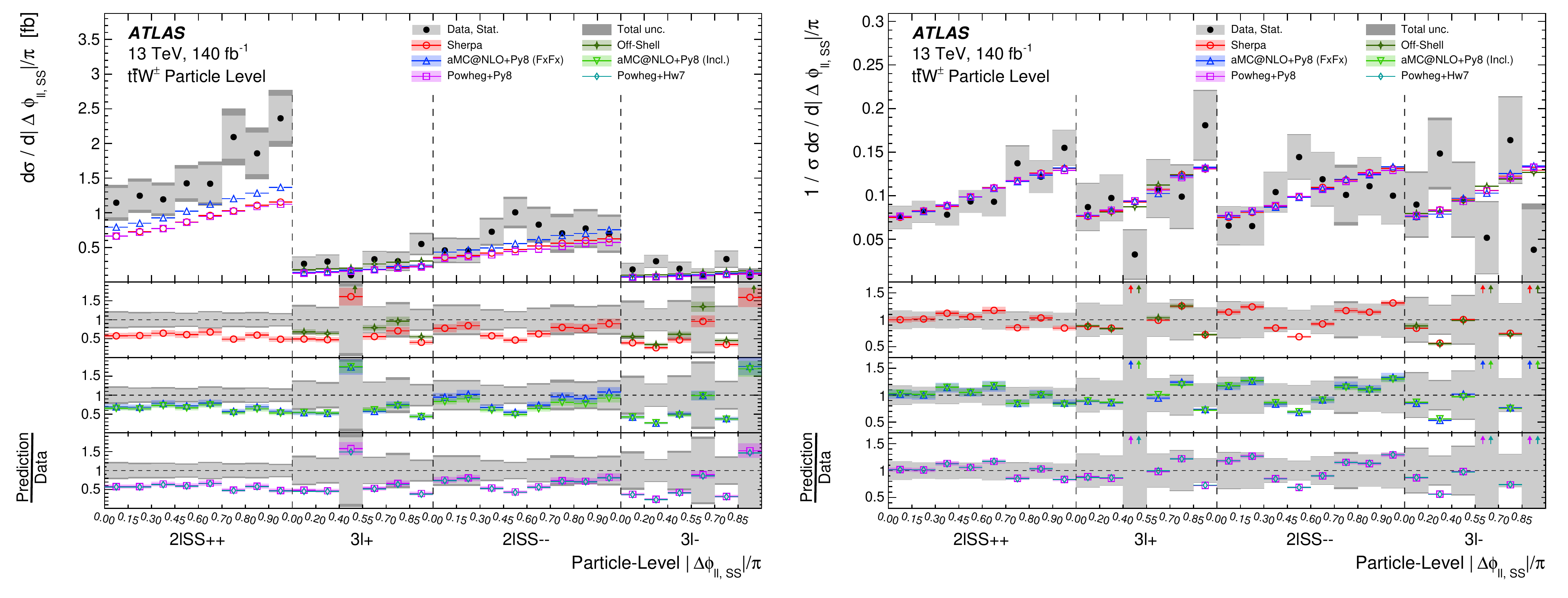}}
\caption{Unfolded distributions of (a) the lepton-charge-combined absolute \xs, the normalised \xs and \rqasym, (b) the corresponding uncertainty breakdowns, and (c) unfolded distributions of the lepton-charge-split absolute and normalised \xs{}s, as a function of \DPhillSS. The coloured markers overlaid upon the \xs measurements show different theoretical predictions. The breakdown of statistical-only uncertainties is shown in light grey in the bottom panels, and combined statistical and systematic uncertainties are shown in dark grey. The dotted/dashed/full lines in (b) show the relative contributions of grouped sources of systematic uncertainty. The triangles indicate points that are outside the vertical range of the figure.}
\label{fig:main:Unfolding:DPhillSS}
\end{figure}

Figure~\ref{fig:main:Unfolding:DPhillSS}(a) shows the
measured absolute \xs, normalised \xs and \rqasym as a function of \DPhillSS.
The normalised \xs measurements show generally good agreement between
data and the predictions in the overall shape of the distributions.
However, there is some deviation between the data and the predictions for \rqasym,
which is expected to have a constant value across \DPhillSS.
The \xs observed in data is larger than the predictions at low and
high \DPhillSS in the \twopp channel, whereas in the \twomm channel
the \xs observed in data is larger than the predictions at intermediate values
of \DPhillSS. This can be seen in
Figure~\ref{fig:main:Unfolding:DPhillSS}(c) and is reflected in the
$\chi^2$ $p$-value of 0.41 with respect to the \Sherpa prediction in Table~\ref{tab:unfolded_chi2_2lSS_asym}.

\begin{figure}[htbp]
\centering
\subfloat[]{\includegraphics[width=0.90\textwidth]{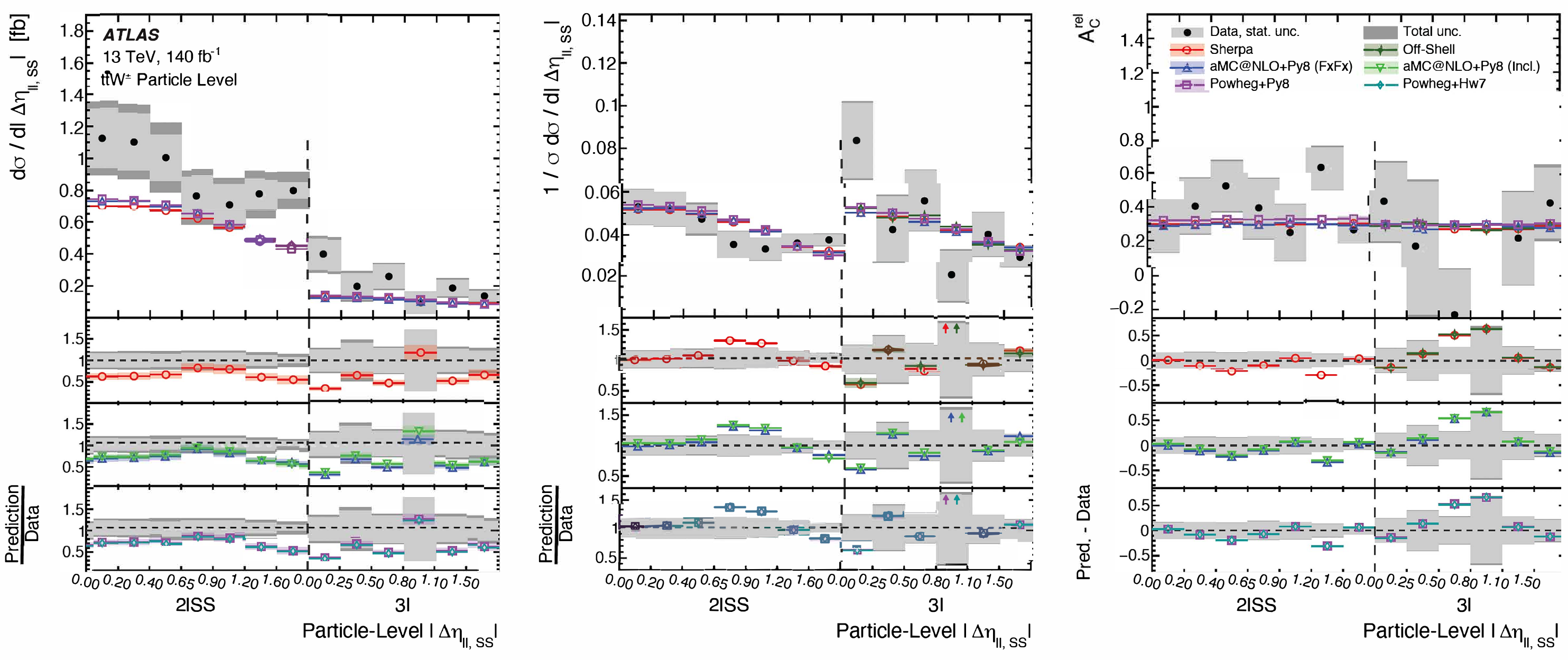}}\\
\subfloat[]{\includegraphics[width=0.90\textwidth]{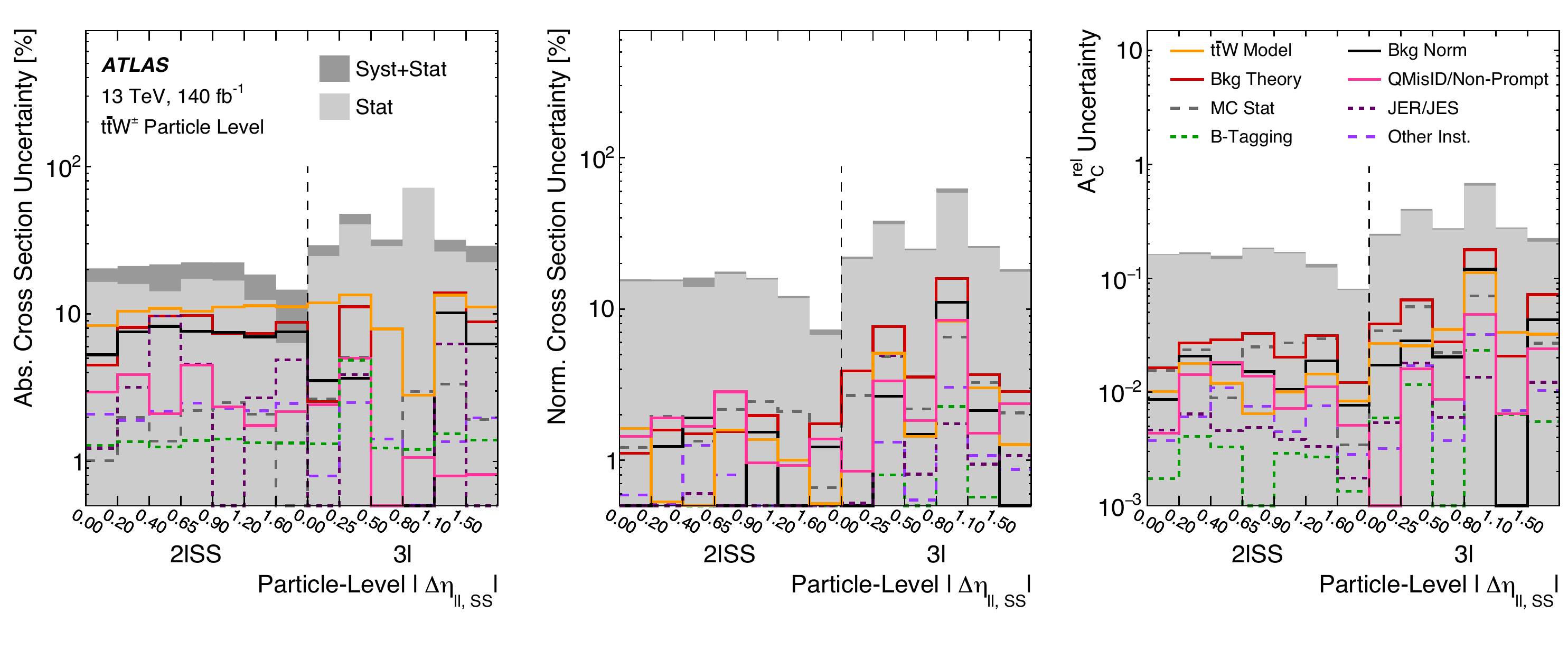}}\\
\subfloat[]{\includegraphics[width=0.90\textwidth]{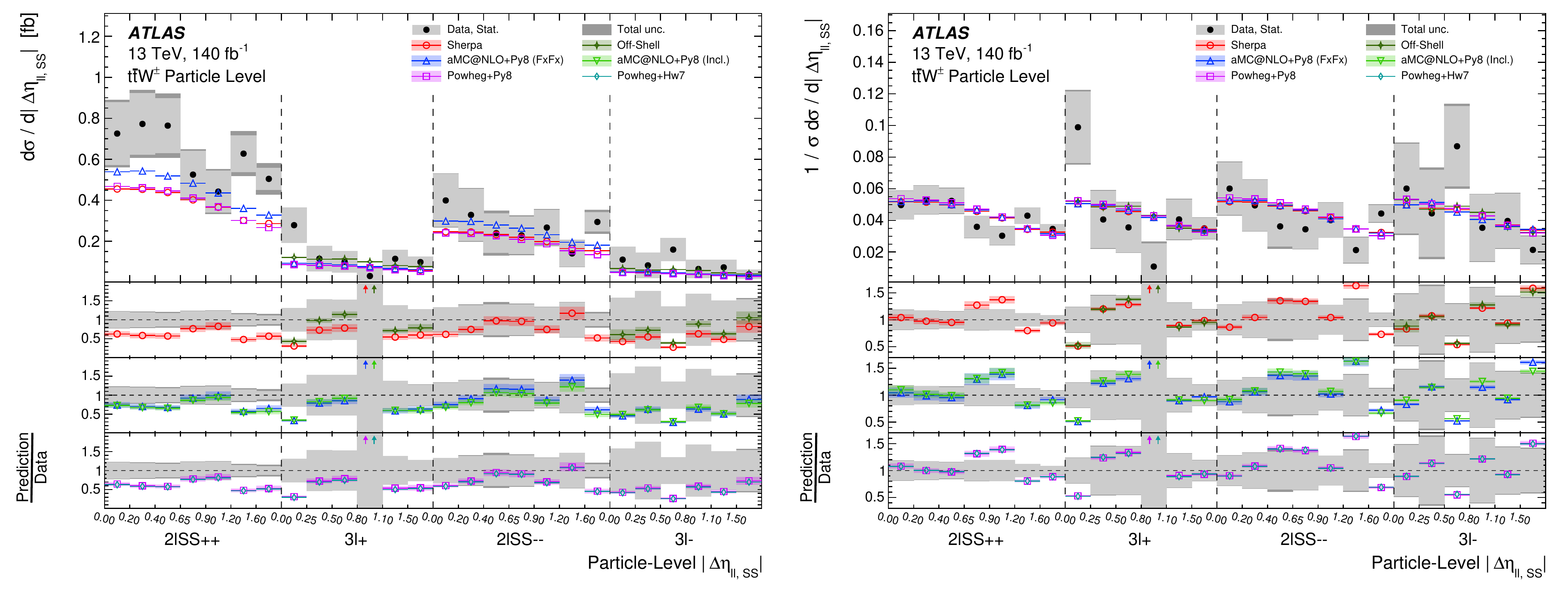}}
\caption{Unfolded distributions of (a) the lepton-charge-combined absolute \xs, the normalised \xs and \rqasym, (b) the corresponding uncertainty breakdowns, and (c) unfolded distributions of the lepton-charge-split absolute and normalised \xs{}s, as a function of \DEtallSS. The coloured markers overlaid upon the \xs measurements show different theoretical predictions. The breakdown of statistical-only uncertainties is shown in light grey in the bottom panels, and combined statistical and systematic uncertainties are shown in dark grey. The dotted/dashed/full lines in (b) show the relative contributions of grouped sources of systematic uncertainty. The triangles indicate points that are outside the vertical range of the figure.}
\label{fig:main:Unfolding:DEtallSS}
\end{figure}
 
Figure~\ref{fig:main:Unfolding:DEtallSS}(a) shows the measured
absolute \xs, normalised \xs and \rqasym as a function of \DEtallSS.
Discrepancies between the measurements in data and the theoretical
predictions can be seen in the \xs and the \rqasym distributions for
both the \llSS and \lll channels. The data show higher values of the
measured \xs compared to the theoretical predictions at high and low \DEtallSS values in the \llSS channel,
for which the $\chi^2$ $p$-value with respect to the \Sherpa prediction for the normalised \xs
measurement is 0.32 as shown in Table~\ref{tab:chi22lSSnorm}. The \rqasym measured in data is larger than the
predictions for intermediate \DEtallSS values in the \llSS channel,
whereas the opposite is seen in the \lll channel, where there are
very different trends in data and predictions for the \threep
and \threem channels, as shown in
Figure~\ref{fig:main:Unfolding:DEtallSS}(c). However, these trends are
not statistically significant.

In summary, all absolute \xs measurements show a significant
normalisation effect, as expected from the larger-than-predicted result of the inclusive \xs
measurement. Additionally, some observables show mild tension between
data and prediction in parts of the normalised \xs distributions.
However, the observed $\chi^2$ $p$-value for any single distribution
does not indicate a significant disagreement between the data and
theoretical predictions at the current accuracy of the measurements.



\FloatBarrier
 
\section{Conclusion}
\label{sec:conclusion}

Measurements of both the inclusive and differential \xs{}s for
producing a top-quark--antiquark pair in association with a $W$ boson
($t\bar{t}W$) are presented. The measurements are based on $\sqrt{s} =
13\,\mathrm{TeV}$ proton--proton collision data with an integrated
luminosity of $140\,\mathrm{fb}^{-1}$, recorded from $2015$ to $2018$
with the ATLAS detector at the Large Hadron Collider. The signal
regions are split into final states with two same-sign or three
isolated leptons (electrons or muons), and are further split according
to their total lepton charge. Additional background-enriched regions
are used in the fit to improve the modelling of several leading
backgrounds.  The inclusive \ttW production \xs is found to be
$880\pm80$~fb, which is consistent with the previous measurement from
CMS and other analyses probing similar \ttW-enriched regions.
Improved theoretical calculations have led to increases in the
predicted \xs through the corrections for missing higher-order QCD and
EWK effects.  Some tension between the predictions and the data
remains.  The measurement and the SM NNLO theoretical prediction of
$745\pm 50\,\textrm{(scale)}\pm 13\,\textrm{(2-loop approx.)}\pm
19\,\textrm{(PDF, \alphas)}$~fb are compatible at the level of
$1.4\sigma$.  The ratio of \xs{}s from different charges of the
associated $W$ boson is measured to be $R(\ttW) = 1.96\pm0.22$ and the
relative charge asymmetry is measured to be $\rqasym = 0.33\pm0.05$,
both are dominated by statistical uncertainties, but agree
with theoretical predictions.
 
Differential absolute and normalised \xs measurements are performed for several particle-level observables.
The compatibility between data and predictions for the overall normalisation of the absolute \xs measurements
is consistent with the inclusive measurement result. The latest off-shell fixed-order predictions do not address this
observed tension, concluding that the latter does not originate from the additional single-resonant and non-resonant
contributions that this calculation includes on top of the double-resonant contribution contained in the predictions
matched to parton showers.
 
The compatibility of the data with theoretical predictions from different
MC generators is tested; for the normalised \xs measurements, compatibility is
demonstrated by $\chi^2$ $p$-values above 0.3 for the \Sherpa predictions
for almost all observables.


\section*{Acknowledgements}


We thank CERN for the very successful operation of the LHC and its injectors, as well as the support staff at
CERN and at our institutions worldwide without whom ATLAS could not be operated efficiently.
 
The crucial computing support from all WLCG partners is acknowledged gratefully, in particular from CERN, the ATLAS Tier-1 facilities at TRIUMF/SFU (Canada), NDGF (Denmark, Norway, Sweden), CC-IN2P3 (France), KIT/GridKA (Germany), INFN-CNAF (Italy), NL-T1 (Netherlands), PIC (Spain), RAL (UK) and BNL (USA), the Tier-2 facilities worldwide and large non-WLCG resource providers. Major contributors of computing resources are listed in Ref.~\cite{ATL-SOFT-PUB-2023-001}.
 
We gratefully acknowledge the support of ANPCyT, Argentina; YerPhI, Armenia; ARC, Australia; BMWFW and FWF, Austria; ANAS, Azerbaijan; CNPq and FAPESP, Brazil; NSERC, NRC and CFI, Canada; CERN; ANID, Chile; CAS, MOST and NSFC, China; Minciencias, Colombia; MEYS CR, Czech Republic; DNRF and DNSRC, Denmark; IN2P3-CNRS and CEA-DRF/IRFU, France; SRNSFG, Georgia; BMBF, HGF and MPG, Germany; GSRI, Greece; RGC and Hong Kong SAR, China; ISF and Benoziyo Center, Israel; INFN, Italy; MEXT and JSPS, Japan; CNRST, Morocco; NWO, Netherlands; RCN, Norway; MEiN, Poland; FCT, Portugal; MNE/IFA, Romania; MESTD, Serbia; MSSR, Slovakia; ARRS and MIZ\v{S}, Slovenia; DSI/NRF, South Africa; MICINN, Spain; SRC and Wallenberg Foundation, Sweden; SERI, SNSF and Cantons of Bern and Geneva, Switzerland; MOST, Taipei; TENMAK, T\"urkiye; STFC, United Kingdom; DOE and NSF, United States of America.
 
Individual groups and members have received support from BCKDF, CANARIE, CRC and DRAC, Canada; PRIMUS 21/SCI/017 and UNCE SCI/013, Czech Republic; COST, ERC, ERDF, Horizon 2020, ICSC-NextGenerationEU and Marie Sk{\l}odowska-Curie Actions, European Union; Investissements d'Avenir Labex, Investissements d'Avenir Idex and ANR, France; DFG and AvH Foundation, Germany; Herakleitos, Thales and Aristeia programmes co-financed by EU-ESF and the Greek NSRF, Greece; BSF-NSF and MINERVA, Israel; Norwegian Financial Mechanism 2014-2021, Norway; NCN and NAWA, Poland; La Caixa Banking Foundation, CERCA Programme Generalitat de Catalunya and PROMETEO and GenT Programmes Generalitat Valenciana, Spain; G\"{o}ran Gustafssons Stiftelse, Sweden; The Royal Society and Leverhulme Trust, United Kingdom.
 
In addition, individual members wish to acknowledge support from Chile: Agencia Nacional de Investigación y Desarrollo (FONDECYT 1190886, FONDECYT 1210400); China: National Natural Science Foundation of China (NSFC - 12175119, NSFC 12275265); EU: H2020 European Research Council (H2020-MSCA-IF-2020: HPOFHIC - 10103); European Union: European Research Council (ERC - 948254), Horizon 2020 Framework Programme (MUCCA - CHIST-ERA-19-XAI-00), Italian Center for High Performance Computing, Big Data and Quantum Computing (ICSC, NextGenerationEU), Marie Sklodowska-Curie Actions (EU H2020 MSC IF GRANT NO 101033496); France: Agence Nationale de la Recherche (ANR-20-CE31-0013, ANR-21-CE31-0022), Investissements d'Avenir Idex (ANR-11-LABX-0012), Investissements d'Avenir Labex (ANR-11-LABX-0012); Germany: Baden-Württemberg Stiftung (BW Stiftung-Postdoc Eliteprogramme), Deutsche Forschungsgemeinschaft (DFG - CR 312/5-1); Italy: Istituto Nazionale di Fisica Nucleare (FELLINI G.A. n. 754496, ICSC, NextGenerationEU); Japan: Japan Society for the Promotion of Science (JSPS KAKENHI 22H01227, JSPS KAKENHI JP21H05085, JSPS KAKENHI JP22H04944); Netherlands: Netherlands Organisation for Scientific Research (NWO Veni 2020 - VI.Veni.202.179); Norway: Research Council of Norway (RCN-314472); Poland: Polish National Agency for Academic Exchange (PPN/PPO/2020/1/00002/U/00001), Polish National Science Centre (NCN 2021/42/E/ST2/00350, NCN UMO-2019/34/E/ST2/00393, UMO-2020/37/B/ST2/01043, UMO-2021/40/C/ST2/00187); Slovenia: Slovenian Research Agency (ARIS grant J1-3010); Spain: BBVA Foundation (LEO22-1-603), Generalitat Valenciana (Artemisa, FEDER, IDIFEDER/2018/048), La Caixa Banking Foundation (LCF/BQ/PI20/11760025), Ministry of Science and Innovation (RYC2019-028510-I, RYC2020-030254-I), PROMETEO and GenT Programmes Generalitat Valenciana (CIDEGENT/2019/023, CIDEGENT/2019/027); Sweden: Swedish Research Council (VR 2022-03845), Knut and Alice Wallenberg Foundation (KAW 2017.0100, KAW 2018.0157, KAW 2019.0447); Switzerland: Swiss National Science Foundation (SNSF - PCEFP2\_194658); United Kingdom: Leverhulme Trust (Leverhulme Trust RPG-2020-004); United States of America: Neubauer Family Foundation.


\clearpage
\printbibliography
 
\clearpage
 
\begin{flushleft}
\hypersetup{urlcolor=black}
{\Large The ATLAS Collaboration}

\bigskip

\AtlasOrcid[0000-0002-6665-4934]{G.~Aad}$^\textrm{\scriptsize 102}$,
\AtlasOrcid[0000-0002-5888-2734]{B.~Abbott}$^\textrm{\scriptsize 120}$,
\AtlasOrcid[0000-0002-1002-1652]{K.~Abeling}$^\textrm{\scriptsize 55}$,
\AtlasOrcid[0000-0001-5763-2760]{N.J.~Abicht}$^\textrm{\scriptsize 49}$,
\AtlasOrcid[0000-0002-8496-9294]{S.H.~Abidi}$^\textrm{\scriptsize 29}$,
\AtlasOrcid[0000-0002-9987-2292]{A.~Aboulhorma}$^\textrm{\scriptsize 35e}$,
\AtlasOrcid[0000-0001-5329-6640]{H.~Abramowicz}$^\textrm{\scriptsize 151}$,
\AtlasOrcid[0000-0002-1599-2896]{H.~Abreu}$^\textrm{\scriptsize 150}$,
\AtlasOrcid[0000-0003-0403-3697]{Y.~Abulaiti}$^\textrm{\scriptsize 117}$,
\AtlasOrcid[0000-0003-0762-7204]{A.C.~Abusleme~Hoffman}$^\textrm{\scriptsize 137a}$
% ...
\bigskip
\\

% ...
$^{29}$Physics Department, Brookhaven National Laboratory, Upton NY; United States of America.\\
% ...
$^{35}$$^{(a)}$Facult\'e des Sciences Ain Chock, R\'eseau Universitaire de Physique des Hautes Energies - Universit\'e Hassan II, Casablanca;$^{(b)}$Facult\'{e} des Sciences, Universit\'{e} Ibn-Tofail, K\'{e}nitra;$^{(c)}$Facult\'e des Sciences Semlalia, Universit\'e Cadi Ayyad, LPHEA-Marrakech;$^{(d)}$LPMR, Facult\'e des Sciences, Universit\'e Mohamed Premier, Oujda;$^{(e)}$Facult\'e des sciences, Universit\'e Mohammed V, Rabat;$^{(f)}$Institute of Applied Physics, Mohammed VI Polytechnic University, Ben Guerir; Morocco.\\
% ...
$^{49}$Fakult\"{a}t Physik , Technische Universit{\"a}t Dortmund, Dortmund; Germany.\\
% ...
$^{55}$II. Physikalisches Institut, Georg-August-Universit\"{a}t G\"ottingen, G\"ottingen; Germany.\\
% ...
$^{102}$CPPM, Aix-Marseille Universit\'e, CNRS/IN2P3, Marseille; France.\\
% ...
$^{117}$Department of Physics, New York University, New York NY; United States of America.\\
% ...
$^{120}$Homer L. Dodge Department of Physics and Astronomy, University of Oklahoma, Norman OK; United States of America.\\
% ...
$^{137}$$^{(a)}$Departamento de F\'isica, Pontificia Universidad Cat\'olica de Chile, Santiago;$^{(b)}$Millennium Institute for Subatomic physics at high energy frontier (SAPHIR), Santiago;$^{(c)}$Instituto de Investigaci\'on Multidisciplinario en Ciencia y Tecnolog\'ia, y Departamento de F\'isica, Universidad de La Serena;$^{(d)}$Universidad Andres Bello, Department of Physics, Santiago;$^{(e)}$Instituto de Alta Investigaci\'on, Universidad de Tarapac\'a, Arica;$^{(f)}$Departamento de F\'isica, Universidad T\'ecnica Federico Santa Mar\'ia, Valpara\'iso; Chile.\\
% ...
$^{150}$Department of Physics, Technion, Israel Institute of Technology, Haifa; Israel.\\
$^{151}$Raymond and Beverly Sackler School of Physics and Astronomy, Tel Aviv University, Tel Aviv; Israel.\\
% ...
$^{a}$ Also Affiliated with an institute covered by a cooperation agreement with CERN.\\
$^{b}$ Also at An-Najah National University, Nablus; Palestine.\\
$^{c}$ Also at Borough of Manhattan Community College, City University of New York, New York NY; United States of America.\\
$^{d}$ Also at Center for High Energy Physics, Peking University; China.\\
$^{e}$ Also at Center for Interdisciplinary Research and Innovation (CIRI-AUTH), Thessaloniki; Greece.\\
$^{f}$ Also at Centro Studi e Ricerche Enrico Fermi; Italy.\\
% ...

\end{flushleft}


\end{document}